\documentclass[twocolumn]{aastex62}
\usepackage{graphicx, natbib, hyperref, float} 
\usepackage[flushleft]{threeparttable} 

\hypersetup{linkcolor=red,citecolor=blue,filecolor=cyan,urlcolor=blue}

\received{xx} \revised{xx}
\accepted{xx}
\submitjournal{ApJ}

\shorttitle{ASPECS: Properties of faint dusty galaxies} \shortauthors{Aravena et al.}

   \def\lsim{\mathrel{\rlap{\lower 3pt \hbox{$\sim$}} \raise 2.0pt \hbox{$<$}}} \def\gsim{\mathrel{\rlap{\lower 3pt
\hbox{$\sim$}} \raise 2.0pt \hbox{$>$}}}
\newcommand{\angstrom}{\mbox{\normalfont\AA}}

\begin{document}

\title{The ALMA Spectroscopic Survey in the Hubble Ultra Deep Field: The nature of the faintest dusty star-forming galaxies}

\author[0000-0002-6290-3198]{Manuel Aravena}\email{manuel.aravenaa@mail.udp.cl} \affil{N\'{u}cleo de Astronom\'{\i}a, Facultad de Ingenier\'{\i}a y Ciencias, Universidad Diego Portales, Av. Ej\'{e}rcito 441, Santiago, Chile}

\author[0000-0002-3952-8588]{Leindert Boogaard} \affil{Leiden Observatory, Leiden University, P. O. Box 9513, NL2300 RA Leiden, The Netherlands}

\author[0000-0003-3926-1411]{Jorge G\'onzalez-L\'opez} \affil{N\'{u}cleo de Astronom\'{\i}a, Facultad de Ingenier\'{\i}a y Ciencias, Universidad Diego Portales, Av. Ej\'{e}rcito 441, Santiago, Chile}

\author[0000-0002-2662-8803]{Roberto Decarli} \affil{INAF Osservatorio di Astrofisica e Scienza dello Spazio, via Gobetti 93/3, I-40129, Bologna, Italy}

\author[0000-0003-4793-7880]{Fabian Walter} \affil{Max Planck Institute f\"ur Astronomie, K\"onigstuhl 17, D-69117 Heidelberg, Germany} \affil{National Radio Astronomy Observatory, Pete V. Domenici Array Science Center, P.O. Box O, Socorro, NM 87801, USA}
\author[0000-0001-6647-3861]{Chris L. Carilli} \affil{National Radio Astronomy Observatory, Pete V. Domenici Array Science Center, P. O. Box O, Socorro, NM 87801, USA}
\author[0000-0003-3037-257X]{Ian Smail} \affil{Centre for Extragalactic Astronomy, Department of Physics, Durham University, South Road, Durham, DH1 3LE, UK}
\author[0000-0003-4678-3939]{Axel Weiss} \affil{Max-Planck-Institut f\"ur Radioastronomie, Auf dem H\"ugel 69, D-53121 Bonn, Germany}
\author[0000-0002-9508-3667]{Roberto J. Assef} \affil{N\'{u}cleo de Astronom\'{\i}a, Facultad de Ingenier\'{\i}a, Universidad Diego Portales, Av. Ej\'{e}rcito 441, Santiago, Chile}
%
%
\author[0000-0002-8686-8737]{Franz Erik Bauer} \affil{Instituto de Astrof\'{\i}sica, Facultad de F\'{\i}sica, Pontificia Universidad Cat\'olica de Chile Av. Vicu\~na Mackenna 4860, 782-0436 Macul,
Santiago, Chile} \affil{Millennium Institute of Astrophysics (MAS), Nuncio Monse{\~{n}}or S{\'{o}}tero Sanz 100, Providencia, Santiago, Chile} 
%
%
\author[0000-0002-4989-2471]{Rychard J. Bouwens} \affil{Leiden Observatory, Leiden University, P. O. Box 9513, NL2300 RA Leiden, The Netherlands}
%
%
\author[0000-0002-3583-780X]{Paulo C.~Cortes} \affil{Joint ALMA Observatory - ESO, Av. Alonso de C\'ordova, 3104, Santiago, Chile} \affil{National Radio Astronomy Observatory, 520 Edgemont Rd., Charlottesville, VA 22903, USA} 
\author[0000-0003-2027-8221]{Pierre Cox} \affil{Sorbonne Universit\'e, UPMC Universit\'e Paris 6 and CNRS, UMR 7095, Institut d'Astrophysique de Paris, 98bis boulevard Arago, F-75015 Paris, France}
\author[0000-0001-9759-4797]{Elisabete da Cunha} \affil{International Centre for Radio Astronomy Research, University of Western Australia, 35 Stirling Hwy., Crawley, WA 6009, Australia} \affil{Research School of Astronomy and Astrophysics, The Australian National University, Canberra, ACT 2611, Australia} \affil{ARC Centre of Excellence for All Sky Astrophysics in 3 Dimensions (ASTRO 3D)}
\author[0000-0002-3331-9590]{Emanuele Daddi} \affil{Laboratoire AIM, CEA/DSM-CNRS-Universite Paris Diderot, Irfu/Service d'Astrophysique, CEA Saclay, Orme des Merisiers, F-91191 Gif-sur-Yvette cedex, France}
\author[0000-0003-0699-6083]{Tanio D\'iaz-Santos} \affil{N\'{u}cleo de Astronom\'{\i}a, Facultad de Ingenier\'{\i}a y Ciencias, Universidad Diego Portales, Av. Ej\'{e}rcito 441, Santiago, Chile} \affil{Chinese Academy of Sciences South America Center for Astronomy (CASSACA), National Astronomical Observatories, CAS, Beijing 100101, China} \affil{Institute of Astrophysics, Foundation for Research and Technology-Hellas (FORTH), Heraklion, GR-70013, Greece}
%
%
%
\author[0000-0003-4268-0393	]{Hanae Inami} \affil{Hiroshima Astrophysical Science Center, Hiroshima University, 1-3-1 Kagamiyama, Higashi-Hiroshima, Hiroshima 739-8526, Japan}
\author[0000-0001-5118-1313]{Rob Ivison} \affil{European Southern Observatory, Karl-Schwarzschild-Strasse 2, D-85748, Garching, Germany} 
%
%
%
\author[0000-0001-8695-825X]{Mladen Novak} \affil{Max Planck Institute f\"ur Astronomie, K\"onigstuhl 17, D-69117 Heidelberg, Germany}
\author[0000-0003-1151-4659]{Gerg\"{o} Popping} \affil{European Southern Observatory, Karl-Schwarzschild-Strasse 2, D-85748, Garching, Germany}
%
\author[0000-0001-9585-1462]{Dominik Riechers} \affil{Cornell University, 220 Space Sciences Building, Ithaca, NY 14853, USA} \affil{Max Planck Institute f\"ur Astronomie, K\"onigstuhl 17, D-69117 Heidelberg, Germany}
%
%
%
%
%
%
%
\author[0000-0001-5434-5942]{Paul van der Werf} \affil{Leiden Observatory, Leiden University, P. O. Box 9513, NL2300 RA Leiden, The Netherland}
\author[0000-0002-9258-1468]{Jeff Wagg} \affil{SKA Organization, Lower Withington Macclesfield, Cheshire SK11 9DL, UK}
%
%
%
%
%
%
\begin{abstract} 
We present a characterization of the physical properties of a sample of 35 securely detected, dusty galaxies in the deep ALMA 1.2 mm image obtained as part of the ALMA Spectroscopic Survey in the Hubble Ultra Deep Field (ASPECS) Large Program. This sample is complemented by 26 additional sources identified via an optical/infrared source positional prior. Using their well-characterized spectral energy distributions, we derive median stellar masses and star formation rates (SFR) of $4.8\times10^{10}~M_\sun$ and 30 $M_\sun$ yr$^{-1}$, respectively, and interquartile ranges of $(2.4-11.7)\times10^{10}~M_\sun$ and $20-50~M_\sun$ yr$^{-1}$. We derive a median spectroscopic redshift of 1.8 with an interquartile range $1.1-2.6$, significantly lower than submillimeter galaxies detected in shallower, wide-field surveys. We find that 59\%$\pm$13\%, 6\%$\pm$4\%, and 34\%$\pm$9\% of our sources are within, above, and below $\pm0.4$ dex from the SFR$-$stellar-mass relation or main sequence (MS), respectively. The ASPECS galaxies closely follow the SFR$-$molecular gas mass relation and other previously established scaling relations, confirming a factor of five increase of the gas-to-stellar-mass ratio from $z=0.5$ to $2.5$ and a mild evolution of the gas depletion timescales with a typical value of 0.7 Gyr at $z=1-3$. ASPECS galaxies located significantly below the MS, a poorly exploited parameter space, have low gas-to-stellar-mass ratios of $\sim0.1-0.2$ and long depletion timescales $>1$ Gyr. Galaxies along the MS dominate the cosmic density of molecular gas at all redshifts. Systems above the MS have an increasing contribution to the total gas reservoirs from $z<1$ to $z=2.5$, while the opposite is found for galaxies below the MS. 
\end{abstract} \keywords{galaxies: evolution --- galaxies: ISM --- galaxies: star-formation --- galaxies: statistics --- submillimeter: galaxies}

\section{Introduction}

%
%
A major focus of galaxy evolution studies in the past few decades has been to understand the physical mechanisms that drive the growth of galaxies, starting from cold atomic hydrogen
through the cold interstellar medium (ISM; e.g. where the bulk of dense molecular hydrogen, H$_2$, resides) to star formation. A critical measurement has been the determination of the evolution of the cosmic star formation rate (SFR) density, establishing an important framework to understand the key epochs in which the different physical mechanisms are taking place \citep[e.g.,][]{madau14}. Due to the direct link between the reservoirs of molecular gas in the universe and star formation activity, understanding this cosmic evolution of galaxies requires measurements of the cold ISM through cold dust and molecular gas observations \citep[e.g.,][]{carilli13}.

To study these processes, several approaches have been developed to select a variety of galaxy populations in which signatures of the ISM content can be observed.

Large (sub)millimeter continuum surveys with bolometer cameras on single-dish telescopes have been conducted over the past two decades, covering significant contiguous areas of the sky, ranging from a few tens of arcmin$^2$ to thousands of deg$^2$ \citep[e.g.,][]{barger98, hughes98, bertoldi00, eales00, cowie02,scott02, coppin06, bertoldi07, greve08, scott08, weiss09, austermann10, vieira10, negrello10, aretxaga11, hatsukade11, scott12, mocanu13, geach17, wang17, simpson19}. By construction, these surveys tend to detect the redshifted far-IR emission from galaxies at $z>1$ and thus trace the dust reservoirs heated by ultraviolet (UV) radiation from star formation or active galactic nucleus (AGN) activity, hence preferentially selecting galaxies with high SFRs and/or substantial dust (and molecular gas) reservoirs \citep[e.g.,][]{casey14}. These efforts yielded the discovery of a population of luminous dusty star-forming galaxies (DSFGs) at high redshift that were not accounted for in previous optical cosmological surveys. These galaxies, commonly called submillimeter galaxies (SMGs), have SFRs $>200$ $M_\sun$ yr$^{-1}$ and assembled a significant fraction of their stellar content with $M_{\rm star}\sim10^{10-11}\ M_\sun$ \citep[e.g.,][]{simpson14, swinbank14, dacunha15, chang18, zavala18, dudzevicute20}. 

An important step to understanding the physical mechanisms of galaxy growth has been the determination that the bulk of star-forming galaxies, typically selected through their optical colors, form a broad correlation or `sequence' between their stellar masses and SFRs, representing what has been called the ``main-sequence'' (MS) of star formation \citep[e.g.,][]{brinchmann04, daddi07, elbaz07, elbaz11, noeske07, peng10, rodighiero10, whitaker12, whitaker14, schreiber15, speagle14}. Galaxies below this sequence are usually called `passive' galaxies, while those above it are called ``starbursts''. Follow-up observations of dust continuum and molecular gas, through $^{12}$CO line emission, in optically selected MS galaxies across the stellar mass versus SFR diagram yielded a revolution in the study of galaxy evolution: these star-forming MS galaxies were found to host large amounts of molecular gas, yielding bright detections of CO line emission \citep[e.g.,][]{daddi10a}. This targeted approach enabled the study of galaxies with faint dust continuum emission that blind bolometer surveys were unable to explore. Most significantly, these studies allowed for the determination of scaling relations between various fundamental parameters, including their stellar masses, specific SFRs, molecular gas depletion timescales, and molecular gas fractions, revealing for some of these parameters clear signs of evolution with redshift \citep[e.g.,][]{daddi10b, tacconi10, tacconi13, tacconi18, sargent14, genzel10, genzel15, leroy13, saintonge11a, saintonge13, saintonge16,santini14, papovich16, schinnerer16, scoville16, scoville17, wiklind19, liu19, freundlich19}. These observations have established a basis for an observational and theoretical framework of galaxy evolution.

Despite the important progress made, dust continuum surveys have only been able to detect directly the (sub)millimeter brightest galaxies, missing the general population of (dusty) star-forming galaxies through cosmic time. Targeted CO/dust observations across the stellar mass vs. SFR diagram, on the other hand, have focused on galaxies preselected through their optical colors, potentially missing a significant fraction of the galaxy population \citep[e.g.,][]{aravena19}.

The advent of the Atacama Large Millimeter/submillimeter Array (ALMA) has opened a new window for studying cold dust and molecular gas in the general population of
star-forming galaxies, enabling us for the first time to produce sensitive blank-field dust continuum and CO line emission searches over significant contiguous areas of the sky
($\sim1-50$ arcmin$^2$). These efforts have mostly been done in well-studied cosmological deep fields in order to take advantage of the wealth of multiwavelength data \citep[e.g.,][]{aravena16b, franco18, gonzalezlopez17, dunlop17, hatsukade18, pavesi18, riechers19}, as well as fields known to be located in galaxy protoclusters at high redshift \citep[e.g.,][]{umehata18}.
\begin{figure*}[ht]
\centering 
\includegraphics[scale=1.6]{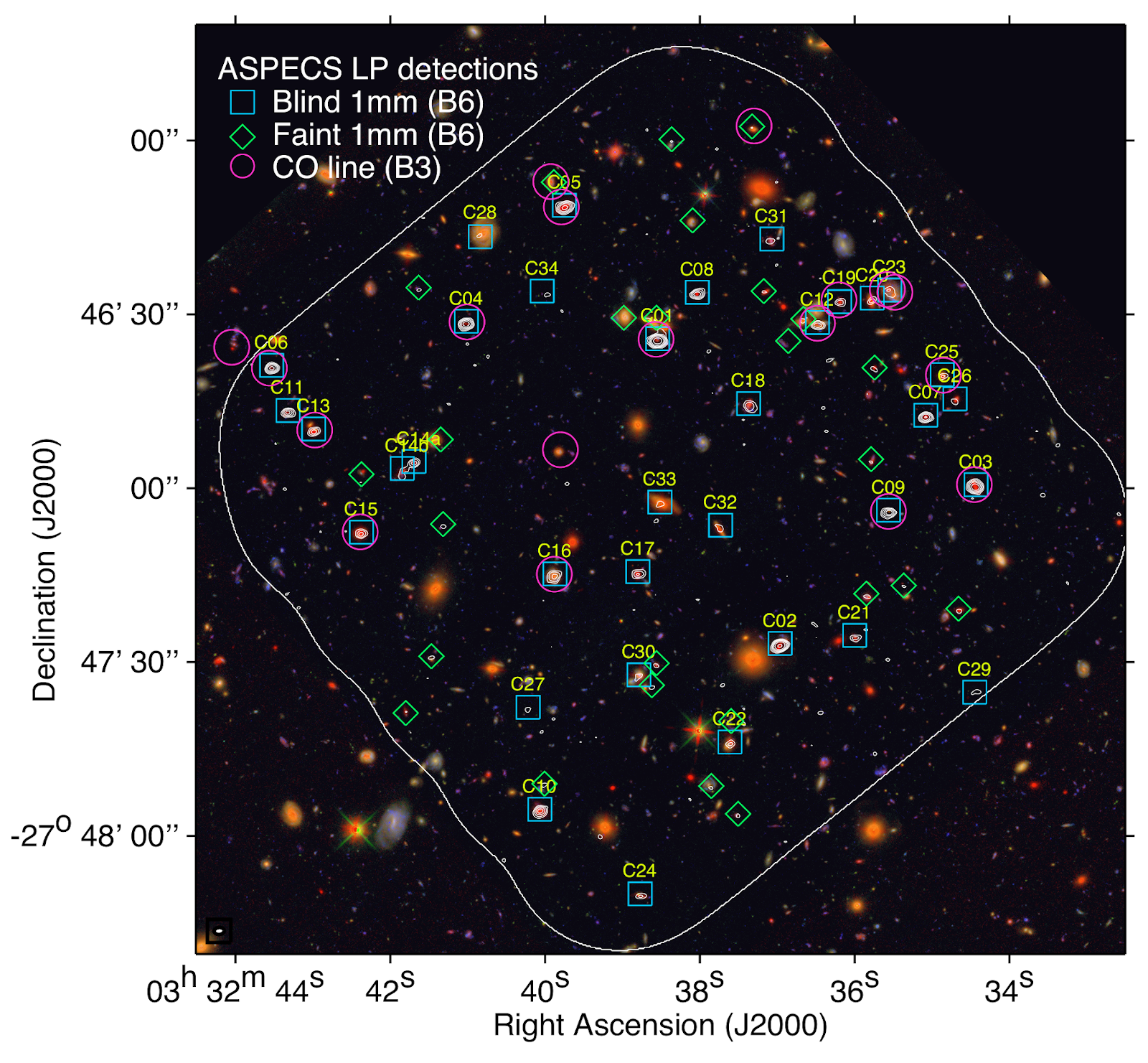} 
\caption{Optical/NIR color image of the HUDF (F450W, F850LP, F160W), with contours overlaid representing the ALMA 1.2 mm emission unveiled by the ASPECS LP survey. Contours are shown at $3, 5, 7, 10$ and $20\sigma$ level, with $\sigma=9.3\mu$Jy beam$^{-1}$. The significant 1.2 mm continuum sources detected by the ASPECS survey are highlighted by cyan squares (main sample) and green diamonds (secondary sample), while the locations of the CO line emitters from the ALMA band 3 part of this program are shown as magenta circles. Some faint sources are better recovered in the tapered, lower-resolution image (not shown here) and thus not necessarily visible through the white contours presented here. }\label{fig:image_hudf} 
\end{figure*}

The ALMA Spectroscopic Survey (ASPECS) in the Hubble Ultra Deep Field (HUDF) is a simultaneous blank-field CO line and dust continuum survey of distant galaxies performed with ALMA. In a first stage, the ASPECS pilot program targeted a region of $\sim$1 arcmin$^2$ in the HUDF, spectroscopically scanning the full ALMA bands 3 (3 mm) and 6 (1 mm) to search for CO line emission from galaxies at $0<z<6$, [CII] line emission at $6<z<8$ and dust continuum \citep[e.g.,][]{walter16, aravena16b, aravena16c, decarli16a,decarli16b, carilli16, bouwens16}. The ASPECS large program (LP) builds upon the observational strategy and results presented by the ASPECS pilot observations, extending the covered area of the HUDF to 4.6 arcmin$^2$, roughly comprising the Hubble eXtremely Deep Field (XDF), the region of the HUDF with the deepest near-IR (NIR) observations \citep{illingworth13, koekemoer13}. The first results of the ASPECS LP based on the ALMA band 3 observations are presented in a series of recent papers \citep[][]{gonzalezlopez19, decarli19,aravena19, boogaard19, popping19, uzgil19, inami20}.

In this paper, we present the physical properties of the faint dusty sources detected in the sensitive ASPECS 1.2 mm continuum map using the wealth of ancillary multiwavelength data available in the HUDF. In Section 2, we briefly summarize the ASPECS observations and ancillary data used. In Section 3, we present our main results and describe the sample of millimeter sources, their multi-wavelength counterparts, redshifts, and spectral energy distributions (SEDs). In Section 4, we study the location of our sources with respect to the main sequence of star formation and compare the properties of the dust continuum sources with galaxies in the field and previous ISM studies. Finally, in Section 5, we summarize the conclusions of this work. Throughout this paper, we assume a standard $\Lambda$CDM cosmology with $H_0=70$ km s$^{-1}$ Mpc$^{-1}$, $\Omega_\Lambda=0.7$ and $\Omega_{M}=0.3$. All magnitudes are presented in the AB system.

\begin{table*} 
\centering 
\caption{Main sample of sources detected in the ASPECS 1.2 mm continuum map.}\label{tab:1} 
\begin{tabular}{lccccccccc}
\hline 
ID & R.A. & Decl. & S/N & $f$ & $S_{\rm 1.2mm}$ & $S_{\rm 3mm}$ & CO ID & OIR? & Reference\\
  & (J2000)            & (J2000)    &          &        & ($\mu$Jy)       & ($\mu$Jy)     &       & (Y/N) &   \\ 
(1) & (2) & (3)  & (4) & (5)     & (6)    & (7)    &   (8)    & (9) & (10)  \\
\hline 
C01  &  03:32:38.54  &  $-27$:46:34.6  &   67.6  & 1.0    &  $752\pm 24$  &  $32.5\pm 3.8$  &   1  & Y & CM \\ 
C02  & 03:32:36.96  &  $-27$:47:27.2  &   44.1  &  1.0   &  $432\pm 10$  &  $29.6\pm6.3$  &  $\ldots$  & Y  & 3D \\ 
C03  &  03:32:34.43  &  $-27$:46:59.8  &   30.7  & 1.0    &  $430\pm 23$  & $<20.0$  &   4  &  Y   & CM \\ 
C04  &  03:32:41.02  &  $-27$:46:31.6  &   26.8  & 1.0    &  $316\pm 12$  &  $22.7\pm4.2$  &   3  &   Y  &  CO \\ 
C05  &  03:32:39.75  &  $-27$:46:11.6  &  23.2  & 1.0    &  $461\pm 28$  &  $27.4\pm4.6$  &   5  &    Y & CM \\ 
C06  &  03:32:43.53  &  $-27$:46:39.2  &   22.6  & 1.0    &  $1071\pm 47$  &  $46.5\pm7.1$  &   7  &  Y  & CO \\ 
C07  & 03:32:35.08  &  $-27$:46:47.8  &   20.1  &  1.0   &  $233\pm 12$  &  $<20.0$  &  $\ldots$  &  Y   & CO \\ 
C08  &  03:32:38.03  &  $-27$:46:26.6  &   16.2  & 1.0    &  $163\pm 10$  & $<20.0$  &  $\ldots$  &   Y  & M \\ 
C09  &  03:32:35.56  &  $-27$:47:04.2  &   15.9  & 1.0    &  $155\pm  10$  &  $<20.0$  &  13  &  Y    &  CO \\ 
C10  &  03:32:40.07  &  $-27$:47:55.8  &   13.8 &  1.0   &  $342\pm 34$  &  $<20.0$  &  $\ldots$  &   Y  & M \\ 
C11  &  03:32:43.32  &  $-27$:46:47.0  &   13.6  & 1.0    &  $289\pm 21$  &  $<20.0$  &  $\ldots$  &   Y  & 3D \\ 
C12  & 03:32:36.48  &  $-27$:46:31.8  &   10.7  & 1.0    &  $114\pm 11$  &  $<20.0$  &  15  &   Y  & CM \\ 
C13  &  03:32:42.99  &  $-27$:46:50.2  &    9.7  & 1.0    &  $116\pm 16$  &  $<20.0$ &  10  &   Y   & CM \\ 
C14a  &  03:32:41.69  &  $-27$:46:55.8  & 9.4  & 1.0  &  $ 96\pm  10$  &  $<20.0$  &  $\ldots$  & Y  & CO \\ 
C14b  &  03:32:41.85  &  $-27$:46:57.0  &    9.4  & 1.0&  $ 89\pm 20$  &  $<20.0$  &  $\ldots$  &   Y  & CM \\ 
C15  &  03:32:42.37  &  $-27$:47:08.0  &    8.9  &  1.0   &  $118\pm 13$  &  $<20.0$  &   2  &   Y  & CM \\ 
C16  &  03:32:39.87 &  $-27$:47:15.2  &    8.8  &  1.0   &  $143\pm 18$  &  $<20.0$  &   6  &   Y  & CM \\ 
C17  &  03:32:38.80  &  $-27$:47:14.8  &    8.1  & 1.0    &  $ 97\pm 15$  &  $<20.0$  & $\ldots$  &   Y  & M \\ 
C18  &  03:32:37.37  &  $-27$:46:45.8  &    7.2  &  1.0   &  $107\pm 16$  &  $<20.0$  &  $\ldots$  &   Y  & M \\ 
C19  &  03:32:36.19  &  $-27$:46:28.0  &    6.8  & 1.0   &  $ 85\pm 12$  &  $<20.0$  &  12  &   Y  & CM \\ 
C20  &  03:32:35.77  &  $-27$:46:27.6  &    6.0  & 1.0    &  $ 95\pm 16$  &  $<20.0$  &  $\ldots$  &  Y   & M \\ 
C21  &  03:32:36.00  & $-27$:47:25.8  &    5.5  &  1.0   &  $ 58\pm 11$  &  $<20.0$  &  $\ldots$  &   Y  & 3D \\ 
C22  &  03:32:37.61  &  $-27$:47:44.2  &    5.5  & 1.0    &  $ 59\pm 11$  &  $<20.0$  & $\ldots$  &   Y  & M \\ 
C23  &  03:32:35.55  &  $-27$:46:26.2  &    5.4  &  1.0   &  $148\pm 30$  &  $<20.0$  &   8  &    Y & CM \\ 
C24  &  03:32:38.77  &  $-27$:48:10.4  &    5.4  & 1.0    & $135\pm 25$  &  $<20.0$  &  $\ldots$  &  Y   & 3D  \\ 
C25  &  03:32:34.87  &  $-27$:46:40.8  &    5.4  &  1.0   &  $ 90\pm 17$  &  $<20.0$  &  14  &   Y  &  CM  \\ 
C26  &  03:32:34.70  & $-27$:46:45.0  &    4.3  &  0.5   &  $ 65\pm 15$  &  $<20.0$  &  $\ldots$  &    Y  & M  \\ 
C27  &  03:32:40.22  &  $-27$:47:38.2  &    4.1  &  0.8  &  $ 46\pm 11$  &  $<20.0$  & $\ldots$  &   N  & $\ldots$ \\ 
C28  &  03:32:40.84  &  $-27$:46:16.8  &    3.9  &  0.9  &  $184\pm 46$  &  $<20.0$  &  $\ldots$  &   Y  & M \\ 
C29  &  03:32:34.45  &  $-27$:47:35.6  &    3.5  &  0.8  &  $308\pm 75$  &  $<20.0$  &  $\ldots$  &   N  & $\ldots$ \\ 
C30  &  03:32:38.79  &  $-27$:47:32.6  &    3.5  & 0.8   &  $ 34\pm  10$  &  $<20.0$  &  $\ldots$  &   Y  & M \\ 
C31  &  03:32:37.07 &  $-27$:46:17.4  &    3.5  &  0.8   &  $ 47\pm 12$  &  $<20.0$  &  $\ldots$  &   Y  & M \\ 
C32  &  03:32:37.73  &  $-27$:47:06.8  &    3.5  & 0.8   &  $ 41\pm  10$  &  $<20.0$  & $\ldots$  &  Y   & M \\ 
C33  &  03:32:38.51  &  $-27$:47:02.8  &    3.3  & 0.6   &  $ 42\pm  10$  &  $<20.0$  &  $\ldots$  &   Y  & M \\ 
C34  &  03:32:40.04  &  $-27$:46:26.4  &    3.3  & 0.6   &  $ 39\pm 11$  &  $<20.0$  &  $\ldots$  &   N  & $\ldots$ \\ 
\hline 
\end{tabular}\\
\flushleft \noindent {\bf Note.} Column (1): source name; Columns (2) and (3): Position of the continuum detection in the ALMA 1.2-mm map. Column (4): S/N of the 1.2 mm detection; Column (5): fidelity ($f$) of the 1 mm detection, as defined in the text \citep[for details see][]{gonzalezlopez20}; Column (6): flux density at 1.2 mm, corrected for PB \citep{gonzalezlopez20}. Column (7): flux density at 3 mm \citep{gonzalezlopez19}. Column (8): CO source ID \citep{aravena19,boogaard19}. Column (9): is there an optical counterpart identification for this source? (yes/no). Column (10): redshift code. M: MUSE \citep[spectroscopic redshift;][]{inami17}; CO: CO line confirmed \citep[spectroscopic redshift;][]{aravena19,boogaard19}; CM: CO and MUSE joint redshift determination \citep[spectroscopic redshift;][]{boogaard19}; 3D: 3D-HST \citep[photometric redshift][]{momcheva15, skelton14}; GS: other HST redshifts \citep[photo-z;][]{rafelski15, morris15}. \end{table*}

\section{Observations}

\subsection{The ASPECS LP}

The ASPECS 1.2 mm continuum map yields an unprecedented rms level of $9.3\mu$Jy beam$^{-1}$, being the most sensitive millimeter survey obtained today over a contiguous area of $\sim5$ arcmin$^2$ \citep[][]{gonzalezlopez20}. The depth of these observations yielded the detection of 35 statistically significant sources plus 26 prior-based lower-significance sources (see below) and allowed for the discovery of a flattening of the 1 mm number counts at fluxes $<100$ $\mu$Jy \citep{gonzalezlopez20, popping20}. These results reflect that most of the extragalactic background light (EBL) at this wavelength in the HUDF is resolved by these observations \citep{gonzalezlopez20}, being a direct consequence of the shape and evolution of the 1 mm luminosity function \citep{popping20}. While the area covered by the ASPECS LP program is relatively modest, its depth allows us to reach well beyond the knee of the luminosity function at 1.1 mm and thus access most of the dust content available in the HUDF galaxies.
\subsection{ALMA observations}

The ASPECS LP survey setup and data reduction steps are described in \citet{gonzalezlopez20}, \citet{walter16} and \citet{aravena16b}. Here we repeat the most relevant information for the present study.

ALMA band 6 observations were obtained during Cycles 4 and 5 under excellent to good weather conditions (PWV$\sim0.5-2.0$). Observations were performed in an 85-point mosaic, covering roughly the same region comprised by the ALMA band 3 mosaic of the XDF \citep[see][]{gonzalezlopez20}. Individual pointings overlap each other by about half the ALMA primary beam (PB) at half-power beamwidth (HPBW), i.e., close to Nyquist sampling. Band 6 was scanned using eight frequency tunings, covering 212.0-272.0 GHz with no overlap or gaps between individual spectral windows. The ALMA PB in individual pointings ranges between $30''$ and $23''$ for this frequency range.

The observations were performed using the C43-1 and C43-2 arrays, leading to an angular resolution of $1.53''\times 1.08''$ at the center of the band 6 scan (242 GHz). Calibration was performed via standard observatory procedures with passband and phase calibration determined from nearby quasars and should be accurate within $\pm10\%$. Calibration and imaging were done using the Common Astronomy Software Application package ({\sc CASA}) versions v5.1.1 and v5.4.0-70, respectively. To obtain continuum maps, we collapsed along the frequency axis in the uv-plane and inverted the visibilities using the {\sc CASA} task TCLEAN using natural weighting and mosaic mode. This yielded an image reaching down to a noise level of 9.3$\mu$Jy beam$^{-1}$. A second version of the 1.2 mm map was obtained by tapering the visibilities in order to gain sensitivity for extended sources that were marginally detected in the naturally weighted image. This yielded an image with a resolution of $2.37''\times2.05''$ with a noise level of 11.3$\mu$Jy beam$^{-1}$.

\subsection{Ancillary data}

Our ALMA observations cover roughly the same region as the Hubble XDF. Thus, the ASPECS LP field benefits from the deepest observations obtained with the Hubble Space Telescope (HST) Advanced Camera for Surveys (ACS) and Wide Field Camera 3 (WFC3) through the HUDF09, HUDF12, and Cosmic Assembly Near-infrared Deep Extragalactic Legacy Survey (CANDELS) programs, as well as public photometric and spectroscopic catalogs \citep{coe06, xu07, rhoads09, mclure13, schenker13, bouwens14, skelton14, momcheva15,morris15, inami17}. As in other ASPECS studies, we make use of this optical and infrared coverage of the XDF, including the photometric and spectroscopic redshift information available from \citet{skelton14}. The area covered by the ASPECS LP footprint was observed by the MUSE Hubble Ultra Deep Survey \citep{bacon17}, representing the main optical spectroscopic sample in this area \citep{inami17}. The Multi-Unit Spectroscopic Explorer (MUSE) at the ESO Very Large Telescope provides integral field spectroscopy in the wavelength range $4750-9350$ \angstrom of a $3'\times3'$ region in the HUDF and a deeper $1'\times1'$ region that mostly overlaps with the ASPECS field. The MUSE spectroscopic survey provides spectroscopic redshifts for optically faint galaxies at $i_{\rm 775W}\sim26$ magnitudes (and down to $\sim30$ mag for fainter emission-line galaxies), and thus very complimentary to our ASPECS survey. MUSE covers key spectral lines, including H$\alpha$ $\lambda\lambda$6563, [OIII] $\lambda\lambda$4959, 5007 and [OII] $\lambda\lambda$3726, 3729 at $z<0.36$;  [OII] $\lambda$3726, 3729 at $z<1.5$, a number of absorption features at $z=1.5-2.0$, and Lyman-$\alpha$ at $z=2.9-6.7$ \citep[e.g.][]{boogaard19}. In addition to the HST and MUSE coverage, a wealth of optical and infrared coverage from space and ground-based telescopes is available in this field. This includes Spitzer Infrared Array Camera (IRAC) and Multiband Imaging Photometer (MIPS) imaging, as well as by the Herschel PACS and SPIRE photometry \citep{elbaz11}. From this, we created a master photometric and
spectroscopic catalog of the XDF region as detailed in \citet{decarli19}, which includes $>30$ bands for $\sim7000$ galaxies, 475 of which have spectroscopic redshifts.

\begin{figure}
    \centering
    \includegraphics[scale=0.5]{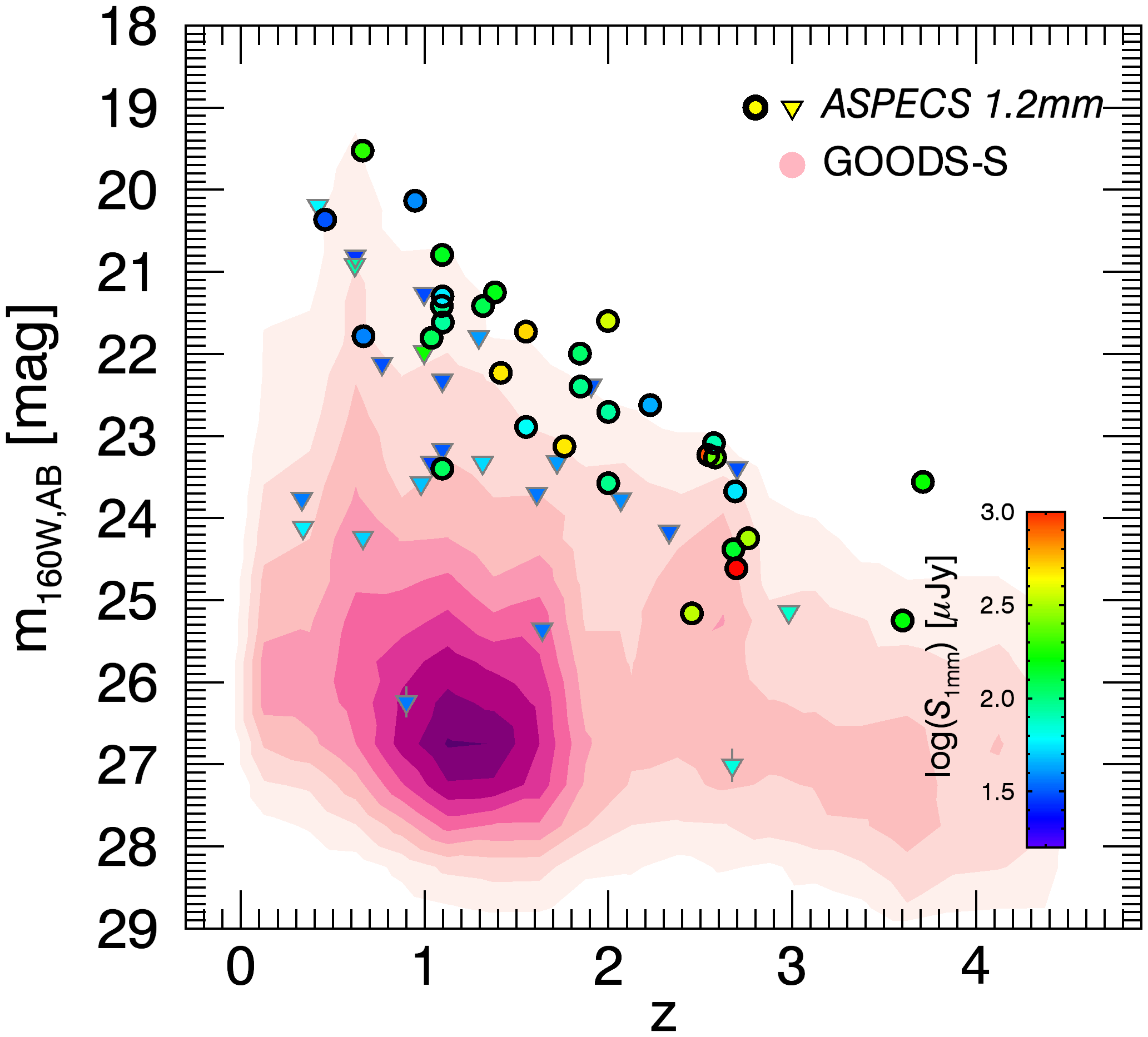}
    \caption{Apparent magnitudes in the HST F160W band as a function of redshift for the ASPECS sources with optical/NIR counterparts. The background contours represent the location of field galaxies from the GOODS-S field for comparison. For display purposes, these are shown as the square root of the number of sources at a given location in this plane. ASPECS galaxies from the main and the faint, prior-based samples are represented by filled circles and triangles, respectively. The coloring of each data point represents the 1.2 mm flux as shown by the color bar. The ASPECS 1 mm galaxies are among the brightest galaxies in the NIR regime at all redshifts in this field.}
    \label{fig:mag_z}
\end{figure}

\section{Results}

\subsection{Main sample}

Details about the 1mm continuum source extraction, fidelity, and completeness analysis, as well as source catalog are described in \citet{gonzalezlopez20}. Here, we provide a brief summary of these procedures.

Source extraction was performed simultaneously in the natural and tapered weighted images, using the LineSeeker code by searching for all pixels with signal-to-noise ratio (S/N) $>2$, and grouping them
into single sources using the DBSCAN algorithm. The noise level was computed from the RMS value in all pixels excluding those with S/N$>5$ \citep[for details see][]{gonzalezlopez19,gonzalezlopez20}.

The fidelity $f$ for each of the extracted sources was computed based on the number of positive ($N_{\rm pos}$) and negative ($N_{\rm neg}$) detections at a given S/N value,
with $f=1-N_{\rm neg}/N_{\rm pos}$. For reference, a fidelity of 50\% is achieved at S/N$=4.3$. A similar procedure on the tapered image yields a 50\% fidelity at S/N$=3.3$.

This extraction yielded a sample of 34 sources, one of which was split into two separate sources based on visual inspection \citep{gonzalezlopez20}, leading to a final high-fidelity sample of 35 sources with fidelities $>0.5$. Only one additional source not significantly detected in the natural weighted image was found in the tapered one, C28, which is found to be associated with a large spiral galaxy at $z=0.622$ \citep[see][]{gonzalezlopez20}. Table \ref{tab:1} lists this main sample, and these sources are highlighted in Fig. \ref{fig:image_hudf}.

\subsection{Multi-wavelength counterparts}

\subsubsection{Optical} 

We searched for matches between the 1.2 mm continuum detections and the HST optical sources in the field, using the catalog from \citet{skelton14}. We identified these optical counterparts within a $1''$ radius from the 1.2 mm position \citep[for details see][]{gonzalezlopez20} with an additional requirement that the probability of chance association $P=1-\mathrm{exp}(-n d^2)$, is less than 5\%. Here, $n$ is the number density of optical sources (with $m_{\rm F160W}<27$) in the neighborhood, and $d$ is the distance between the millimeter and the optical source. Based on this, we expect one to two sources to be a spurious association. The $1''$ radius is well matched to the 1.2 mm map beam size and the typical size of optical sources in the field \citep{aravena16b}. The astrometry of the ALMA and HST images, when corrected for known distortions \citep{dunlop17}, is accurate to within $<0.1''$ thus not representing a major source of possible offsets. While offsets between the optical and (sub)millimeter components of $\sim1''$ are not unusual for bright DSFGs \citep[e.g.,][]{hodge12}, this is typically not the case for for fainter dusty galaxies or ``typical''  galaxies as in the sample studied here \citep[e.g.,][]{daddi10a, franco20}.

Figure \ref{fig:image_hudf} shows the location of the 1.2 mm sources with respect to the optical galaxies in the field. Figure \ref{fig:mag_z} shows the HST F160W magnitudes as a function of redshift for the 1.2 mm sources compared to the optical galaxies in the field. At all redshifts, ASPECS 1.2 mm galaxies are among the brightest. Multi-wavelength cutouts for individual sources are shown in the Appendix \ref{app_b}. We find that from the sample of 35 millimeter detections in the ASPECS field, 32 have clear optical counterparts in the HST images. The three 1.2 mm detections with no optical counterparts have low S/N values and are consistent with the number of spurious sources at these significances, based on the fidelity analysis \citep[see discussion in][]{gonzalezlopez20}. 

\subsubsection{IR} 

Several of our ASPECS 1.2 mm continuum sources have an IR counterpart in Spitzer 24$\mu$m and/or Herschel PACS 100/160$\mu$m catalogs \citep{elbaz11}.  We find that out of the 35 significant ASPECS sources, 25 have a Herschel PACS counterpart within $1''$, and 10 have only upper limits at these wavelengths \citep{gonzalezlopez20}. Due to its poor angular resolution ($18-36''$), we do not attempt to match our sample to Herschel SPIRE sources. Out of these 10 Herschel PACS undetected sources, five have a clear 24$\mu$m and optical counterparts (1mm.C8, 1mm.C11, 1mm.C18, 1mm.C21 and 1mm.C31). From the other five sources (1mm.C9, 1mm.C25, 1mm.C27, 1mm.C29 and 1mm.C34), only two have clear optical counterparts (1mm.C9 and 1mm.C25).

Conversely, out of the 37 Herschel PACS sources in the HUDF, 25 have a clear 1.2 mm counterpart as mentioned above, nine can be associated to a $\sim2.0-3.5\sigma$ 1.2-mm positive blobs in the map, and three have no millimeter match. Most of these nine sources are part of the ASPECS 1mm secondary sample. The other three Herschel sources undetected at 1.2 mm are located near the edge of the ASPECS map, where the sensitivity worsens rapidly.


\begin{figure}
    \centering
    \includegraphics[scale=0.55]{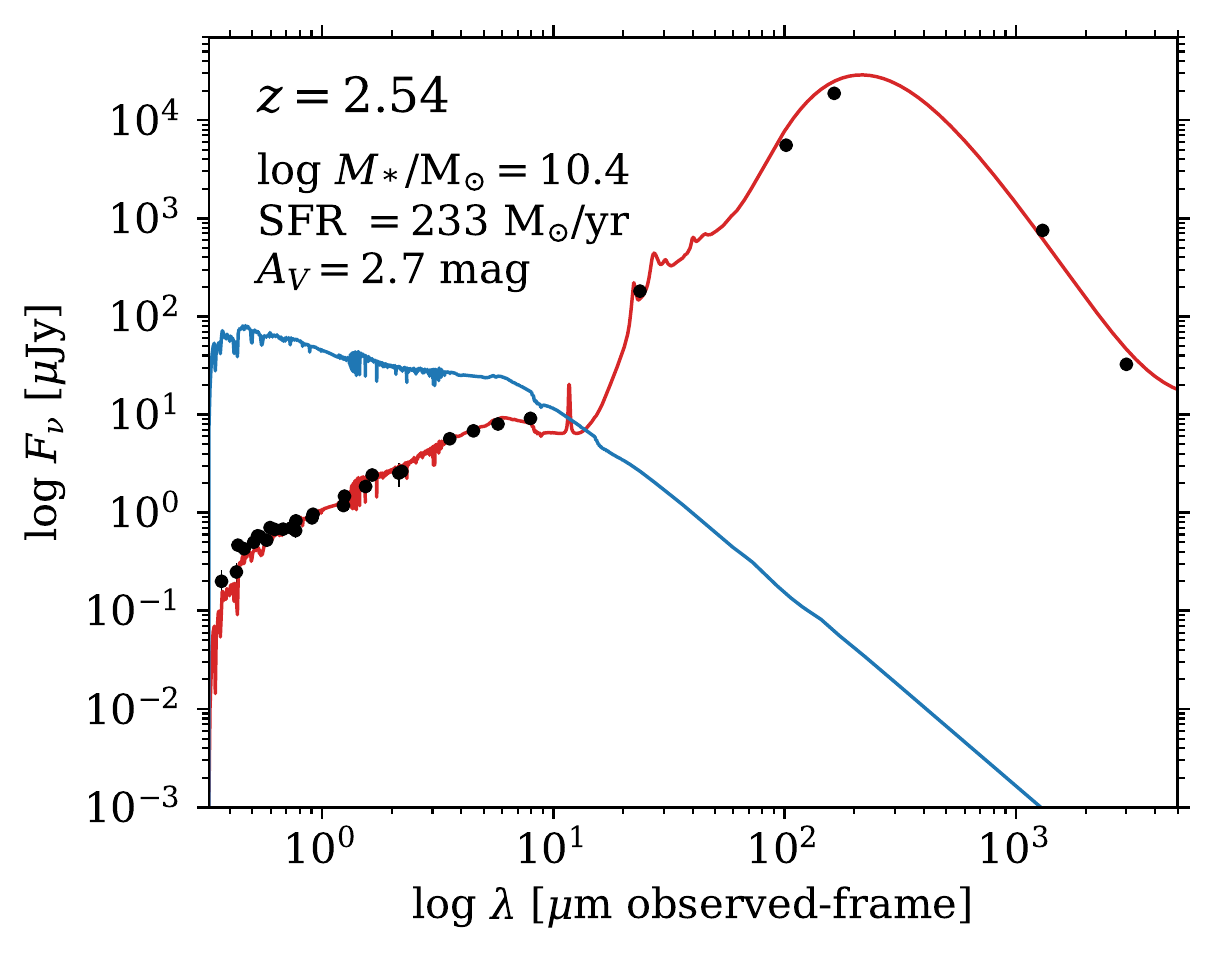}
    \caption{Example of the \textsc{magphys} SED fitting performed on source ASPECS-LP.1mm.C01. The black circles show the observed photometric data, and the red solid curve shows the best-fitted template. The blue curve shows the model-unattenuated stellar emission. The redshift and median values of the posterior likelihood distribution of the stellar mass, SFR, and visual attenuation ($A_{\rm V}$) are given in the left corner.}
    \label{fig:sed1}
\end{figure}

\subsubsection{CO}

It is interesting to check how many of the previously reported ASPECS CO emitters from ALMA band 3 observations \citep{gonzalezlopez19} are detected in the deep ASPECS 1.2 mm map. Of the 18 ASPECS CO line emitters (16 blind plus 2 prior based), 13 have a counterpart in the main sample of dust continuum sources reported here. From the other 5 CO sources (ids 3mm.9, 3mm.11, 3mm.16, 3mm.MP1 and 3mm.MP2), two have associated emission at the $\sim3\sigma$ level (3mm.16 and 3mm.MP1), and are included in the faint prior-based sample (see Section 3.3). One of these, 3mm.9, falls outside the coverage of the ASPECS 1.2-mm mosaic (see Fig. \ref{fig:image_hudf}), while 3mm.MP2 is blended with 3mm.8 corresponding to source C24 in the 1mm map. Hence, only one CO source, 3mm.11, has a formal nondetection at 1.2 mm, yielding an upper limit of 50$\mu$Jy ($5\sigma$). We note that this source identification and redshift have been confirmed through a clear CO(4-3) line detection in the ALMA band 6 cube \citep{boogaard20}.

\subsection{Prior-based millimeter sample}

Millimeter sources at a lower significance level than the main sample, down to a fidelity of 50\%, were extracted based on the existence of an optical or IR counterpart (see above). Optical sources were restricted to have HST F160W magnitudes $<27$ mag, to ensure more reliable associations. This prior based approach relies on the fact that bright optical sources will most likely be associated with massive or star-forming galaxies, which would likely show faint 1.2 mm emission. For this secondary sample, we consider a millimeter source as plausible if the probability of chance association with an optical match is $P<0.05$ and lies within $1''$. Based on this, we expect $\sim1$ source to be a false optical-millimeter match. Similarly, one can look for associations between faint millimeter sources and IR Herschel/PACS sources in the field. Given the low density of both source samples, the likelihood of a chance association is negligible. For additional details on the prior-based selection, we refer the reader to \citet{gonzalezlopez20}.

With this approach we find an additional sample of 25 millimeter sources using an optical prior. Twelve of these have an IR (Spitzer and/or Herschel) counterpart. Only one additional source is found using a Herschel prior alone. Throughout the rest of this paper, we treat this prior-based, faint millimeter sample as the secondary sample.

%
%

\begin{figure*} 
\centering 
\includegraphics[scale=0.8]{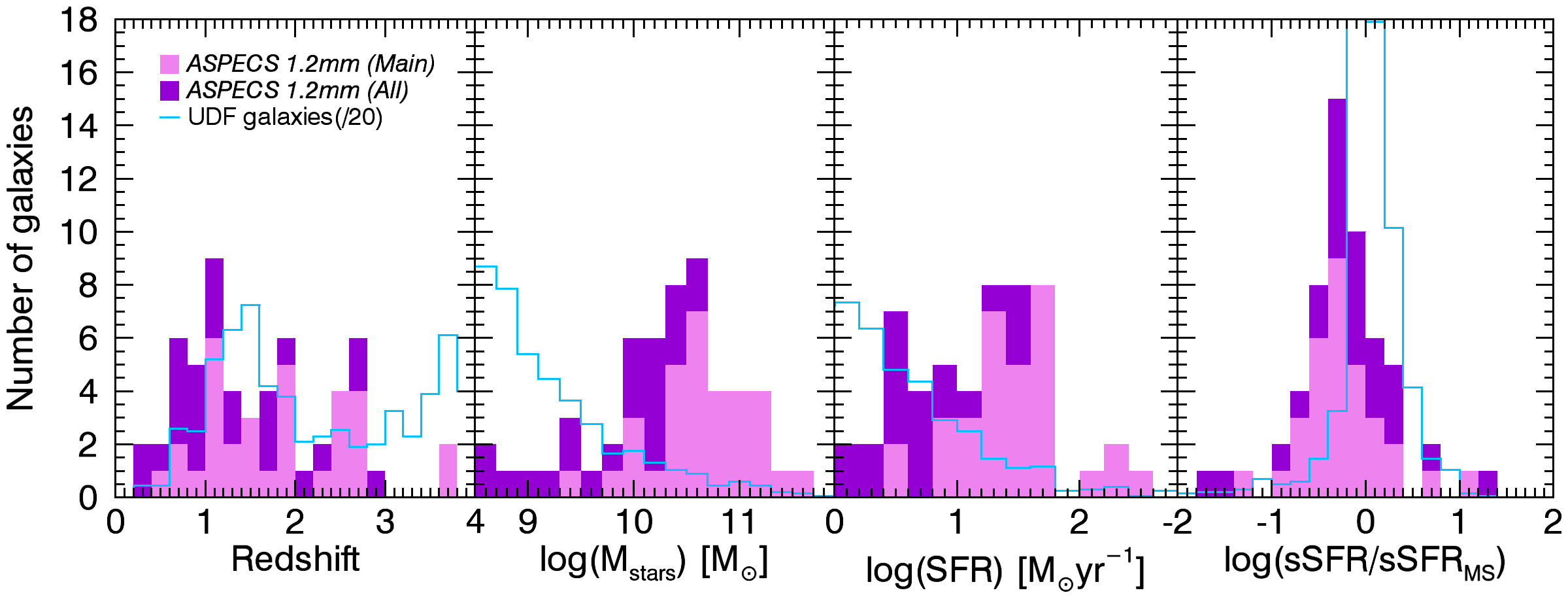} 
\caption{Distribution of properties of the ASPECS dust continuum sources compared to the galaxies in the HUDF. From left to right, we show the distribution of redshift, stellar mass, SFR and the normalized specific SFR ($\Delta_{\rm MS}=$log(sSFR/sSFR$_{\rm MS}$); see Section 4.2). The histograms for the field galaxies have been normalized by a factor of 20, for display purposes, and restricted to galaxies with log(SFR)$>-1$ and log($M_{\rm stars}>8.6$). The main and main+secondary samples of ASPECS 1.2 mm galaxies are shown in pink and purple colors, respectively. Dust-selected galaxies show a clear differences in their overall properties compared to the HUDF galaxies.} \label{fig:histo1}
\end{figure*}

\subsection{SED fitting and derivation of properties}

For all galaxies in the HUDF, we fit their SED using the high-redshift extension of \textsc{magphys} \citep{dacunha08, dacunha15}. We use the available 26 broad- and medium-band filters in the optical and infrared regimes, from the $U$ band to Spitzer IRAC 8$\mu$m, and including the Spitzer MIPS 24 $\mu$m and the ALMA 1.2 mm and 3 mm fluxes and upper limits. For the continuum-detected galaxies, we include the ALMA 1.2 mm and 3 mm and Herschel PACS 100 and 160$\mu$m fluxes or limits. We note that for the field galaxies we do not include the Herschel PACS limits, as the depth of the ALMA maps sets significantly more stringent constraints.

\textsc{magphys} yields estimates for the stellar masses, SFR, specific SFR, and attenuation. To constrain the FIR SED, \textsc{magphys} assumes a two-component graybody approximation, with dust emissivity indexes of 1.5 and 2.0 for the warm and cold components, respectively. Under these assumptions, \textsc{magphys} fits yield estimates of the IR luminosity and the dust masses and temperatures. \textsc{magphys} employs a physically motivated prescription to balance the energy output at different wavelengths. Thus, even in cases with poor constraints on the IR SED, estimates on the IR luminosity and/or dust mass arise from constraints on the dust-reprocessed UV emission, which is well sampled by the UV-to-infrared photometry as demonstrated by \citet{dudzevicute20}. An example of the photometry and SED fits obtained for the ASPECS sources is shown in Fig. \ref{fig:sed1}. The full set of SED fits is presented in the Appendix \ref{app_c}. The properties derived for individual ASPECS 1mm sources are shown in Table \ref{tab:2}. The values listed correspond to the same values used in \citet{gonzalezlopez20} and \citet{boogaard20}.

\begin{figure}
    \includegraphics[scale=0.55]{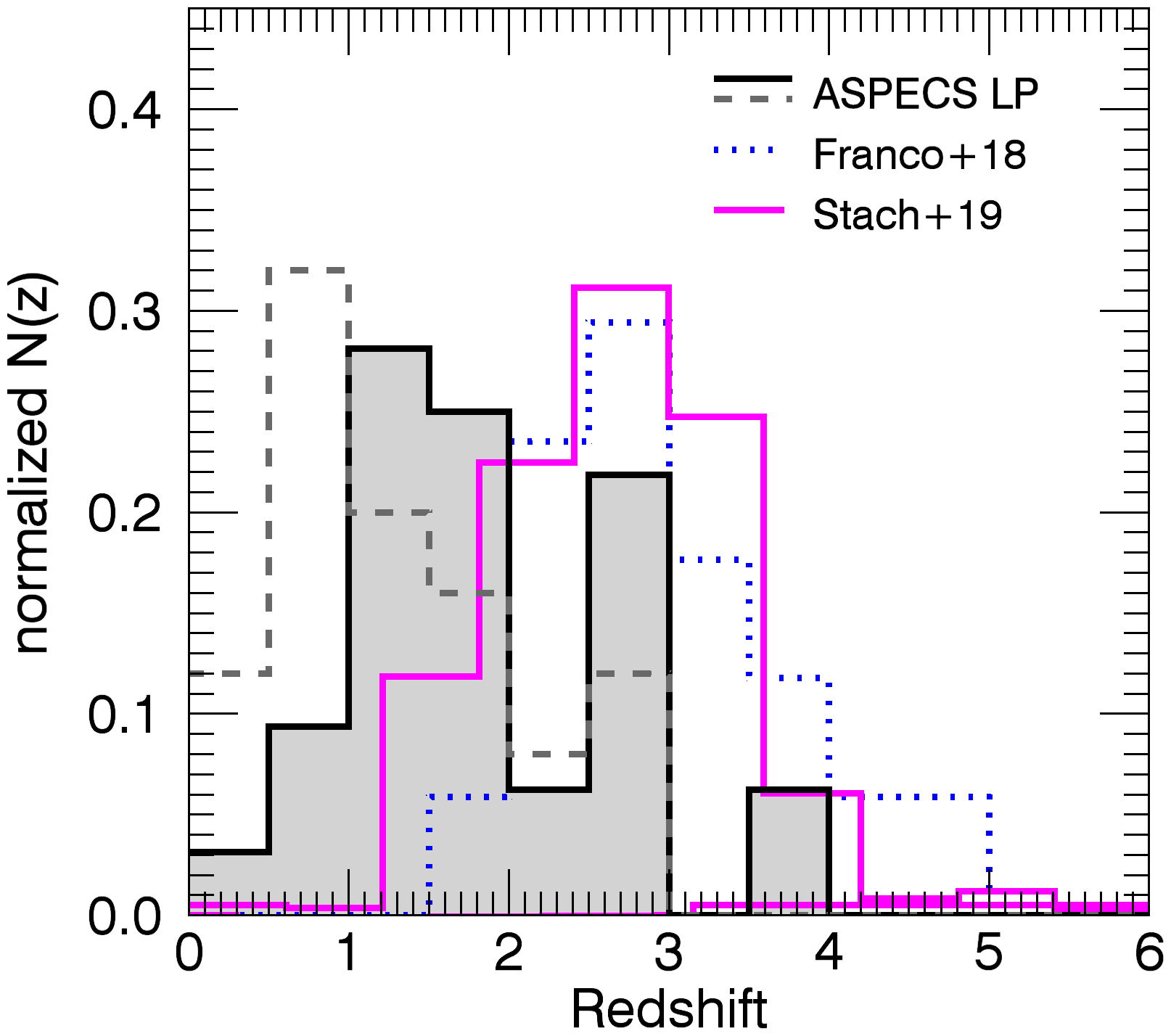} \caption{Redshift distribution of the ASPECS 1 mm galaxies in the main sample compared to recent deep ALMA millimeter  surveys. The solid black and dashed gray histograms show the distribution of ASPECS main and secondary samples, respectively. Two representative comparison samples of dusty galaxies selected from (sub)millimeter surveys with ALMA and SCUBA-2 are shown as blue dotted and solid magenta histograms, respectively \citep{franco18, stach19, dudzevicute20}. All histograms are normalized to the total number of galaxies in each sample. The ASPECS galaxy sample shows lower redshifts than millimeter-brighter galaxies detected in those surveys.} \label{fig:zdistr} \end{figure}
    
Figure \ref{fig:histo1} shows the distributions of redshift, SFR, stellar mass, and specific SFRs for our ASPECS 1.2 mm continuum sources compared to all galaxies in the HUDF. This HUDF galaxy sample was restricted to galaxies with log($M_{\rm stars}$)$>8.5$ and log(SFR)$>-1$, and to sources that fall within the region covered by the ASPECS observations (within PB$>0.1$) and with redshifts that match those found for the ASPECS 1.2 mm sources ($0.5<z<5$). From this, we find that the ASPECS galaxies are among the most massive and star-forming galaxies in the ASPECS field in the specified redshift range. The ASPECS 1.2 mm galaxies span almost 2 dex in stellar mass and SFR. For the main sample, we find a median stellar mass of $4.8\times10^{10} M_\sun$, with an interquartile range of $(2.4-11.7)\times10^{10}\ M_\sun$, and a median SFR of $33\ M_\sun$ yr$^{-1}$, with an interquartile range of $19-51\ M_\sun$ yr$^{-1}$.  These values confirm previous results found for this galaxy population by the ASPECS pilot survey \citep{aravena16b}, which targeted a smaller region of the HUDF at slightly shallower depth. Interestingly, when including sources from the faint, prior-based sample, we find lower redshifts, lower stellar masses, and SFRs compared to the main sample. For this sample, we find a median stellar mass of $0.9\times10^{10} M_\sun$, with an interquartile range $(0.1-1.5)\times10^{10}\ M_\sun$, and a median SFR of $4\ M_\sun$ yr$^{-1}$, with an interquartile range of $3-8\ M_\sun$ yr$^{-1}$. 

Based on the comparison of the stellar masses and SFRs obtained with \textsc{magphys} and \textsc{prospector} \citep[see ][]{bouwens20}, we find that these parameters are precise to within $\sim$0.1-0.2 dex in the stellar-mass range explored in this study. To test the reliability of the physical parameters obtained with \textsc{magphys}, \citet{dudzevicute20} run \textsc{magphys} SED fitting on a sample of $\sim9400$ galaxies from the EAGLE galaxy simulation with SFR$>10$ $M_\sun$ yr$^{-1}$ and $z>0.25$. They find that \textsc{magphys} successfully provides good estimates of all parameters, except for the stellar masses where a significant systematic underestimation of 0.46 dex was found. \citet{dudzevicute20} attribute this difference to variations in the adopted star-formation histories, dust model, and geometry between \textsc{magphys} and the used radiative transfer code. Based on the comparison of the stellar masses and SFRs obtained with the \textsc{magphys} and \textsc{prospector} SED fitting codes in HUDF/ASPECS galaxies, \citet{bouwens20} find that these parameters are precise to within $\sim0.1-0.2$ dex, and \textsc{magphys} stellar masses are lower by $\sim$0.1 dex in the stellar-mass range explored in this study ($\sim10^{9.5-11} M_\sun$). These later results are in good agreement with the comparisons made by \citet{liu19}. To account for these systematic uncertainties, we have added in quadrature a 0.1 dex uncertainty to all the \textsc{magphys} parameters derived.

\section{Analysis and Discussion}

\subsection{Redshifts}
 
The vast majority of identified ASPECS 1.2 mm galaxies have a reliable spectroscopic redshift: out of the 32 ASPECS 1.2 mm galaxies with optical counterparts in the main sample, 28 have an unambiguous spectroscopic redshift available either from the deep ASPECS CO spectroscopy \citep{gonzalezlopez19}, MUSE spectroscopy \citep{boogaard19} and/or from other optical/NIR spectroscopic surveys in the HUDF. The other four sources have a photometric redshift from the 3D-HST survey \citep[e.g.,][]{skelton14,momcheva15}. Similarly, out of the 26 sources in the secondary sample, 21 have a spectroscopic redshift and 5 have a photometric redshift from 3D-HST (see Tables \ref{tab:1} and \ref{tab:3}).

Figure \ref{fig:zdistr} shows the redshift distribution of the ASPECS 1.2 mm continuum sources in the main sample compared to two other galaxy samples detected in previous (sub)millimeter continuum deep maps, including the ALMA survey of the GOODS South field \citep{franco18} and the recent ALMA survey of the SCUBA-2 Cosmology Legacy Survey UKIDSS/UDS field \citep[AS2UDS;][]{stach19, dudzevicute20}. The latter survey, in particular, specifically targets millimeter-brighter galaxies ($S_{\rm 1mm}\geq0.3-0.5$ mJy), most of which correspond to SMGs. 

We find that the main sample of ASPECS 1.2 mm continuum galaxies has a median redshift of 1.85 with an interquartile range of $1.10-2.57$. The secondary sample of ASPECS 1.2 mm galaxies has consistently lower redshifts, yielding a median redshift of 1.10 and an interquartile range of $0.77-1.72$. These redshifts are significantly lower than those found in shallower surveys such as that in the GOODS South field, with a median redshift of 2.92 \citep{franco18} and in the AS2UDS with a median redshift of 2.61 \citep{dudzevicute20}. These results confirm previous indications from the ASPECS pilot survey that these faint DSFGs (with $S_{\rm 1.2mm}<0.1$ mJy) have significantly different redshift distributions and lower median redshifts than brighter SMGs selected from shallower, larger-area surveys at similar wavelengths \citep[e.g.,][]{aravena16b}. Furthermore, this supports previous findings that the redshift distribution of (sub)millimeter surveys is sensitive to depth \citep[e.g.,][]{archibald01, ivison07, bethermin15b, brisbin17}.

\begin{figure*} 
\centering 
\includegraphics[scale=0.6]{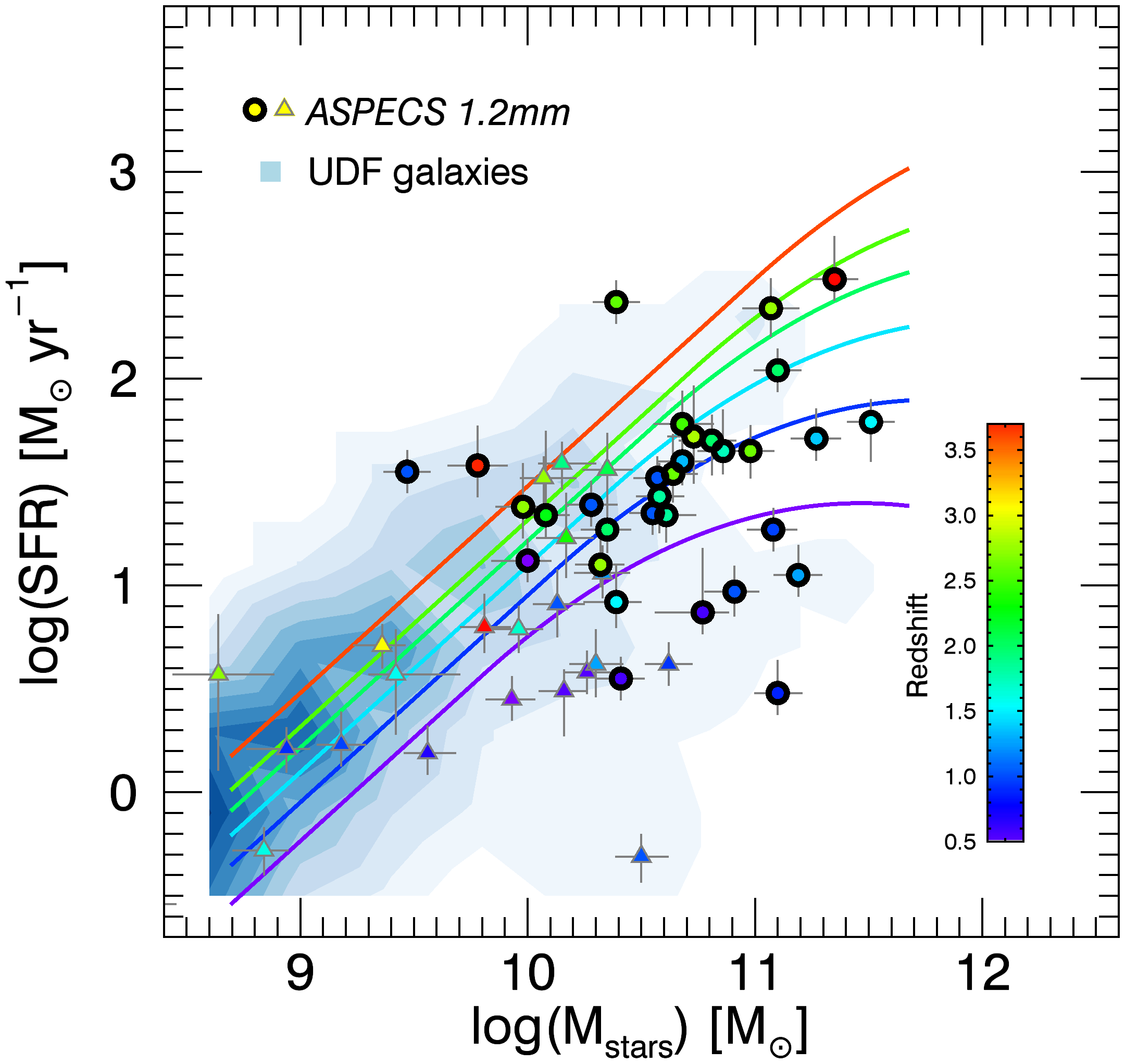}\hspace{2mm}
\includegraphics[scale=0.6]{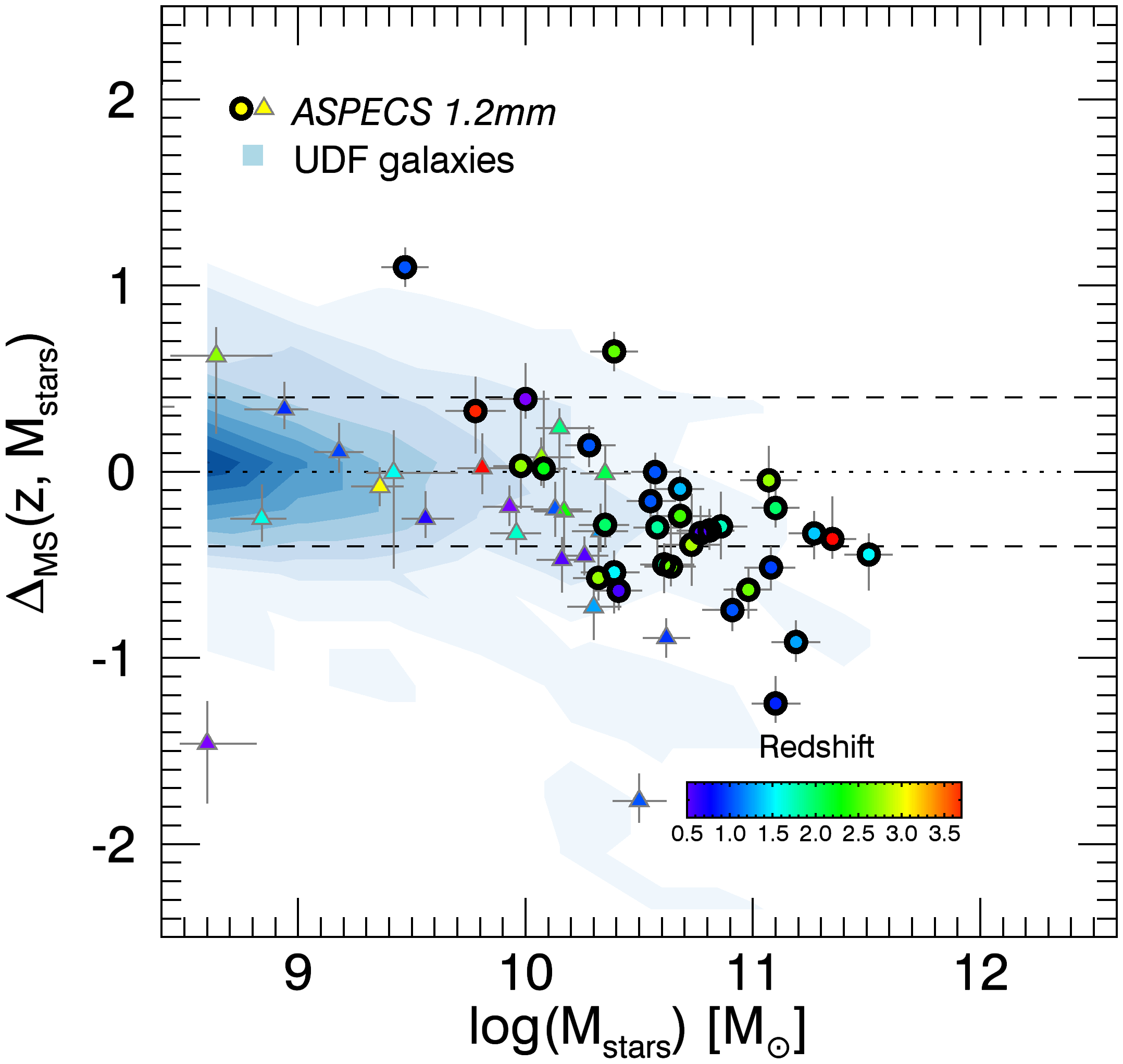} 
\caption{Location of the ASPECS dust continuum sources compared to field galaxies in the HUDF. The ASPECS galaxies are denoted by the colored large circles and triangles, representing galaxies in the main and secondary samples, respectively. ({\it Left:}) Stellar mass versus SFR diagram. The HUDF galaxies are represented by blue contours, which scale as the square root of the number of sources in a given bin element in this diagram. The solid curves represent the observational relationships between SFR and stellar mass as a function of redshift \citep[from][]{schreiber15}. These redshifts are denoted in different colors as shown by the color bar to the right. ({\it Right:}) Offset from the MS ($\Delta_{\rm MS}$) as a function of stellar mass. A significant fraction of the ASPECS 1mm galaxies lie slightly below the MS.} \label{fig:ms}
\end{figure*}

\subsection{Location in stellar mass vs. SFR plane}

\label{sec:ms}
Based on the properties derived through SED fitting, we explore the location of our galaxies in the stellar mass versus SFR diagram. The left panel of Fig. \ref{fig:ms} shows the comparison between the ASPECS sources and HUDF galaxies. The right panel of Fig. \ref{fig:ms} shows the normalized specific SFR, representing the galaxies' offset with respect to the star formation MS, using a prescription for its location as a function of redshift and stellar mass from \citet{schreiber15}.  Here, the specific SFR is defined as sSFR=SFR$/M_{\rm stars}$, and thus the offset from the MS is defined as $\Delta_{\rm MS}={\rm log(sSFR/sSFR}_{\rm MS}(z, M_{\rm stars}))$. 

We define the boundaries of the main sequence to be within $\pm0.4$ dex from the prescription curves given by \citet{schreiber15}. This selection is conservative, thus accounting for possible uncertainties on the stellar mass estimates. With this prescription, we find that most of the UDF galaxies are consistently located within the MS, well centered around the MS with a median $\Delta_{\rm MS}=0.0$ and an interquartile range width of 0.31 dex.

We find that $59\%\pm13\%$ (19 out of the 32) of galaxies from the main sample lie within $\pm0.4$ dex from the MS at the respective redshift and stellar mass. Furthermore, $6\%\pm4\%$ (2 out of 32) and $34\%\pm9\%$ (11 out of 32) of the ASPECS 1.2 mm galaxies are located above and below the MS, respectively. For the secondary sample, we find that $68\%\pm16\%$, $8\%\pm6\%$ and $24\%\pm10\%$ (17, 2, and 6 sources) fall within, above, and below 0.4 dex from the MS, respectively, i.e., the percentages for secondary sample of ASPECS 1.2 mm galaxies are very similar to those obtained for the main sample. We note that the MS limits are broad, covering roughly an order of magnitude around the star-forming ``sequence''. Thus, these results indicate that galaxies below the MS represent a non-negligible fraction of the dusty (gas-rich) galaxies in the ASPECS survey. Such below-MS galaxies, typically associated with ``passive'' (or even ``quenched'') star formation activity, have so far been mostly ignored in observations of cold dust and molecular gas although significant progress has been made in recent years \citep[e.g.,][]{sargent15, suess17, spilker18, gobat18, bezanson19}.


The ASPECS 1.2 mm galaxies in the main sample show $\Delta_{\rm MS}$ median and interquartile range width values of -0.30 and 0.51 dex, respectively. Similarly, the secondary sample shows median and interquartile width values of -0.19 and 0.40 dex, respectively. These values indicate that ASPECS 1.2 mm galaxies are systematically below the nominal MS prescription, with a wider distribution, contrary to the case of the UDF sample, which is well centered around the MS, and suggest the almost lack of an MS for these faint dusty galaxies.

Furthermore, we find a tendency of decreasing $\Delta_{\rm MS}$ with increasing stellar mass, particularly for galaxies with stellar masses above $\sim10^{11}\ M_\sun$, which tend to lie below the defined boundaries of the MS. The fractions of galaxies below the MS in this mass range are similar between the main and secondary sample. This could imply that the most massive ASPECS 1.2 mm galaxies are halting their SFR significantly, faster than field galaxies, while still retaining a significant amount of ISM dust and gas. This would be in line with observations of a flattening of the MS at the highest stellar masses\footnote{However, this effect could also be produced if the adopted MS prescription \citep{schreiber15} does not apply to our sample in this stellar mass range, or if the derived SFR values for our galaxies are underestimated. To explore this, we first used the (different) MS prescription from \citet{speagle14}, and found consistent results when compared to those obtained with the adopted MS prescription, even yielding systematically higher fractions of galaxies below the MS for both the ASPECS 1.2 mm galaxies and the UDF galaxies. Furthermore, with the adopted MS prescription the majority of the field galaxies fall very well aligned with the MS, and since the properties for the UDF and ASPECS galaxies were derived consistently, we do not expect unphysical systematic differences in the SFR values for the ASPECS sample.}.

\begin{figure}[ht] 
\centering
\includegraphics[scale=0.6]{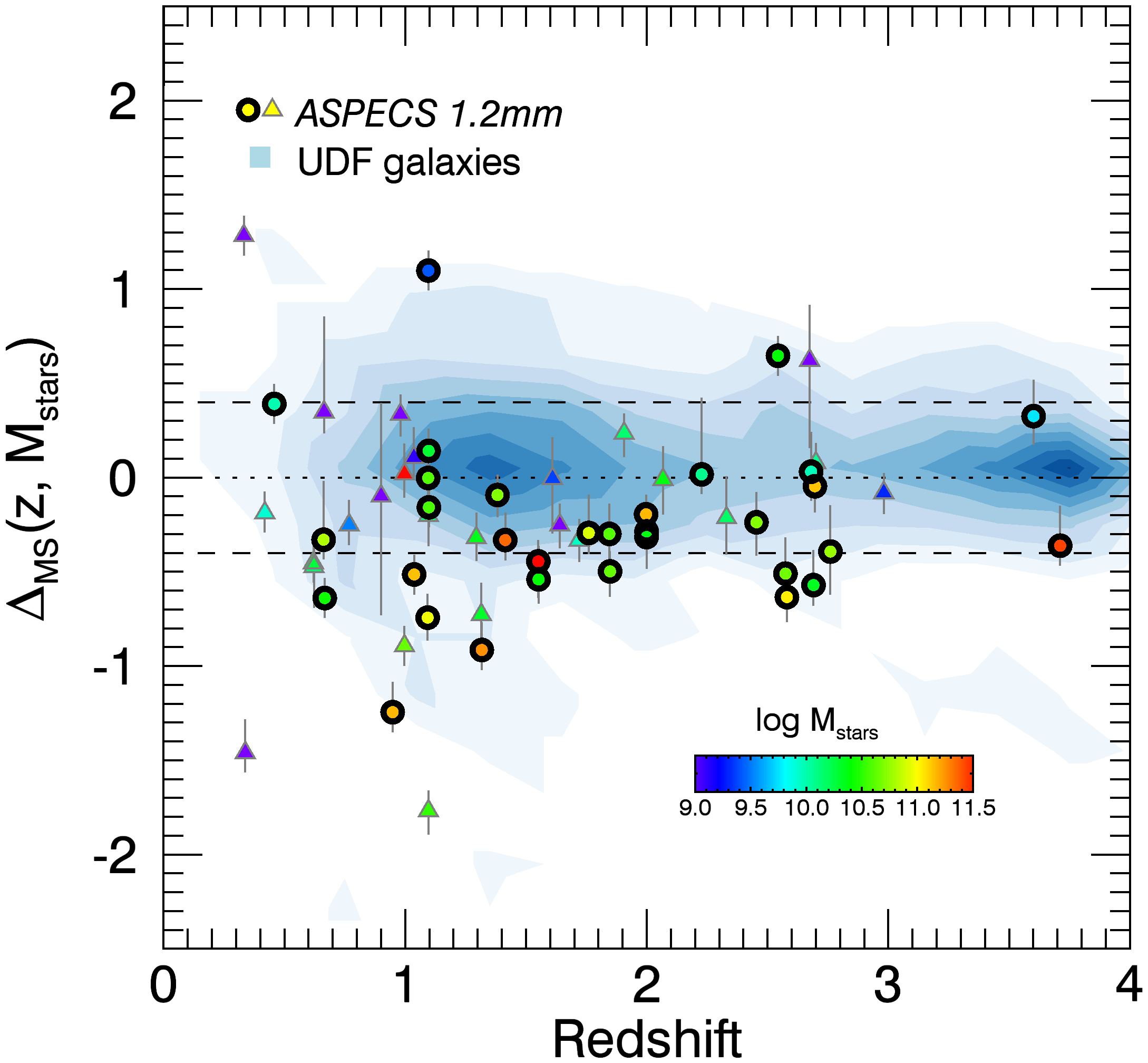}
\caption{Evolution of the normalized sSFR, or offset from the MS ($\Delta_{\rm MS}$), with redshift for the ASPECS 1.2 mm galaxies compared to field UDF galaxies. ASPECS galaxies are color-coded according to their stellar masses as denoted by the horizontal color bar. Circles and triangles represent galaxies in the main and secondary ASPECS samples, respectively. The background contours represent the square root of the number of field galaxies in a specific location of this diagram. The dashed and dotted lines denote the region considered to be consistent with the MS of star formation.} 
\label{fig:evolz2} 
\end{figure}

Figure \ref{fig:evolz2} shows the evolution of $\Delta_{\rm MS}$ with redshift. While the bulk of field galaxies show almost no evolution, there appears to be a mild tendency for the ASPECS 1.2 mm galaxies to have lower $\Delta_{\rm MS}$ at decreasing redshifts. This pattern, however, could be produced by the low number of ASPECS 1.2 mm sources.

\subsection{ISM molecular gas masses}
\label{sec:ism_mass}

The \textsc{magphys} SED fitting routine uses a prescription to balance the energy output at various wavelengths, thus producing estimates of the dust mass ($M_{\rm d}$) based on the best SED fit model. For this estimation, \textsc{magphys} uses a two-component gray body dust SED with varying dust temperature and mass, and fixed dust emissivity indices $\beta=1.5$ and $2.0$ for the warm and cold dust components, respectively. Note that for the majority of our ASPECS galaxies there are Herschel far-IR measurements that improve greatly the completeness of the dust SED, providing better accuracy in the dust masses. Following the approach in \citet{aravena16a}, we used this $M_{\rm d}$ estimate along with an assumption of the molecular gas-to-dust ratio ($\delta_{\rm GDR}$) to compute the molecular gas mass as $M_{\rm mol, SED}=\delta_{\rm GDR} M_{\rm d}$. 

For consistency with previous studies in the ASPECS series \citep[e.g.][]{aravena16b, aravena19, boogaard19, decarli19}, we fix $\delta_{\rm GDR}=200$ and use this SED-based approach to compute the molecular gas masses for the ASPECS 1.2 mm continuum sample throughout. The computed masses are listed in Tables \ref{tab:2} and \ref{tab:4}. 

In the Appendix \ref{app_a}, we present a detailed comparison between three different estimates of the molecular gas mass for our sample (SED-based, 1.2-mm continuum, CO), and the effects of using a metallicity-dependent prescription for $\delta_{\rm GDR}$. Using either this later approach or any of the other methods yields molecular gas estimates that are well within the uncertainties of the measurements and thus fully consistent with the masses used here. 

\begin{figure}
    \centering
    \includegraphics[scale=0.55]{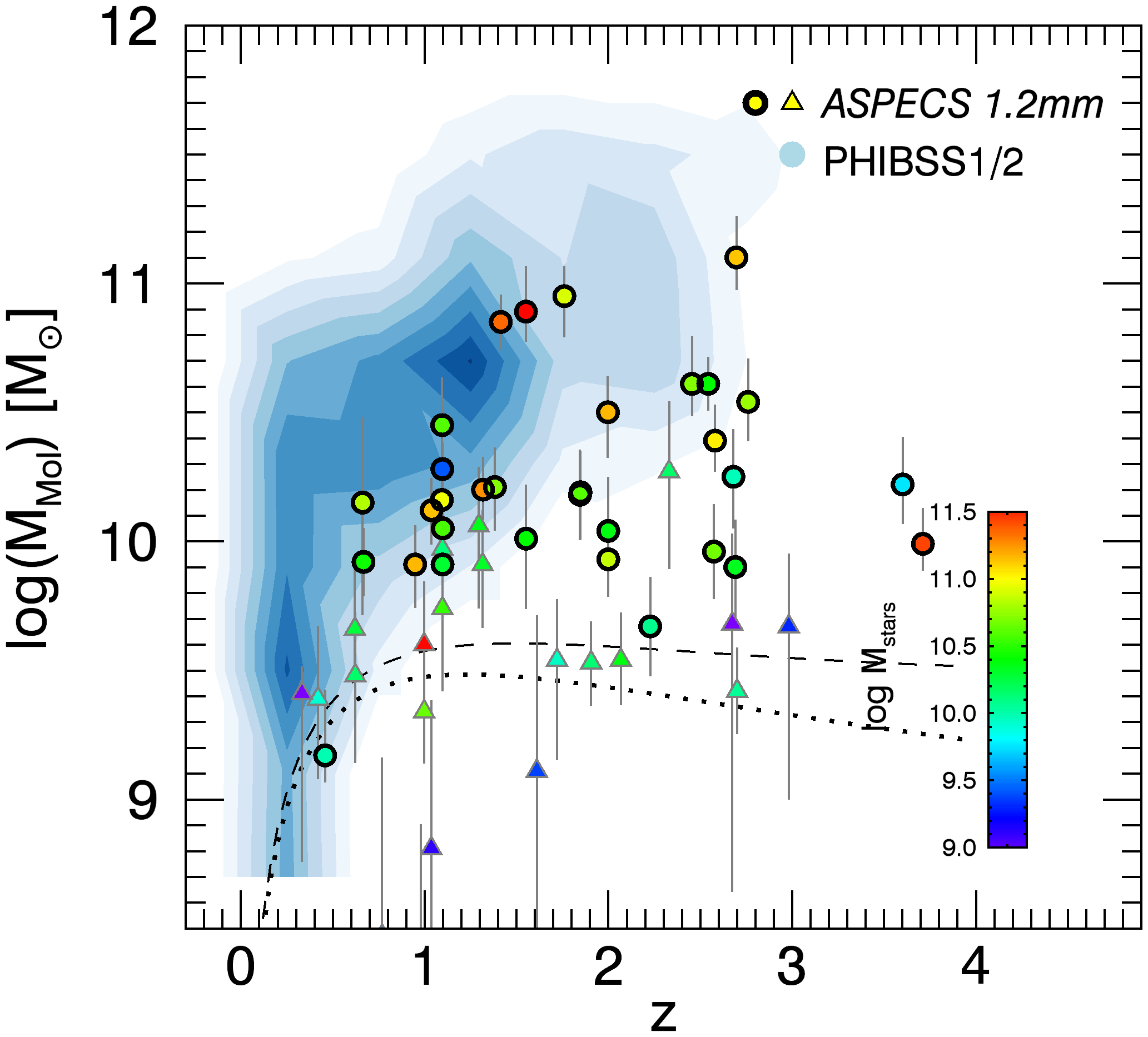}
    \caption{Molecular gas masses as a function of redshift. The ASPECS 1.2 mm galaxies in the main and secondary samples are shown by filled circles and triangles, respectively, with $M_{\rm mol}$ computed using the RJ method. The color of each data point is denoted by its stellar mass as shown by the vertical color bar. The blue background contours represent the number of sources ($\sqrt{N}$) from the PHIBSS1/2 sample \citep{tacconi18, freundlich19}. The dashed and dotted lines represent the value of $M_{\rm mol}$ for a galaxy at redshift $z$ and with a 1.2-mm flux given by the $3\sigma$ level ($28\mu$Jy), assuming a dust temperature of 25 K and 45 K, respectively. The ASPECS 1.2 mm map allows us to reach lower $M_{\rm mol}$ than achieved by most ISM galaxy surveys.}
    \label{fig:mmol_z}
\end{figure}

\begin{figure}
    \centering
 \includegraphics[scale=0.6]{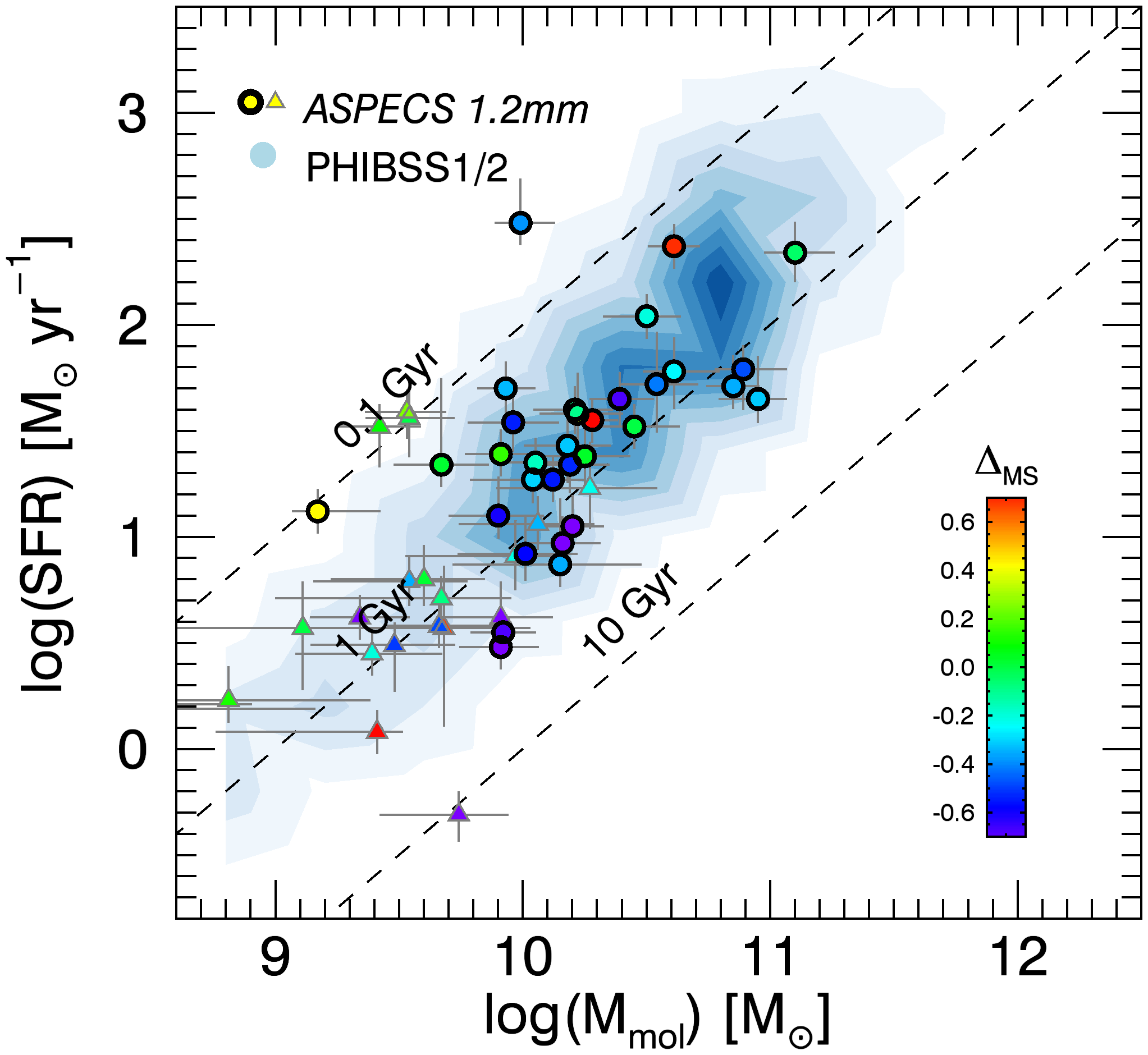} 
    \caption{SFR vs M$_{\rm mol}$ or integrated ``Schmidt-Kennicutt'' diagram for the ASPECS 1mm sources compared to the PHIBSS1/2 sample \citep{tacconi18, freundlich19}. Filled circles and triangles show the location of the ASPECS main and secondary samples, respectively. The background light blue contours represent the square root of the number of PHIBSS1/2 sources in a particular location in the diagram. The ASPECS 1.2 mm galaxies are color-coded according to their distance to the MS (normalized sSFR: $\Delta_{MS}$). The dashed lines represent the location of constant gas depletion timescales.}
    \label{fig:mmol_sfr}
\end{figure}

\begin{figure*}[ht] 
\centering
 \includegraphics[scale=0.6]{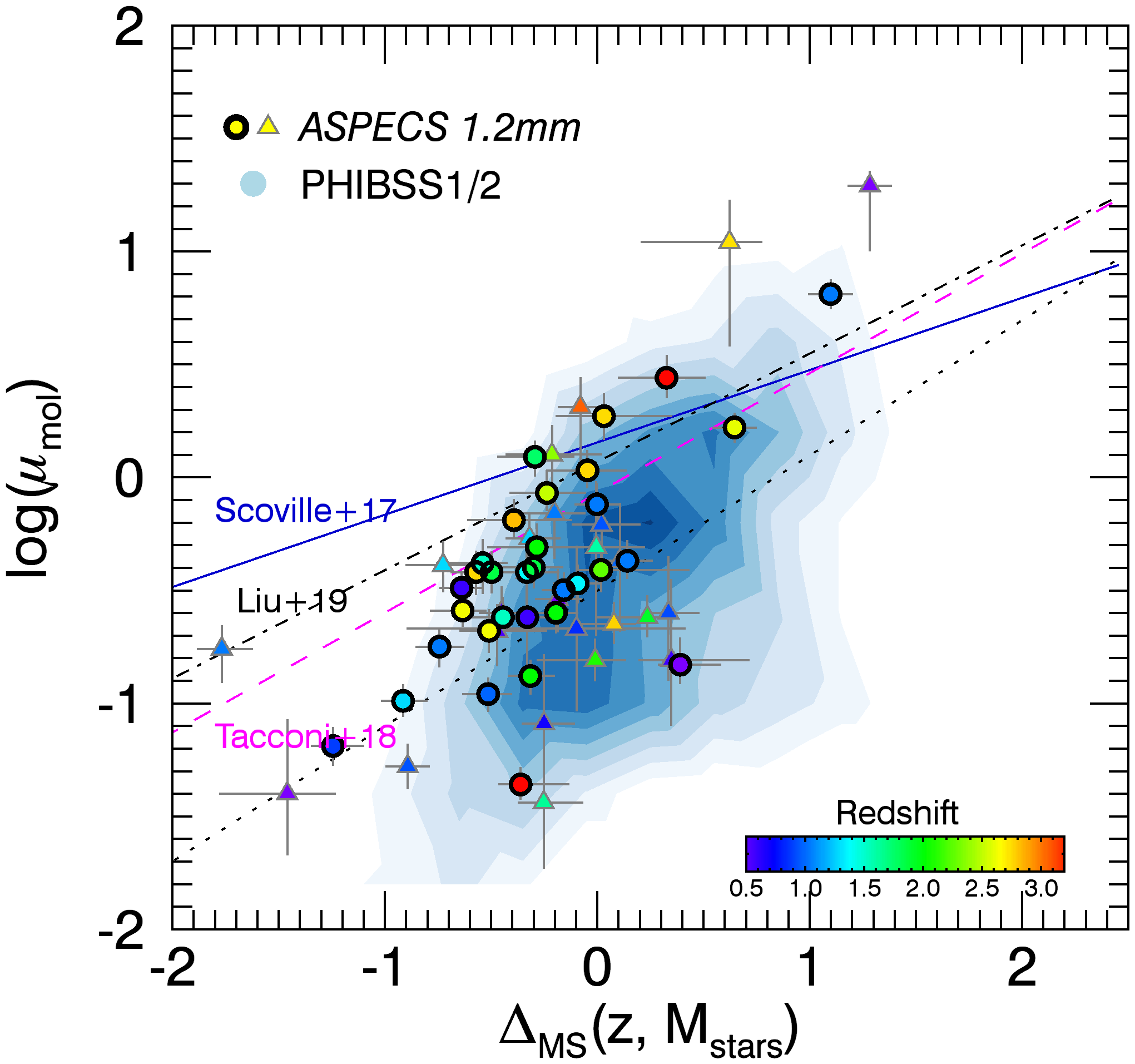} 
 \includegraphics[scale=0.6]{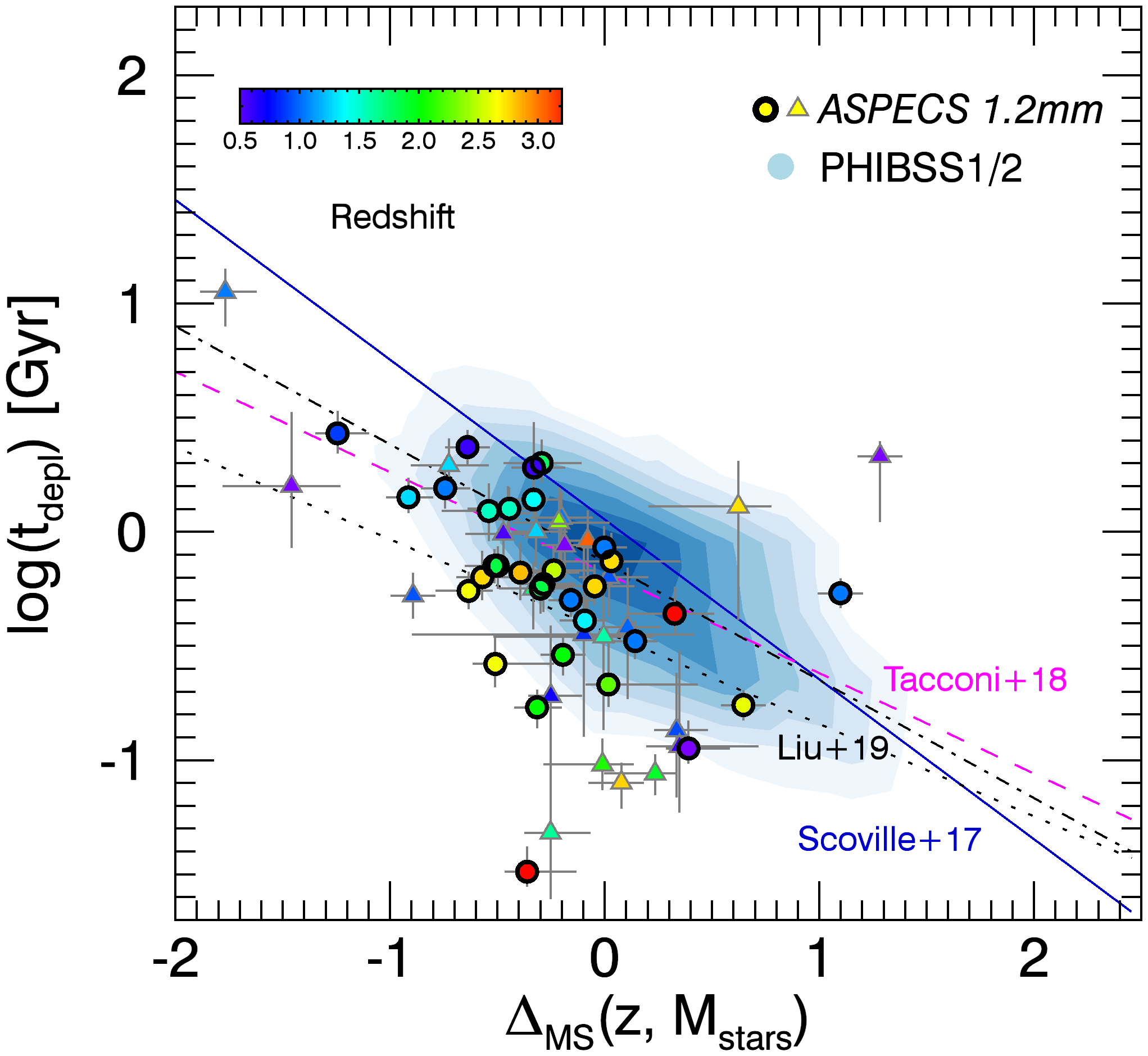} 
 \caption{ISM properties of the ASPECS 1.2 mm sources, compared to the PHIBSS1/2 sample \citep{tacconi18, freundlich19}. Filled circles and triangles show the location of the ASPECS main and faint samples, respectively. The background light blue contours represent the square root of the number of PHIBSS1/2 sources in a particular location in the diagram. The left panel shows the molecular gas to stellar mass ratio ($\mu_{\rm mol}$) as a function of the offset from the MS ($\Delta_{\rm MS}=$ log(sSFR/sSFR$_{\rm MS}$). The right panel shows the molecular gas depletion timescale ($t_{\rm dep}$) as a function of  $\Delta_{\rm MS}$. Individual ASPECS sources are color-coded according to their redshift, as denoted in the horizontal color bar. The blue solid and magenta dashed lines represent the scaling relations predicted by the models of \citet{scoville17} and \citet{tacconi18} at the median stellar mass and redshift of the ASPECS sample. The black dotted and dash-dotted lines show the prescription of \citet{liu19} when applied to the interquartiles range of the stellar masses and redshifts for the ASPECS 1.2 mm galaxies.} \label{fig:tdep} \end{figure*}

\begin{figure*}[ht] 
\centering
\includegraphics[scale=0.6]{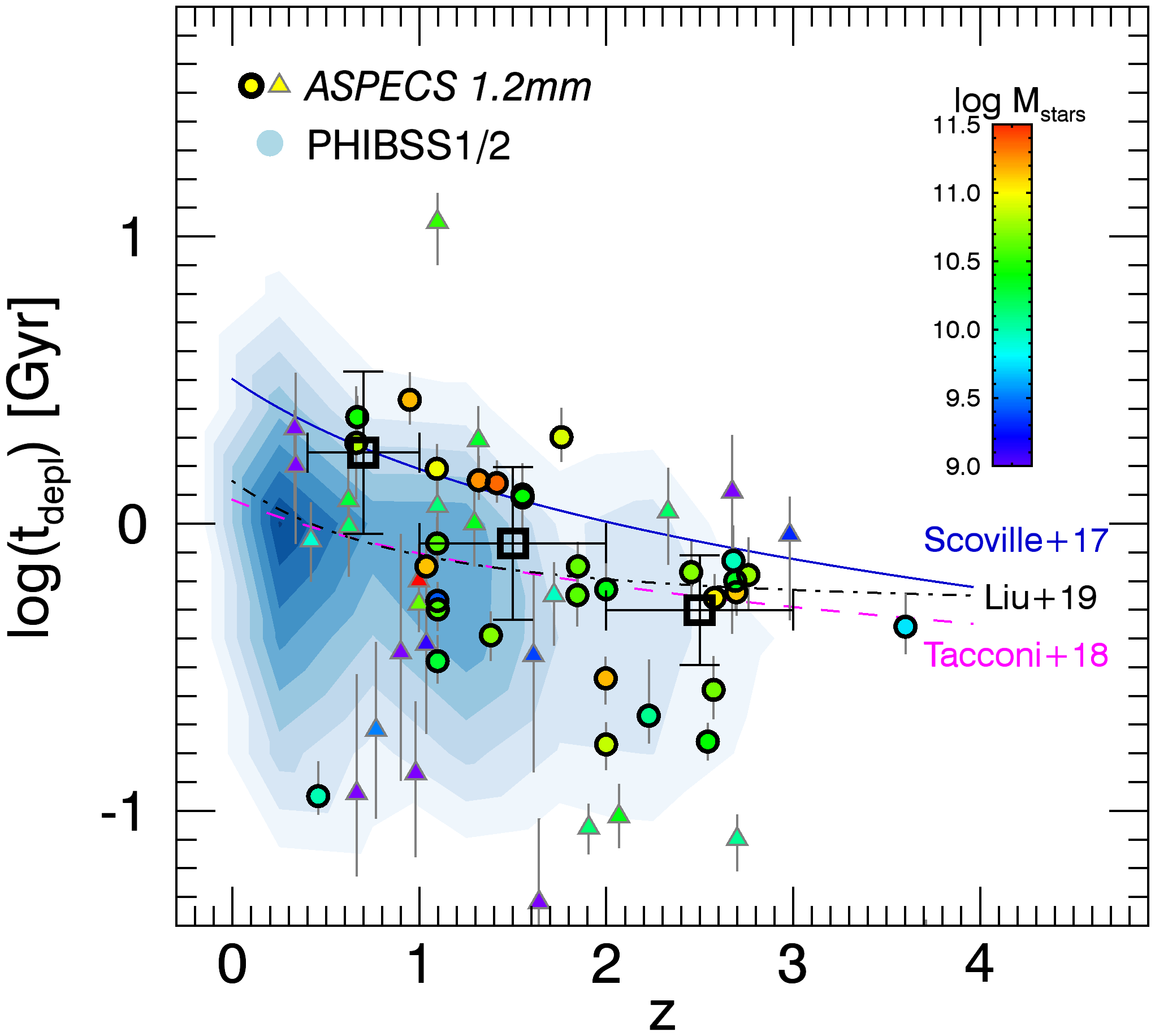}
\includegraphics[scale=0.6]{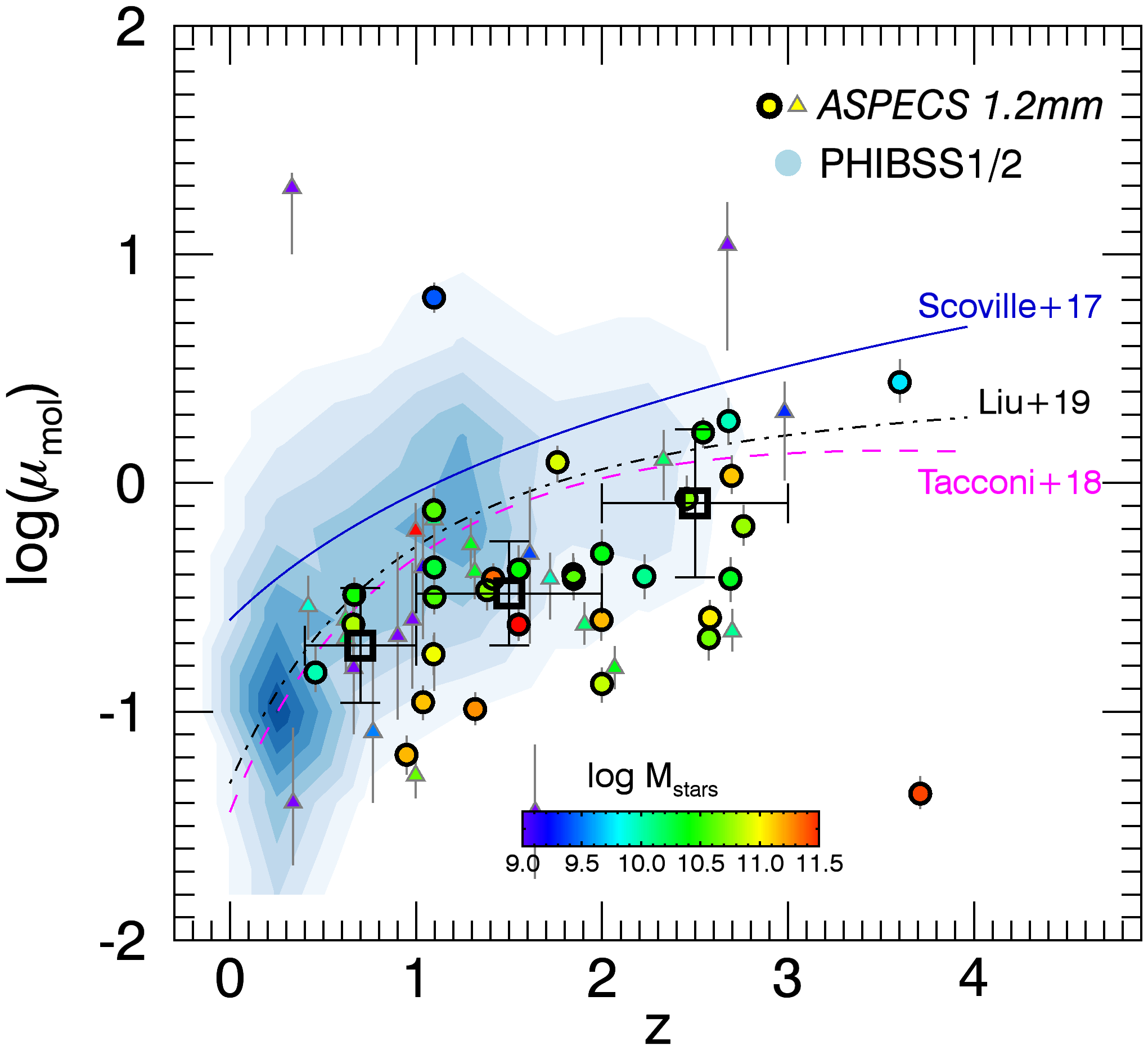} 
\caption{Evolution of the gas depletion timescale and molecular-to-stellar gas mass ratio with redshift for the ASPECS 1.2 mm continuum sources compared to galaxies from the PHIBSS1/2 survey. The background contours represent the square root of the number of PHIBSS1/2 sources in these plots. ASPECS sources in the main and faint samples are shown by circles and triangles, respectively, and are color-coded according to their stellar masses. The open squares show the weighted-averages of $t_{\rm dep}$ and $\mu_{\rm mol}$ for the ASPECS galaxies in broad redshift bins centered $z\sim0.7, 1.5$ and $2.5$, highlighting the evolution of these parameters with redshift. The blue solid and magenta dashed lines represent the scaling relations predicted by the models of \citet{scoville17} and \citet{tacconi18} at the median stellar mass of the ASPECS sample for an MS galaxy. The black dotted and dash-dotted lines show the prescription of \citet{liu19} when applied to the interquartiles range of the stellar masses and redshifts for the ASPECS 1.2 mm galaxies. \citep{tacconi13}. } 
\label{fig:evolz} 
\end{figure*}

\subsection{ISM properties}

In the following, we study and place in context the ISM properties of the ASPECS 1.2 mm galaxies with respect to previous observations of distant galaxies.  

For comparison, we use the compilation of CO line and dust continuum measurements obtained by the IRAM Plateau de Bureau HIgh-z Blue Sequence Survey \citep[PHIBSS;][]{tacconi13} and PHIBSS2 \citep[][]{tacconi18, freundlich19}. These provide stellar masses, SFRs, and molecular gas masses (from dust and CO observations) for a large sample of 1444 massive star-forming galaxies from $z=0$ to 4 \citep[see also][]{liu19}. Figure \ref{fig:mmol_z} shows the molecular gas mass as a function of redshift for the ASPECS 1.2 mm galaxies compared to the PHIBSS1/2 sample. The galaxies detected by the ASPECS 1.2 mm survey have molecular gas masses well below the lower mass envelope of the PHIBSS1/2 sample in this plane. This means that the ASPECS observations reach galaxies with lower molecular gas masses, with a well described $M_{\rm mol}$ selection function as shown by the dashed and dotted lines in this plot.

The relationship between the SFR and molecular gas mass in galaxies is usually termed the global ``Schmidt-Kennicutt'' (SK) relation or ``star-formation'' law and exposes the intimate interplay between star formation and molecular gas. The ratio between these two quantities is usually referred to as `star formation efficiency', defined as SFE$\equiv$SFR/$M_{\rm mol}$. The inverse of this relation, typically termed as the gas depletion timescale and defined as $t_{\rm dep}=M_{\rm mol}$/SFR, is rendered as the time necessary to exhaust all molecular gas reservoir in the galaxy at the current SFR in the absence of feedback mechanisms. 
 
Another key parameter corresponds to the molecular gas fraction, $f_{\rm gas}=M_{\rm mol}/(M_{\rm mol}+M_{\rm stars})$, which corresponds to the fraction of baryons contained in the form of molecular gas in a galaxy. For simplicity and for consistency in the comparison with earlier work, hereafter we conduct our analysis using the molecular-gas-to-stellar-mass ratio, $\mu_{\rm mol}=M_{\rm mol}/M_{\rm stars}$.

Both parameters, $t_{\rm dep}$ and $f_{\rm gas}$ (or $\mu_{\rm mol}$) are thought to follow scaling relations with redshift, sSFR, and stellar mass, as determined from previous targeted star-forming galaxies \citep[e.g.][]{daddi10a, tacconi10, tacconi13, tacconi18, genzel15, scoville17, liu19}. 

Figure \ref{fig:mmol_sfr} shows the SFR as a function of M$_{mol}$ (global SK plot) for the ASPECS 1.2 mm galaxies compared to the PHIBSS1/2 sample. The coloring of each point represents the normalized sSFR or offset from the MS, and the dashed lines show the location of constant molecular gas depletion timescales. Most ASPECS 1.2 mm galaxies in the main sample are located slightly above the line of $t_{\rm dep}\sim1$ Gyr, consistent with the location of the PHIBSS1/2 sources.  Only one of the ASPECS sources is consistent with gas depletion timescales shorter than 0.1 Gyr. The secondary sample shows a large scatter in gas depletion timescales with sources spanning the full range from 0.1 to 10 Gyr, although most of them are consistent with $t_{\rm dep}\sim1$ Gyr. 

The left panel of Fig. \ref{fig:tdep} shows the molecular-gas-to-stellar-mass ratio ($\mu_{\rm mol}$) versus the offset from the MS ($\Delta_{\rm MS}$). The ASPECS 1.2 mm galaxies in the main sample follow the PHIBSS1/2 sources and the expected scaling relations for their typical mass and redshift \citep{liu19}, where galaxies show an increasing $\mu_{\rm mol}$ with increasing $\Delta_{\rm MS}$. It is interesting to note that those galaxies with $\Delta_{\rm MS}<-0.5$ show molecular-gas-to-stellar-mass ratios of $\sim0.1-0.2$ (or molecular gas fractions $f_{\rm gas}\sim0.09-0.17$), consistent with recent measurements based on CO line emission in individual ``quenched'' galaxies and post-starbursts \citep[e.g.,][]{sargent15, suess17, spilker18, gobat18, bezanson19}. This indicates that galaxies below the MS have already exhausted a significant part of their molecular gas reservoirs and are in the process of being quenched.

The right panel of Fig. \ref{fig:tdep} shows the molecular gas depletion timescale versus $\Delta_{\rm MS}$. As with $\mu_{\rm mol}$, the majority of the ASPECS 1.2 mm galaxies in the main sample tend to form a sequence in this plane, where galaxies with larger positive offsets from the MS tend to have shorter depletion timescales and thus are consistent with a ``starburst'' mode of star formation, and conversely, galaxies with negative offsets from the MS have longer depletion timescales. The ASPECS 1.2 mm galaxies are in agreement with the bulk of literature PHIBSS1/2 sources in this plane and follow closely the scaling relations predicted for their range of stellar masses and redshift \citep[e.g.,][]{liu19}. As predicted from the scaling relations, the galaxies with $\Delta_{\rm MS}<-0.5$ show molecular gas depletion timescales of $\sim1-3$ Gyr.


\begin{table*} 
\centering 
\caption{Properties of the main sample of ASPECS 1.2 mm galaxies.} \label{tab:2}
\begin{tabular}{lcccccccccc} 
\hline 
ID & $z$ & $m_{\rm 160}$ & SFR & $M_{\rm stars}$ & $\Delta_{\rm MS}$ &  $T_{\rm d}$ & $L_{\rm IR}$ & $M_{\rm Mol, SED}$ &  $M_{\rm Mol, RJ}$ & $M_{\rm Mol, CO}$\\ 
  &      & (AB mag)     &  ($M_\sun$ yr${-1}$) &  ($10^{10} M_\sun$) &   &   (K)        & ($10^{11} L_\sun$) &  ($10^{10}
M_\sun$) & ($10^{10} M_\sun$) & ($10^{10} M_\sun$) \\ 
(1) & (2) & (3) & (4) & (5) & (6) & (7) & (8) & (9) & (10) & (11) \\ 
\hline\hline 

C01  &  2.543  &  23.2  &  $233_{- 23}^{+ 23}$  &  $ 2.5_{-0.2}^{+ 0.2}$  &  $ 0.65_{- 0.10}^{+ 0.10}$  &  $55_{- 1}^{+ 1}$  &  $79.4_{-  8.0}^{+ 8.0}$  &  $  4.1_{-0.4}^{+0.4}$  &  $ 10.0\pm  0.5$  &  $13.3\pm0.5$  \\
C02  &  1.760  &  23.1  &  $ 45_{-  6}^{+ 21}$  &  $ 7.2_{-  1.0}^{+ 1.1}$  &  $-0.29_{- 0.18}^{+ 0.18}$  &  $41_{- 5}^{+10}$  &  $ 5.7_{-  0.7}^{+ 2.1}$  &  $  9.0_{-2.4}^{+1.3}$  &  $  6.2\pm  0.3$  &  $\ldots$  \\
C03  &  1.414  &  22.2  &  $ 52_{-  6}^{+ 14}$  &  $18.6_{-  2.5}^{+ 2.3}$  &  $-0.33_{- 0.14}^{+ 0.11}$  &  $38_{- 1}^{+ 1}$  &  $ 8.9_{-  1.2}^{+ 1.2}$  &  $  7.1_{-0.9}^{+0.8}$  &  $  6.2\pm  0.3$  & $10.0\pm0.8$ \\
C04  &  2.454  &  25.2  &  $ 61_{- 18}^{+ 20}$  &  $ 4.8_{-  0.9}^{+ 1.2}$  &  $-0.23_{- 0.18}^{+ 0.18}$  &  $38_{- 4}^{+ 9}$  &  $ 8.3_{-  2.6}^{+ 3.0}$  &  $  4.1_{-0.7}^{+1.8}$  &  $  4.2\pm  0.2$  & $5.0\pm0.5$   \\
C05  &  1.551  &  21.7  &  $ 62_{- 20}^{+  8}$  &  $32.4_{-  3.5}^{+ 3.3}$  &  $-0.45_{- 0.18}^{+ 0.11}$  &  $38_{- 2}^{+ 4}$  &  $10.5_{-  3.1}^{+ 1.3}$  &  $  7.7_{-1.1}^{+3.2}$  &  $  6.7\pm  0.4$  & $8.8\pm0.8$   \\
C06  &  2.696  &  24.6  &  $217_{- 44}^{+ 60}$  &  $11.7_{-  2.1}^{+ 2.4}$  &  $-0.05_{- 0.18}^{+ 0.18}$  &  $37_{- 2}^{+4}$  &  $25.7_{-  5.5}^{+ 6.5}$  &  $ 12.5_{-   2.1}^{+  4.4}$  &  $ 14.0\pm  0.7$  &  $11.0\pm1.3$  \\
C07  &  2.580  &  23.3  &  $ 44_{-  8}^{+  9}$  &  $ 9.5_{-  1.1}^{+ 1.0}$  &  $-0.64_{- 0.14}^{+ 0.11}$  &  $40_{- 4}^{+ 5}$  &  $ 7.6_{-  1.1}^{+ 0.9}$  &  $  2.5_{-   0.4}^{+  0.6}$  &  $  3.1\pm  0.2$  &  $\ldots$   \\
C08  &  3.711  &  23.6  &  $300_{- 30}^{+150}$  &  $22.4_{-  2.7}^{+ 2.2}$  &  $-0.36_{- 0.10}^{+ 0.22}$  &  $55_{-1}^{+ 9}$  &  $33.9_{-  3.4}^{+16}$  &  $  1.0_{- 0.1}^{+  0.2}$  &  $  1.9\pm  0.1$  &  $\ldots$  \\
C09  &  3.601  &  25.2  &  $ 38_{-  9}^{+ 17}$  &  $ 0.6_{-  0.1}^{+ 0.1}$  &  $ 0.33_{- 0.22}^{+ 0.18}$  &  $40_{- 6}^{+12}$  &  $ 4.0_{-  1.2}^{+ 2.0}$  &  $  1.7_{- 0.4}^{+  0.7}$  &  $  1.9\pm  0.1$  &  $\ldots$  \\
C10  &  1.997  &  21.6  &  $109_{- 11}^{+ 11}$  &  $12.6_{-  1.3}^{+ 1.3}$  &  $-0.20_{- 0.10}^{+ 0.10}$  &  $48_{- 2}^{+ 2}$  &  $26.9_{-  2.8}^{+ 2.7}$  &  $  3.2_{-1.0}^{+  0.8}$  &  $  4.8\pm  0.5$  & $2.0\pm0.3$   \\
C11  &  2.760  &  24.2  &  $ 53_{- 21}^{+ 35}$  &  $ 5.4_{-  0.8}^{+ 1.1}$  &  $-0.39_{- 0.22}^{+ 0.27}$  &  $37_{- 5}^{+ 9}$  &  $ 6.6_{-  2.3}^{+ 3.9}$  &  $  3.5_{-   0.8}^{+  1.3}$  &  $  3.7\pm  0.3$  & $\ldots$   \\
C12  &  1.096  &  23.4  &  $ 36_{-  4}^{+  4}$  &  $ 0.3_{-  0.1}^{+ 0.1}$  &  $ 1.09_{- 0.10}^{+ 0.10}$  &  $35_{- 1}^{+ 2}$  &  $ 4.6_{-  0.5}^{+ 0.5}$  &  $  1.9_{-   0.2}^{+  0.2}$  &  $  1.6\pm  0.1$  &  $1.5\pm0.2$  \\
C13  &  1.037  &  21.8  &  $ 18_{-  2}^{+  2}$  &  $12.0_{-  1.8}^{+ 1.3}$  &  $-0.52_{- 0.11}^{+ 0.11}$  &  $35_{- 1}^{+2}$  &  $ 4.4_{-  0.5}^{+ 0.4}$  &  $  1.3_{-   0.3}^{+  0.3}$  &  $  1.6\pm  0.2$  &   $3.7\pm0.4$ \\
C14a  &  1.999  &  23.6  &  $ 50_{- 14}^{+ 10}$  &  $ 6.5_{-  1.5}^{+ 0.8}$  &  $-0.32_{- 0.10}^{+ 0.11}$  &  $51_{- 3}^{+ 1.0}$  &  $ 8.7_{-  2.4}^{+ 1.8}$  &  $  0.8_{-   0.1}^{+  0.2}$  &  $  1.3\pm  0.1$  &   $\ldots$ \\
C14b  &  1.999  &  22.7  &  $ 19_{-  3}^{+  2}$  &  $ 2.2_{-  0.3}^{+ 0.2}$  &  $-0.29_{- 0.11}^{+ 0.14}$  &  $37_{- 4}^{+ 8}$  &  $ 1.6_{-  0.2}^{+ 0.2}$  &  $  1.1_{- 0.5}^{+  0.6}$  &  $  1.3\pm  0.3$  &  $\ldots$  \\
C15  &  1.317  &  21.4  &  $ 11_{-  1}^{+  3}$  &  $15.5_{-  1.9}^{+ 1.7}$  &  $-0.91_{- 0.10}^{+ 0.11}$  &  $36_{- 2}^{+ 5}$  &  $ 3.4_{-  0.4}^{+ 0.4}$  &  $  1.6_{- 0.2}^{+0.3}$  &  $  1.7\pm  0.2$  &  $4.7\pm0.4$  \\
C16  &  1.095  &  20.8  &  $ 33_{-  3}^{+  3}$  &  $ 3.7_{-  0.4}^{+ 0.4}$  &  $-0.01_{- 0.10}^{+ 0.10}$  &  $45_{- 9}^{+ 5}$  &  $ 3.4_{-  0.3}^{+ 0.3}$  &  $  2.8_{-0.7}^{+  1.2}$  &  $  2.0\pm  0.2$  &  $3.3\pm0.4$  \\
C17  &  1.848  &  22.4  &  $ 22_{-  4}^{+  2}$  &  $ 4.1_{-  0.4}^{+ 0.9}$  &  $-0.50_{- 0.14}^{+ 0.11}$  &  $37_{- 5}^{+10}$  &  $ 2.2_{-  0.3}^{+ 0.5}$  &  $  1.5_{-   0.5}^{+  0.5}$  &  $  1.4\pm  0.2$  & $\ldots$   \\
C18  &  1.845  &  22.0  &  $ 27_{-  8}^{+  9}$  &  $ 3.8_{-  0.4}^{+ 0.5}$  &  $-0.30_{- 0.18}^{+ 0.14}$  &  $39_{- 6}^{+ 10}$  &  $ 2.9_{-  0.7}^{+ 1.3}$  &  $  1.5_{-0.4}^{+  0.6}$  &  $  1.5\pm  0.2$  &  $\ldots$  \\
C19  &  2.574  &  23.1  &  $ 34_{-  7}^{+ 16}$  &  $ 4.4_{-  0.6}^{+ 0.5}$  &  $-0.50_{- 0.10}^{+ 0.27}$  &  $51_{- 9}^{+ 8}$  &  $ 4.1_{-  1.1}^{+ 1.6}$  &  $  0.9_{-   0.3}^{+  0.4}$  &  $  1.1\pm  0.2$  &  $1.8\pm0.2$  \\
C20  &  1.093  &  21.4  &  $  9_{-  2}^{+  2}$  &  $ 8.1_{-  1.6}^{+ 1.0}$  &  $-0.74_{- 0.11}^{+ 0.11}$  &  $34_{- 3}^{+ 4}$  &  $ 1.8_{-  0.2}^{+ 0.3}$  &  $  1.4_{-   0.4}^{+  0.4}$  &  $  1.3\pm  0.2$  &  $\ldots$  \\
C21  &  2.690  &  23.7  &  $ 13_{-  2}^{+  5}$  &  $ 2.1_{-  0.3}^{+ 0.2}$  &  $-0.58_{- 0.11}^{+ 0.22}$  &  $37.6_{- 5.7}^{+11.1}$  &  $ 1.29_{-  0.3}^{+ 0.51}$  &  $  0.8_{-   0.3}^{+  0.3}$  &  $  0.8\pm  0.1$  &  $\ldots$  \\
C22  &  1.097  &  21.3  &  $ 25_{-  3}^{+  4}$  &  $ 1.9_{-  0.2}^{+ 0.3}$  &  $ 0.15_{- 0.11}^{+ 0.10}$  &  $48_{-10}^{+ 9}$  &  $ 2.34_{-  0.3}^{+ 0.3}$  &  $  0.8_{-   0.2}^{+  0.3}$  &  $  0.8\pm  0.1$  &  $\ldots$  \\
C23  &  1.382  &  21.3  &  $ 40_{-  6}^{+  5}$  &  $ 4.8_{-  0.5}^{+ 0.5}$  &  $-0.10_{- 0.11}^{+ 0.10}$  &  $37_{- 1}^{+ 2}$  &  $ 5.0_{-  1.0}^{+ 0.6}$  &  $  1.6_{-   0.4}^{+  0.5}$  &  $  2.1\pm  0.4$  &  $1.7\pm0.3$  \\ 
C24  &  2.680  &  24.4  &  $ 24_{-  6}^{+ 12}$  &  $ 1.0_{-  0.2}^{+ 0.2}$  &  $ 0.03_{- 0.22}^{+ 0.32}$  &  $38_{- 6}^{+11}$  &  $ 2.5_{-  0.7}^{+ 1.6}$  &  $  1.8_{-   0.6}^{+  0.8}$  &  $  1.8\pm  0.3$  & $\ldots$   \\
C25  &  1.098  &  21.6  &  $ 22_{-  2}^{+  4}$  &  $ 3.5_{-  0.4}^{+ 0.5}$  &  $-0.16_{- 0.10}^{+ 0.10}$  &  $35_{- 1}^{+ 2}$  &  $ 2.6_{-  0.3}^{+ 0.5}$  &  $  1.1_{-   0.2}^{+  0.3}$  &  $  1.2\pm  0.2$  &  $2.4\pm0.3$  \\
C26  &  1.552  &  22.9  &  $  8_{-  2}^{+  3}$  &  $ 2.5_{-  0.6}^{+ 0.3}$  &  $-0.54_{- 0.22}^{+ 0.11}$  &  $42_{- 7}^{+ 9}$  &  $ 1.1_{-  0.2}^{+ 0.4}$  &  $  1.0_{-   0.5}^{+  0.5}$  &  $  0.9\pm  0.2$  &  $\ldots$  \\
C28  &  0.662  &  19.5  &  $  7_{-  1}^{+  7}$  &  $ 5.9_{-  2.1}^{+ 0.9}$  &  $-0.33_{- 0.10}^{+ 0.14}$  &  $34_{- 5}^{+11}$  &  $ 0.6_{-  0.1}^{+ 0.9}$  &  $  1.4_{-   0.9}^{+  1.5}$  &  $  2.0\pm  0.5$  &   $\ldots$ \\
C30  &  0.458  &  20.4  &  $ 13_{-  1}^{+  1}$  &  $ 1.0_{-  0.3}^{+ 0.1}$  &  $ 0.39_{- 0.10}^{+ 0.18}$  &  $41_{- 1}^{+ 8}$  &  $ 1.1_{-  0.1}^{+ 0.2}$  &  $  0.1_{-   0.0}^{+  0.1}$  &  $  0.3\pm  0.1$  &  $\ldots$   \\
C31  &  2.227  &  22.6  &  $ 22_{-  2}^{+ 31}$  &  $ 1.2_{-  0.1}^{+ 0.1}$  &  $ 0.01_{- 0.10}^{+ 0.41}$  &  $50_{-11}^{+ 8}$  &  $ 1.9_{-  0.2}^{+ 4.0}$  &  $  0.5_{-   0.2}^{+  0.2}$  &  $  0.7\pm  0.2$  &   $\ldots$  \\
C32  &  0.667  &  21.8  &  $  4_{-  0}^{+  0}$  &  $ 2.6_{-  0.3}^{+ 0.3}$  &  $-0.63_{- 0.10}^{+ 0.10}$  &  $35_{- 5}^{+ 9}$  &  $ 0.6_{-  0.1}^{+ 0.1}$  &  $  0.8_{-   0.2}^{+  0.2}$  &  $  0.4\pm  0.1$  &  $\ldots$  \\
C33  &  0.948  &  20.1  &  $  3_{-  0}^{+  1}$  &  $12.6_{-  1.3}^{+ 1.5}$  &  $-1.24_{- 0.10}^{+ 0.14}$  &  $32_{- 3}^{+ 6}$  &  $ 0.7_{-  0.1}^{+ 0.4}$  &  $  0.8_{-   0.2}^{+  0.2}$  &  $  0.6\pm  0.1$  &  $\ldots$  \\
 \hline\hline \end{tabular}

\flushleft \noindent {\bf Note.} For all \textsc{magphys}-derived parameters, an additional 0.1 dex error has been added in quadrature to the original MAGPHYS uncertainties. This is particularly important in cases with excellent SED fits, where low uncertainty values are produced due to the discrete sampling spacing of the underlying SED templates. Column (1): source ID, ASPECS-LP.1mm.xx. Column (2): best redshift available (see Table \ref{tab:1} for redshift references). Column (3): AB magnitude in the F160W HST band. Columns (4)-(8): SFR, stellar mass, normalized specific SFR ($\Delta_{\rm MS}$), dust temperature ($T_{\rm d}$) and IR luminosity ($L_{\rm IR}$), derived from \textsc{magphys} SED fitting. Column (9): molecular gas mass derived from the dust mass delivered by \textsc{magphys} and a gas-to-dust ratio $\delta_{\rm GDR}=200$. Column (10): molecular gas mass obtained from the 1.2 mm flux and the calibrations from \citet{scoville14}. Column (11): molecular gas mass obtained from the CO line emission detected by ASPECS 3mm spectroscopy. To convert the CO $J>1$ to the ground transition, we use the average line ratios derived for the ASPECS sample itself \citep[][]{boogaard20}, with $r_{21}=0.83\pm0.12$, $r_{31}=0.58\pm0.10$ and $r_{41}=0.30\pm0.08$ for galaxies at $z<1.7$ and $r_{21}=1.02\pm0.18$, $r_{31}=0.92\pm0.17$ and $r_{41}=0.76\pm0.16$ for galaxies at $z>1.7$ (see Appendix \ref{app_a}).  For consistency with previous ASPECS work, a fixed $\alpha_{\rm CO}=3.6$ $M_\odot$ (K km s$^{-1}$)$^{-1}$ is used. This represents a systematic underestimation of $<$0.1 dex with respect to values obtained when using a metallicity-dependent $\alpha_{\rm CO}$ scheme.

\end{table*}

\subsection{Evolution of the ISM}

Figure \ref{fig:evolz} shows the evolution with redshift of the gas depletion timescales and molecular gas ratio for the ASPECS 1.2 mm galaxies, compared with the PHIBSS1/2 sample and with the standard scaling relations for these parameters \citep[e.g.,][]{scoville17, tacconi18, liu19}. We find that the ASPECS 1.2 mm galaxies exhibit the mild evolution of $t_{\rm dep}$ with redshift as expected from previous observations and models. As seen previously for the ASPECS CO-selected galaxies \citep{aravena19}, the ASPECS 1.2 mm galaxies tend to be slightly above the bulk of PHIBSS1/2 sources; however, the average $t_{\rm dep}$ obtained for the main sample in different redshift bins (open black squares) are in agreement with the evolution of $t_{\rm dep}$ for MS galaxies predicted by \citet{tacconi18} and \citet{liu19} at the median stellar mass and redshift of the ASPECS 1.2 mm sample. Two of the ASPECS galaxies show significantly lower $t_{\rm dep}$ values than those expected for MS galaxies at their respective redshifts, yet consistent with the spread of values shown by the PHIBBS1/2 galaxies. Overall, the mild evolution of the average gas depletion timescales yields a typical $t_{\rm dep}\sim0.7$ Gyr in the redshift range $z=1-3$.

Similarly, the ASPECS 1.2 mm galaxies follow and support the general trend of increasing the molecular gas ratio by roughly a factor of 5.0 from $z=0.5$ to $z\sim3$. As with $t_{\rm dep}$, we find that the ASPECS sources in the main sample lie slightly below the region occupied by the PHIBSS1/2 galaxies, although the average $\mu_{\rm mol}$ points fall within the uncertainties from the values predicted for MS galaxies by \citet{tacconi18} and \citet{liu19} at the median stellar mass and redshift of the ASPECS 1.2 mm sample. The lower $\mu_{\rm mol}$ values compared to the PHIBSS1/2 sample argue for a milder evolution of this parameter for ASPECS 1.2 mm galaxies. We also find that one and two ASPECS sources from the main and secondary samples, respectively, have log($\mu_{\rm mol})>0.8$, consistent with a dominant molecular phase component and well above the general trend paved by the PHIBSS1/2 galaxies and the ASPECS main sample sources. 

\subsection{Molecular gas budget}

One of the most important results of ``molecular deep field'' observations, as exemplified by the ASPECS project, corresponds to the determination of the cosmic molecular gas density as a function of redshift ($\rho_{\rm H2}(z)$), as it provides a measurement of the amount of material available to support star formation \citep[e.g.,][]{walter14, decarli16a, decarli19, riechers19}. Measuring this parameter requires observations of molecular gas through CO line or dust continuum emission over a significantly large, well-defined cosmological volume. Current studies from the various surveys present a consistent picture for the evolution of  $\rho_{\rm H2}$ out to $z\sim3$, showing minor changes of this parameter between $z=3$ to 1 and a steep decline from $z=1$ to 0, thus following very closely the evolution of the cosmic SFR density in this redshift range \citep{decarli16a, riechers19, decarli19, magnelli20}.  It is thus interesting to ask, what is the nature of the galaxies that dominate $\rho_{\rm H2}$ with redshift? Or equivalently, what is the contribution of different galaxy types to $\rho_{\rm H2}$?

We here follow the same approach introduced in \citet{aravena19} and compute the fraction of $\rho_{\rm H2}$ contributed by galaxies within, above and below the MS, at three redshift bins. From the full sample of ASPECS 1.2 mm galaxies (main and secondary samples), we select sources with $\Delta_{\rm MS}>0.4$, $\Delta_{\rm MS}<-0.4$ and $-0.4<\Delta_{\rm MS}<0.4$ to be above, below ,and within the MS, respectively. We subdivided each of these samples into three redshift bins $z=0.4-1.0, 1.0-2.0$ and $2.0-3.0$, and added up the derived dust-based, molecular gas masses for all galaxies in each redshift bin and galaxy type. Each of these measurements was thus divided by the total molecular gas mass, obtained from the value of $\rho_{\rm H2}$ (i.e., $\rho_{\rm H2}(z)\times{\rm Volume}(z)$) for that redshift bin from the ASPECS survey \citep{decarli19}, thus yielding the fraction of $\rho_{\rm H2}$ contributed from galaxies in a particular galaxy class (or cosmic molecular gas budget). The addition of the molecular gas masses from all galaxies in a particular redshift bin (divided by $\rho_{\rm H2}(z)$) yields to full contribution from the individually detected galaxies to the cosmic molecular gas budget. The uncertainties in the molecular gas budget value are computed as the sum in quadrature of the individual molecular gas mass values and the statistical uncertainty, which follows a binomial distribution, scaled to the total molecular gas in that redshift bin.

\begin{figure}
    \centering
    \includegraphics[scale=0.55]{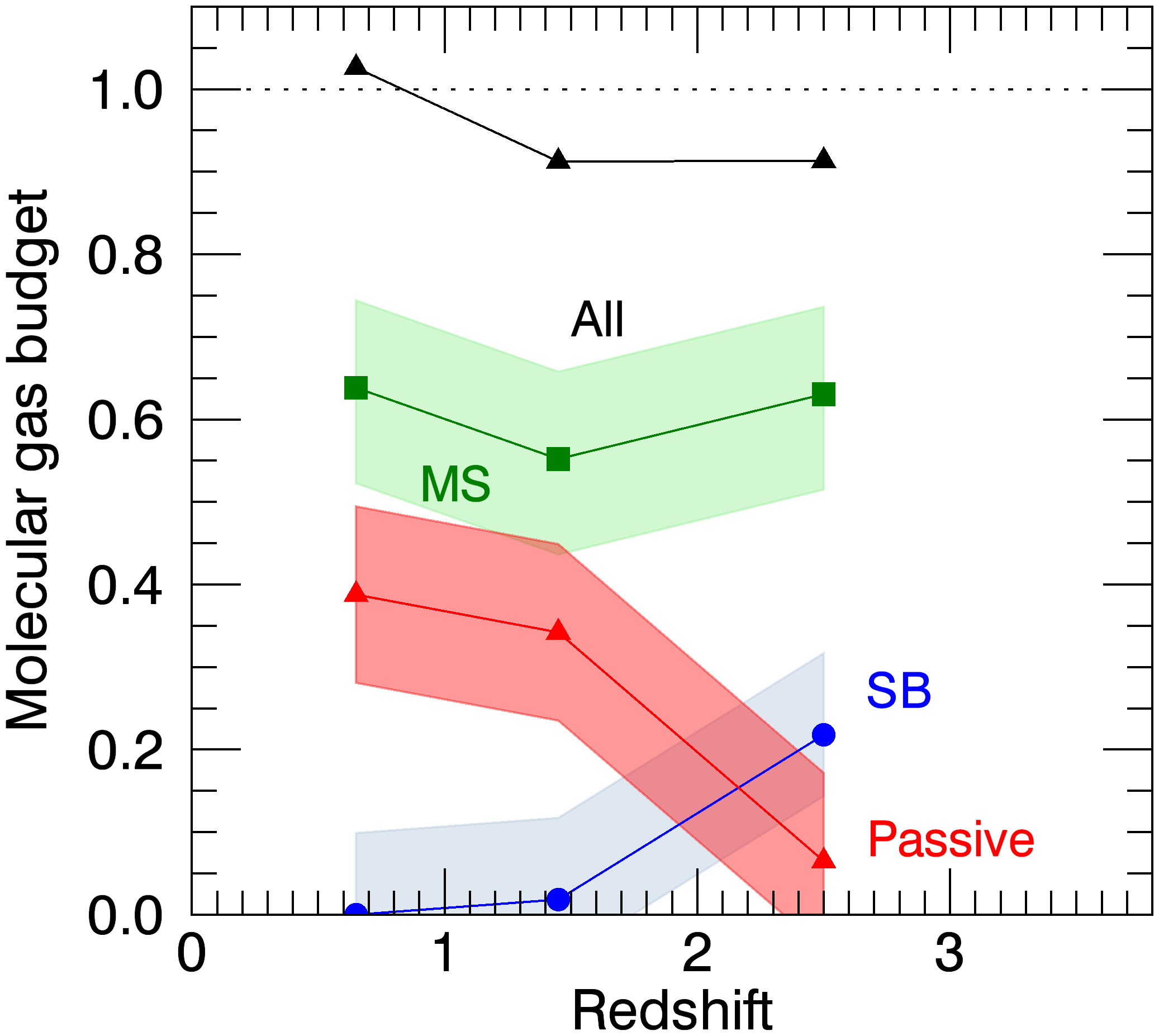}
    \caption{Contribution to the total molecular gas budge from galaxies above (starburst), in or below (passive) the MS (defined in Section \ref{sec:ms}) as a function of redshift inferred from the ASPECS 1.2 mm dust continuum observations survey. The blue, green and red data points and lines represent galaxies above, on and below the MS, respectively. The black curve shows the contribution of all the ASPECS 1.2 mm galaxies considered here to the total molecular gas at each redshift. Each data point is computed from the sum of molecular gas masses (estimated from the dust continuum) of all galaxies in that redshift bin and galaxy type. The redshift measurement of each point is computed as the average redshift from all galaxies in that bin. The shaded region represents the uncertainties of each measurement. Galaxies below the MS appear to have an increasing contribution to the cosmic density of molecular gas at lower redshifts.}
    \label{fig:gasbudget}
\end{figure}

Figure \ref{fig:gasbudget} shows the results from this procedure. The dotted horizontal line represents a fraction of $\rho_{\rm H2}$ equal to 100\% at each redshift. We find that the contribution to $\rho_{\rm H2}$ from all galaxies studied here to the molecular gas budget corresponds to almost 100\% at $z<2.0$ decreasing to 80\% at $z\sim2.5$, implying that the galaxies in this study account for almost all the cosmic density of molecular gas measured in the HUDF. This is in good agreement with the fact that the ASPECS 1.2 mm galaxies, including the secondary sample, make up close to 100\% of the EBL expected in the HUDF at this wavelength. Furthermore, we find that the dominant population of $\rho_{\rm H2}$ at all redshifts corresponds to MS galaxies. There is a strong evolution with increasing redshift. Galaxies above the MS, which contribute with about 20\% of $\rho_{\rm H2}$ at $z\sim2.5$, suffer a steep decline at $z<1$ dropping their contribution to $\sim0\%$. Meantime, the contribution from galaxies within the MS stays relatively constant at $\sim60\%$ of $\rho_{\rm H2}$ from $z\sim2.5$ to $z\sim0$. Surprisingly, galaxies below the MS increase their contribution to $\rho_{\rm H2}$ from $\sim5\%$ at $z\sim2.5$ to $\sim40\%$ at $z<1$. This result suggests an overall cessation of star formation in galaxies on and above the MS from $z\sim2.5$ to $z<1$, which coincides with an increased abundance of below-MS galaxies. These galaxies, which ceased their star formation activity, would still retain a significant portion of their dust and molecular gas reservoirs, contributing an important fraction to $\rho_{\rm H2}$ at $z<1$.
We note that since the galaxies in this study make up most of $\rho_{\rm H2}$ at $z<3$, there is little room for a significantly larger contribution from above or below the MS galaxies that we might be missing as a results of the limited areal coverage of the ASPECS field (i.e. massive galaxies above the MS are less abundant and thus would only be found in larger-area surveys).

These trends are in agreement with previous results obtained from the ASPECS CO sample alone presented by \citet{aravena19}, and thus the physical interpretation given there is also applicable here. Our results follow consistently the contributions from galaxies above and in the MS to the cosmic SFR density as a function of redshift \citep[e.g.,][]{sargent12, schreiber15}. Furthermore, our results are in apparent disagreement with the large molecular gas masses found for massive early-type galaxies at $z\sim1.8$ \citep[e.g. below the MS galaxies][]{gobat18}. However, it seems plausible that while galaxies below the MS might have larger molecular gas reservoirs at $z>1.5$, they are less abundant and thus represent a minor fraction of $\rho_{\rm H2}$. At lower redshifts, the average molecular gas content of galaxies below the MS is lower; however, this population is more numerous, thus contributing a larger fraction of $\rho_{\rm H2}$. 


\section{Conclusions and closing remarks}

In this paper, we have measured and analyzed the properties of a sample of 61 faint, dust-selected galaxies found in the deep ALMA 1.2 mm map of the HUDF as part of the ASPECS LP; 35 and 26 of these sources form the main and secondary samples, respectively. The integrated emission from these individual galaxies makes up most of the 1.2 mm light measured in the HUDF \citep{gonzalezlopez20}, and thus accounts for almost all the dust emission found in this region of the sky. As such, these galaxies represent a unique opportunity to study the evolution of the dust and molecular gas mass with cosmic time in a close to statistically complete fashion. The main results from this study can be summarized as follows:

\begin{itemize}
    \item Out of the 35 sources blindly detected at 1.2 mm, 32 have clear optical counterparts in the HST images and 25 have Herschel PACS counterparts. Additionally, we select 26 millimeter sources using optical and IR priors. Out of the 18 CO line emitters detected by the 3mm line spectroscopy \citep{gonzalezlopez19, boogaard19, aravena19}, 15 have a 1 mm continuum counterpart at the $>3\sigma$ level. Most of these sources have an accurate spectroscopic redshift from MUSE or ALMA spectroscopy.
    
    \item For all sources with reliable counterparts, we derived physical properties using \textsc{magphys}. We find a large range in stellar masses and SFRs for the ASPECS galaxies in the main sample, with median values of $4.3\times10^{10}$ $M_\sun$ and $30$ $M_\sun$ yr$^{-1}$, and interquartile ranges of and interquartile ranges of $(2.4-11.7)\times10^{10}\ M_\sun$ and $20-50$ $M_\sun$ yr$^{-1}$, respectively. 
   
    \item The ASPECS 1.2 mm galaxies in the main sample are found to have a median redshift of $1.8$, with an interquartile range of $1.1-2.6$. The median redshift for the ASPECS galaxies is thus significantly lower than those found for brighter SMGs.
    
    \item Overall, 59\% of the ASPECS 1 mm galaxies in the main sample are consistent with the MS of star formation at their respective redshift, with 34\% and 6\% lying more than 0.4 dex below and above the MS, respectively. We find similar fractions for the secondary ASPECS sample. A wider spread in $\Delta_{\rm MS}$ with respect to UDF galaxies and systematically lower values are found for ASPECS galaxies. A tentative trend of $\Delta_{\rm MS}$ with stellar mass is found; however no trend is found with redshift. These results point to a relevant role of massive below-MS galaxies as molecular gas reservoirs.
 
    \item The ASPECS 1.2 mm galaxies follow tightly the relationship between SFR and $M_{\rm mol}$, consistent with that found for previous samples. 
    
    \item We find that the ASPECS 1.2 mm galaxies follow the trends of molecular gas depletion timescale and molecular-to-stellar ratio with $\Delta_{\rm MS}$ expected from standard scaling relations. In particular, we find that sources significantly below the MS, with $\Delta_{\rm MS}<-0.5$, have $\mu_{\rm mol}\sim0.1-0.2$, consistent with recent findings from the literature.

    \item The ASPECS 1.2 mm galaxies consistently follow the evolution of $t_{\rm dep}$ and $\mu_{\rm mol}$ with redshift expected from standard scaling relations. Our observations support the mild evolution of $t_{\rm dep}$ with redshift, yielding a typical depletion timescale of 0.7 Gyr at $z=1-3$, and confirm the decrease by a factor of 5 in the molecular-gas-to-stellar mass ratio from $z=3$ to $z=0.5$.
    
    \item We find substantial evidence for a changing contribution from different classes of galaxies to the cosmic density of molecular gas as a function of redshift. While star-forming MS galaxies appear to dominate at all redshifts, galaxies below and above the MS significantly increase/decrease their contribution with decreasing redshift, from $z\sim2.5$ to $z<1$. This is attributed to a higher abundance of passive galaxies at lower redshifts even though they are expected to have higher molecular gas reservoirs at $z>1$, hinting at a cessation of star-formation activity in passive galaxies at lower redshifts.
\end{itemize}

The ASPECS LP survey has enabled a complementary view of the evolution of the ISM content through cosmic time in a unique flux-limited, dust-mass-selected sample of galaxies. Overall, the derived properties of these galaxies are consistent with standard scaling relations previously established through targeted observations of molecular gas and dust. The large fraction of sources classified to be below the MS, as well as the increasingly important role of these galaxies in the cosmic molecular gas density ($\rho_{\rm H2}$), indicates that this population of galaxies has so far been overlooked. 

Despite the importance of this complementary approach, and progress made so far, expanding significantly the current ASPECS LP observations beyond the XDF/HUDF (in either band-3 or 6) will be difficult with current instrumentation and facilities. Particularly, adding $>3$ times more areal coverage to the current ASPECS footprint at similar depth ($\sim10-13\mu$Jy at 1.2 mm continuum) will require $>500$ hr of observing time with ALMA. Given the already large number of sources ($>1000$) for which global ISM properties have been derived \citep[e.g.,][]{tacconi18, liu19}, future observational efforts should concentrate on understanding the physics involved in galaxy build up, through high-resolution observations of molecular gas and dust continuum. Similarly, it will be critical to understand the accuracy and applicability of molecular gas mass estimators in the ASPECS galaxies through dedicated observations of the CO(1-0) and [CI] emission lines as additional key tracers of the ISM.

\acknowledgements 
MA has been supported by the grant ``CONICYT + PCI + INSTITUTO MAX PLANCK DE ASTRONOMIA MPG190030'' and ``CONICYT+PCI+REDES 190194''. J.G-L. acknowledges partial support from ALMA-CONICYT project 31160033. F.W. and M.N. acknowledge funding from the ERC Advanced Grant 'Cosmic Gas'. FEB acknowledges support from CONICYT grants CATA-Basal AFB-170002 (FEB), FONDECYT Regular 1190818 (FEB), 1200495 (FEB) and Chile's Ministry of Economy, Development, and Tourism's Millennium Science Initiative through grant IC120009, awarded to The Millennium Institute of Astrophysics, MAS (FEB). T.D-S. acknowledges support from ALMA-CONICYT project 31130005 and FONDECYT project 1151239. T.D-S. acknowledges support from the CASSACA and CONICYT fund CAS-CONICYT Call 2018. D.R. acknowledges support from the National Science Foundation under grant numbers AST-1614213 and AST-1910107 and from the Alexander von Humboldt Foundation through a Humboldt Research Fellowship for Experienced Researchers. I.R.S. acknowledges support from STFC (ST/P000541/1). L. H. I. acknowledges support from JSPS KAKENHI Grant Number JP19K23462. The National Radio Astronomy Observatory is a facility of the National Science Foundation operated under cooperative agreement by Associated Universities, Inc. This paper makes use of the following ALMA data: 2016.1.00324. ALMA is a partnership of ESO (representing its member states), NSF (USA) and NINS (Japan), together with NRC (Canada), NSC and ASIAA (Taiwan), and KASI (Republic of Korea), in cooperation with the Republic of Chile. The Joint ALMA Observatory is operated by ESO, AUI/NRAO and NAOJ.

\bibliographystyle{apj} \bibliography{dustpecs}

\appendix
\label{app4}
\label{app_a}
\section{Comparison of the various molecular gas mass estimates}
The availability of a well-sampled multi-wavelength SED, dust continuum and CO line measurements for the ASPECS 1.2 mm sources allows us to compute the molecular gas mass of our galaxies in various ways. In the following, we describe the methods used to compute molecular gas masses and thereby present a comparison between these estimates. These various estimates are listed in Tables \ref{tab:2} and \ref{tab:4}.

\subsection{CO-based estimate}

CO line detections ($J_{\rm up}=2-4$) are available for a subsample of our sources from the ASPECS 3 mm line scan, and the associated flux densities can be used to obtain molecular gas masses $M_{\rm mol, CO}$ \citep[for details see ][]{aravena19, boogaard19}. In short, the CO line flux densities are used to obtain CO line luminosities ($L'_{\rm CO[J\rightarrow(J-1)]}$).  The luminosities are converted into the ground-state CO(1-0) line luminosities ($L'_{\rm CO(1-0)}$) using the average CO line ratios derived from the extended sample of ASPECS CO galaxies at $z\sim1.5$ and $z\sim2.5$ \citep{boogaard20, riechers20}. We use the following line ratios \citep[for details see][]{boogaard20}: for galaxies at $z=1.0-1.6$, we use $r_{21}=0.83\pm0.12$, $r_{31}=0.58\pm0.10$ and $r_{41}=0.30\pm0.08$; for galaxies at $z=2.0-2.7$, we use $r_{21}=1.02\pm0.18$, $r_{31}=0.92\pm0.17$ and $r_{41}=0.76\pm0.16$. These line ratios are consistent with a higher excitation (close to local thermal equilibrium up to $J<4$) than the average CO spectral line energy distribution (SLED) of (typically more massive) BzK galaxies at $z=1.5$ previously studied by \citet{daddi15}. We thus  convert the line luminosities to molecular gas masses using $M_{\rm mol, CO}=\alpha_{\rm CO} L'_{\rm CO(1-0)}$, where $\alpha_{\rm CO}$ is the CO-luminosity-to-gas-mass conversion factor, assumed to be equal to $3.6\pm0.8\ M_\sun$ (K km s$^{-1}$ pc$^2$)$^{-1}$ \citep{daddi10a}. 

In the next section, we discuss the effect of assuming a metallicity ($Z$) dependence of $\alpha_{\rm CO}$. As explained in \citet{aravena19}, assuming a metallicity-dependent $\alpha_{\rm CO}$ yields a factor +0.1 dex larger $M_{\rm mol, CO}$. Since only a handful of our sources have any estimate of their metallicity based on optical spectroscopy \citep{boogaard19}, we use the range of stellar masses of our galaxies and the well-known stellar-mass$-$metallicity (MZ) relation to provide a rough estimate of $Z$. For this, we adopt the MZ relationship computed by \citet{genzel15}, who combined the MZ relations at different redshifts presented by \citet{erb06}, \citet{maiolino08}, \citet{zahid14} and \citet{wuyts14}. This prescription, also recently used by \citet{tacconi18}, is given by

\begin{equation}
Z=12+\log(O/H) = a-0087 (M_{\rm stars}-b)^2,
\end{equation}
where $a=8.74$ and $b=10.4+4.46 {\rm log}(1+z) -1.78 {\rm log}(1+z)^2$. Similarly, we adopt a metallicity-dependent $\alpha_{\rm CO}$ prescription from \citet{bolatto13}, with the form
\begin{equation}
\alpha_{\rm CO} (Z) = 4.36\times0.67\times {\rm exp} (0.36\times10^{(8.67-Z)}),
\end{equation}
where $12+\log(O/H)=Z$ is the galaxy metallicity. 

\subsection{\textsc{magphys} SED Dust-based Estimate}

The \textsc{magphys} SED fitting routine uses a prescription to balance the energy output at various wavelengths, thus producing estimates of the dust mass ($M_{\rm d}$) based on the best SED fit model (for additional details see Section \ref{sec:ism_mass}). We thus used $M_{\rm d}$ together with an assumption of the molecular gas-to-dust ratio ($\delta_{\rm GDR}$) to compute the molecular gas mass as $M_{\rm mol, SED}=\delta_{\rm GDR} M_{\rm d}$. 

The value of $\delta_{\rm GDR}$ is known to depend on $Z$ based on observations of local galaxies and simulations out to higher redshifts \citep[][]{popping17, coogan19}. We adopt the broken power-law form of the $\delta_{\rm GDR}-$Z relation prescribed by \citet{remyruyer14}. For the mass range of the ASPECS galaxies, this form is given by

\begin{equation}
log(\delta_{\rm GDR}) = 2.21 - 1.0[12+\log(O/H) - 8.69].
\end{equation}

For the median stellar mass and redshift of the ASPECS sample, $\sim10^{10.6}\ M_\sun$ and $z\sim1.8$, respectively, we find metallicities in the range 12+log$_{10}$(O/H)$=8.4-8.6$. For this metallicity range, we thus expect  $\delta_{\rm GDR}\sim200$ \citep[e.g.][]{remyruyer14, devis19}. We note that using the $\delta_{\rm GDR}$-Z relationship found by \citet{leroy11}, would yield $\delta_{\rm GDR}=130$ for the median stellar mass and redshift values for the ASPECS sample, corresponding to a factor of $\approx0.18$ dex lower.

\subsection{Rayleigh-Jeans Dust-based Estimate}

Various studies have shown that the cold dust emission from distant galaxies in the Rayleigh-Jeans (RJ) long-wavelength regime can be used as a reliable tracer of the ISM molecular gas mass, under reasonable assumptions on the dust properties \citep{scoville14, scoville16, hughes17,kaasinen19}. Although the method comes in different flavors, the dust temperatures are expected to have little effect on the dust content measurements, and it is assumed to be fixed at $T_{\rm d}=25$ K, which corresponds to the typical value for local star forming galaxies. Following \citet{scoville14}, the dust emissivity index is assumed fixed at $\beta=1.8$, which corresponds to the value measured in our Galaxy. Instead of assuming a dust-to-gas ratio, as usually done for molecular gas estimates based on dust continuum emission (see below), this method was independently calibrated to a value of $L_{\nu850\mu{\rm m}}/M_{\rm mol}$, expected to be fairly constant for a relatively ample range of galaxy properties, and using a fixed $\alpha_{\rm CO}$ of $6.5\ M_\sun$ (K km s$^{-1}$ pc$^2$)$^{-1}$ \citep{scoville16, scoville17}.  Hereafter, we refer to this method as ``RJ''. Thus, assuming that most of the ISM content is molecular, the molecular gas mass as a function of the measured flux density $S_{\nu}$ (in mJy) is thus given by 

\begin{equation} 
M_{\rm mol, RJ} = 1.2(1+z)^{-4.8}\left(\frac{\nu_{\rm obs}}{350}\right)^{-3.8} \frac{\Gamma_{\rm 0}}{\Gamma_{\rm RJ}} S_\nu D_{\rm L}^2, 
\end{equation}

\noindent where $D_{\rm L}$ is the luminosity distance at the source redshift $z$, in units of Gpc; $\nu_{\rm obs}$ is the observed frequency, which in our case corresponds to 242 GHz; and $\Gamma_{\rm RJ}$ is a factor to correct for the deviation from the RJ limit at increasing redshifts.

\subsection{Comparison}

\begin{figure*}[t] 
\centering 
\includegraphics[scale=0.5]{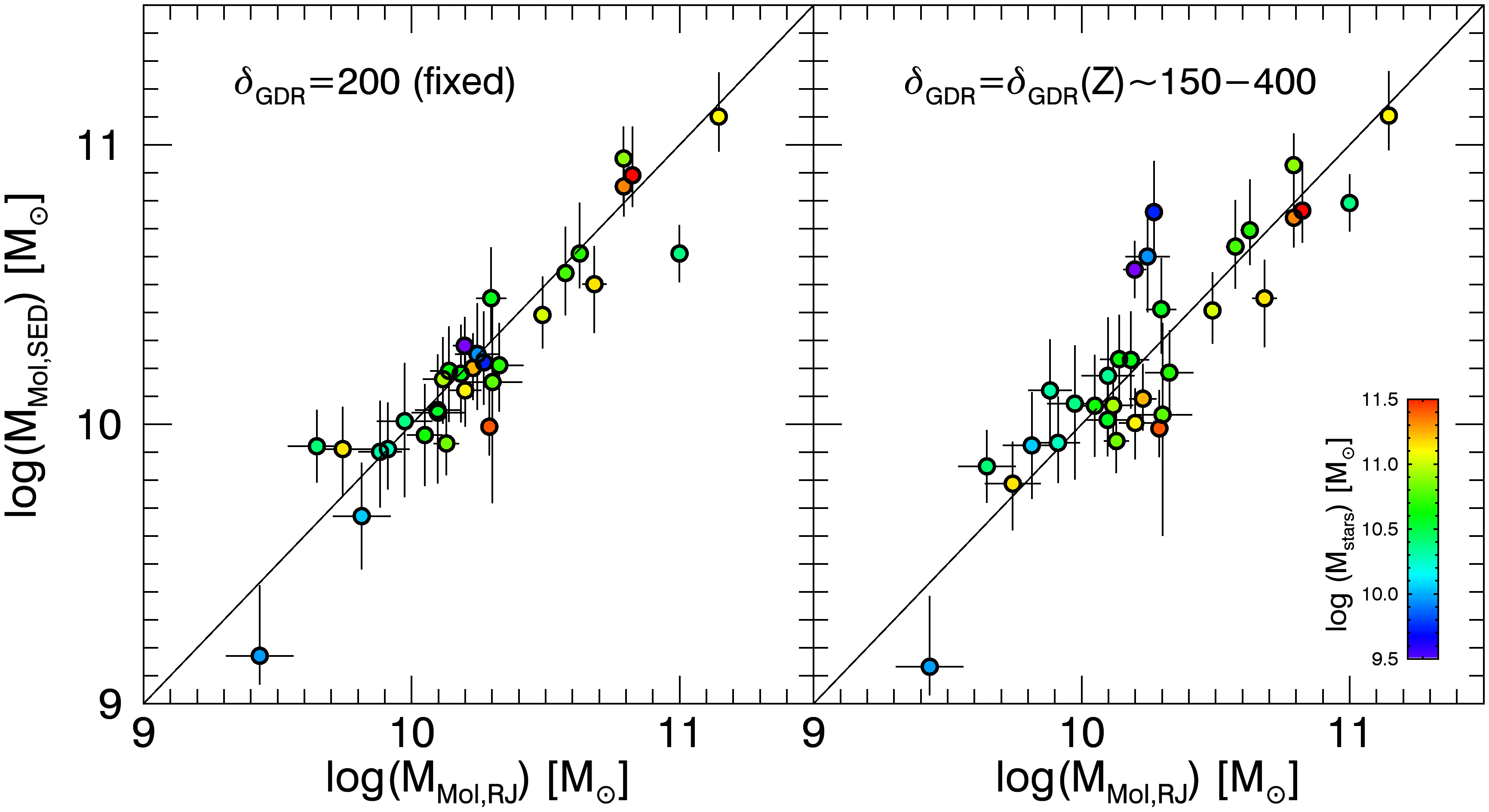} 
\caption{Comparison between the molecular gas mass estimates using the RJ-based method, from a single ALMA band 6 measurement, and the SED-based approach, from \textsc{magphys} fitting, for the main sample of ASPECS 1.2-mm sources. The left panel shows the case when a fixed value for the gas-to-dust mass ratio $\delta_{\rm GDR}=200$ is assumed. The right panel shows the case when a metallicity-dependent $\delta_{\rm GDR}$ is used. In this later case, the metallicity is inferred from the stellar masses. } \label{fig:gasmass1}
\end{figure*}

\begin{figure*}[t] 
\centering 
\includegraphics[scale=0.5]{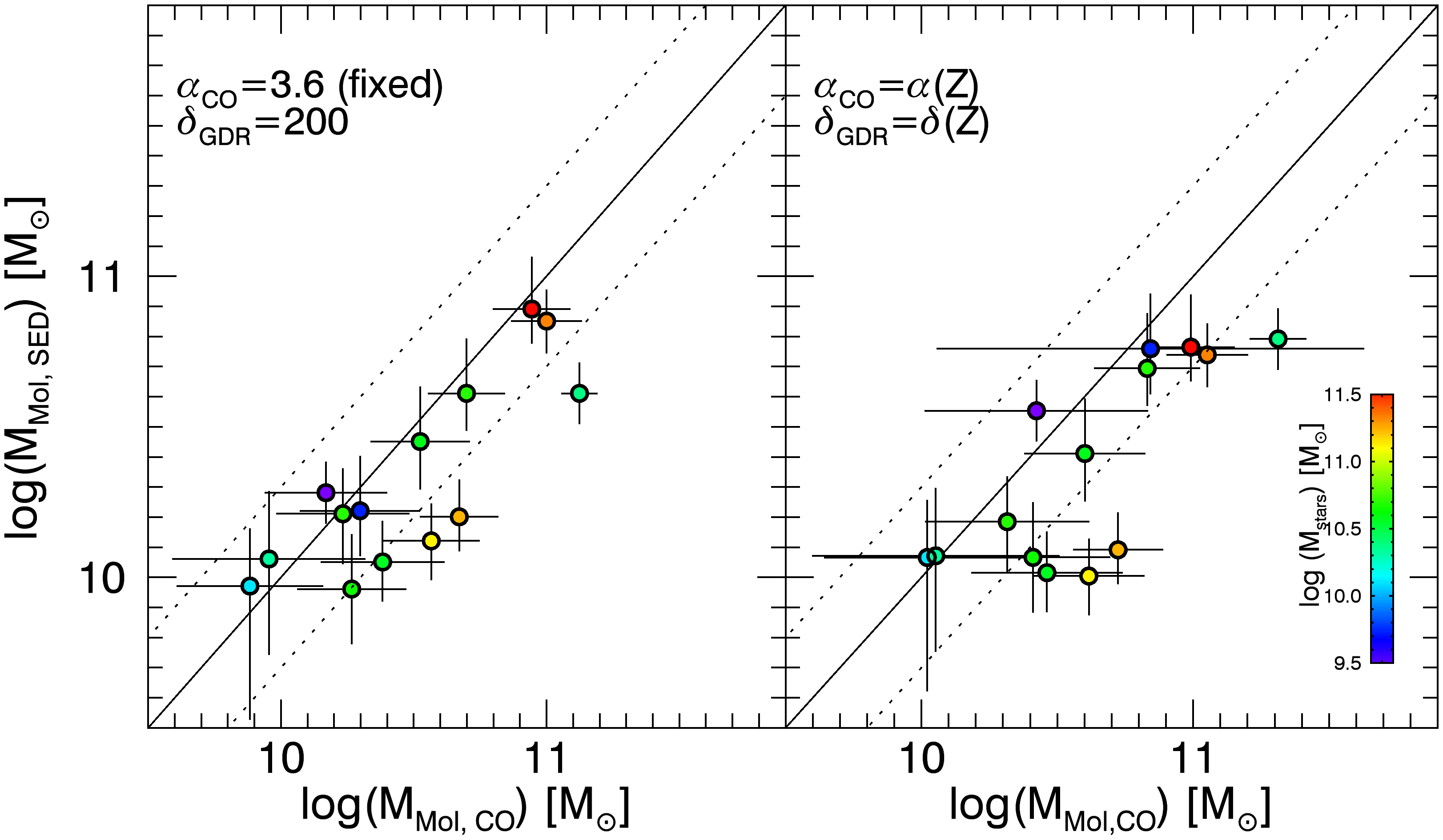} 
\caption{Comparison between the molecular gas mass values obtained from SED fitting method and the CO-based estimates for CO line and dust-continuum-detected galaxies in the ASPECS LP survey. CO measurements are reported by \citet{aravena19} and \citet{boogaard19}; however their conversion to the ground-state CO transition has been recently revised for our sample to be close to LTE and thus the CO luminosities ($L'_{\rm CO}$) are roughly equal for the $J<4$ CO transitions \citep{boogaard20}. The left panel shows the case when $\delta_{\rm GDR}$ and $\alpha_{\rm CO}$ are kept fixed at standard values. The right panel shows the case with a metallicity-dependent approach for both parameters. The dotted lines highlight differences of 0.3 dex with respect to the linear 1:1 scaling shown by the solid line.} \label{fig:gasmass2} 
\end{figure*}

Each of the three methods described in the previous section has its own uncertainties owing to the assumption of various conversion factors or metallicity-dependent prescriptions. Here, we briefly contrast the results obtained by these methods, applied on the ASPECS 1.2 mm galaxy sample. 

Figure \ref{fig:gasmass1} shows the comparison of molecular gas mass estimates for the ASPECS 1.2 mm sources, between the RJ and \textsc{magphys} SED methods, using two different
assumptions for $\delta_{\rm GDR}$ in the latter approach. In the first case (left panel), we use a fixed value for $\delta_{\rm GDR}$ of 200, following the expected typical (median) metallicity for our sample, as well as previous estimates for the ASPECS pilot survey \citep{aravena16b}. In the second case (right panel), we compute a metallicity-dependent $\delta_{\rm GDR}$ individually for each source. Here, we use the stellar mass of each source, the MZ relation, and the adopted prescription of $\delta_{\rm GDR}(Z)$. 

From this comparison, we find that there is remarkable agreement in the molecular gas mass values obtained through the RJ and \textsc{magphys} methods when using a fixed $\delta_{\rm GDR}=200$ (Fig. \ref{fig:gasmass1}, left panel). This agreement is somewhat expected since both estimates essentially come from the same dust measurements, even though the \textsc{magphys} method includes more information from the SED photometry.

These methods use different combinations of ($\alpha_{\rm CO}$, $\delta_{\rm GDR}$, $\beta$) parameter values. For the \textsc{magphys} SED method, we assume $\delta_{\rm GDR}=200$ and emissivity index $\beta=1.5$, whereas the RJ method uses $\alpha_{\rm CO}=6.5$ and $\beta=1.8$. The large $\alpha_{\rm CO}$ factor used in the RJ approach implies a factor of $0.26$ dex larger molecular gas masses, compared to assuming $\alpha_{\rm CO}=$3.6 for example. However, this is compensated by the larger $\delta_{\rm GDR}$, compared to a typical $\delta_{\rm GDR}=100$, and lower $\beta$ value used in the \textsc{magphys} method, which yield 0.3 dex and $\sim$0.05 dex (assuming $z\sim1$) larger \textsc{magphys} molecular gas mass estimates.

If we now, instead of fixing the parameters, allow the $\delta_{\rm GDR}$ value to vary with metallicity individually for each source (Fig. \ref{fig:gasmass1}, right panel), we find little variation in the relationship, yielding a very similar scatter and location of individual sources, well within the uncertainties of individual measurements. This indicates that using a fixed or metallicity-dependent $\delta_{\rm GDR}$ does not affect the resulting molecular gas masses significantly. Moreover, this hints that the most massive ASPECS galaxies have relatively uniform metallicities \citep[as previously found by ][]{boogaard19}.


As a reference, we also compare the \textsc{magphys} SED-based molecular gas mass estimates with the CO-based molecular gas masses for those 1.2 mm continuum sources that were also previously detected in CO line emission in the ASPECS 3mm line scan \citep{gonzalezlopez19, aravena19, boogaard19}. We note that the \textsc{magphys} method mostly relies on the assumed prescription for $\delta_{\rm GDR}$, whereas the CO-based molecular gas masses are subject to assumptions for $\alpha_{\rm CO}$ and the CO SLED shape (CO excitation). As mentioned earlier, we use the average CO SLED obtained for a significant fraction of the galaxies in our sample based on recent observations obtained with the VLA \citep{riechers20} and by the ASPECS 3 mm and 1 mm line survey \citep{boogaard20}.

By fixing $\delta_{\rm GDR}$ and $\alpha_{\rm CO}$ to the canonical values of 200 and 3.6, respectively, we find a clear correlation between both estimates (Fig. \ref{fig:gasmass2}, left panel), indicating that both methods yield consistent molecular gas masses. Allowing both parameters to vary with metallicity (i.e., with stellar mass) as shown in Fig. \ref{fig:gasmass2} (right panel), we find a slightly larger scatter, yet the results compared with the fixed-metallicity approach are qualitatively similar.

In summary, we find that there is overall good agreement between the various methods to compute the molecular gas masses, supporting our choice for the \textsc{magphys} SED method. Using the RJ method for our 1.2 mm continuum sample would yield mostly negligible differences in the molecular gas masses. The use of a metallicity-based approach using standard prescriptions for the mass-metallicity, $\delta_{\rm GDR}-Z$ and $\alpha_{\rm CO}-Z$ relations does not produce a substantial difference compared to using (well informed) fixed parameters for $\delta_{\rm GDR}$ and $\alpha_{\rm CO}$. The use of metallicity-based estimates yields mostly an increase of the scatter when comparing different estimates.

\newpage
\section{Multi-wavelength thumbnails}
\label{app_b}
Figure \ref{fig:postage1} shows multi-wavelength image cutouts centered at the location of the ASPECS 1.2 mm sources.
\begin{figure}[H] 
\centering 
\includegraphics[scale=2.3]{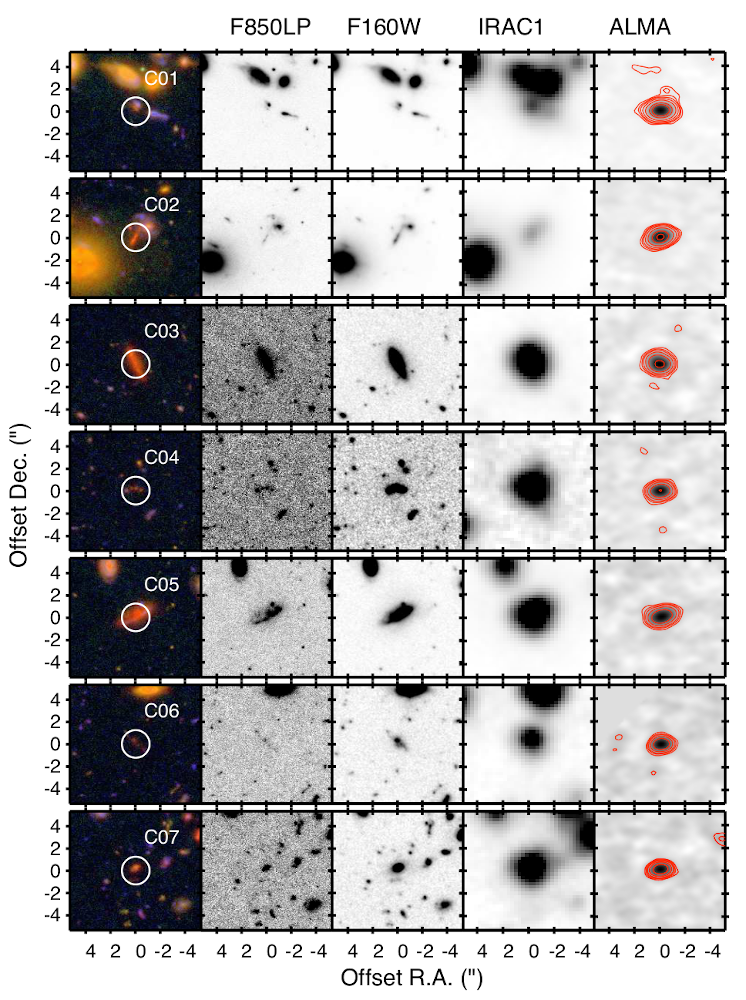} 
\caption{Multi-wavelength postage stamps toward the 35 ALMA dust continuum detections from the ASPECS LP survey. From left to right, we show an optical/NIR false-color composite (F435W/F850LP/F105W) and individual images in the F850LP and F160W bands, the IRAC channel 1, and ALMA at 1.2 mm.} 
\label{fig:postage1} 
\end{figure}

\begin{figure}[H]
    \centering 
    \includegraphics[scale=2.3]{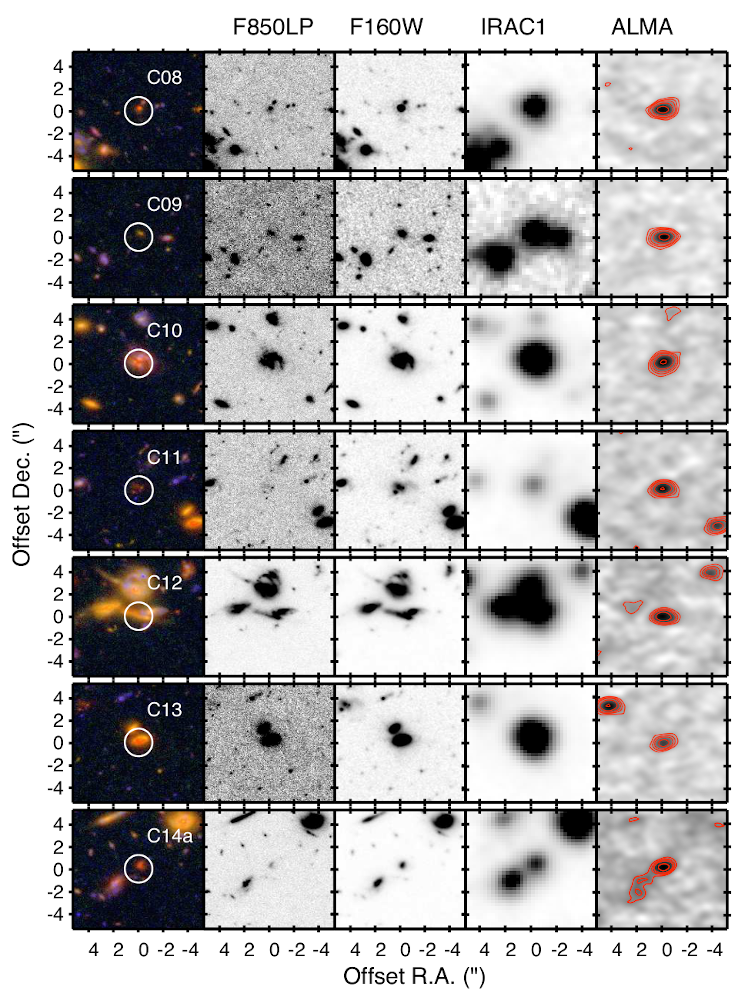} 
    \figurenum{\ref{fig:postage1}} 
    \caption{Continued.}
    \end{figure}

\begin{figure}[H] 
\centering 
\includegraphics[scale=2.3]{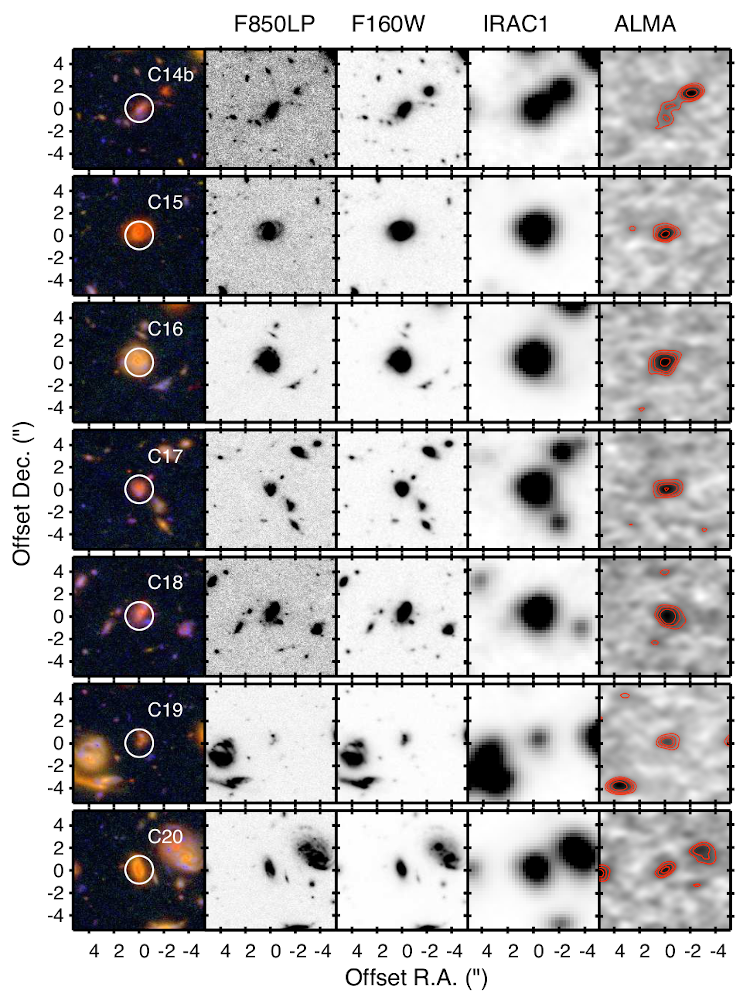} 
\figurenum{\ref{fig:postage1}} 
\caption{Continued.}
\end{figure}

\begin{figure}[H] 
\centering
    \includegraphics[scale=2.3]{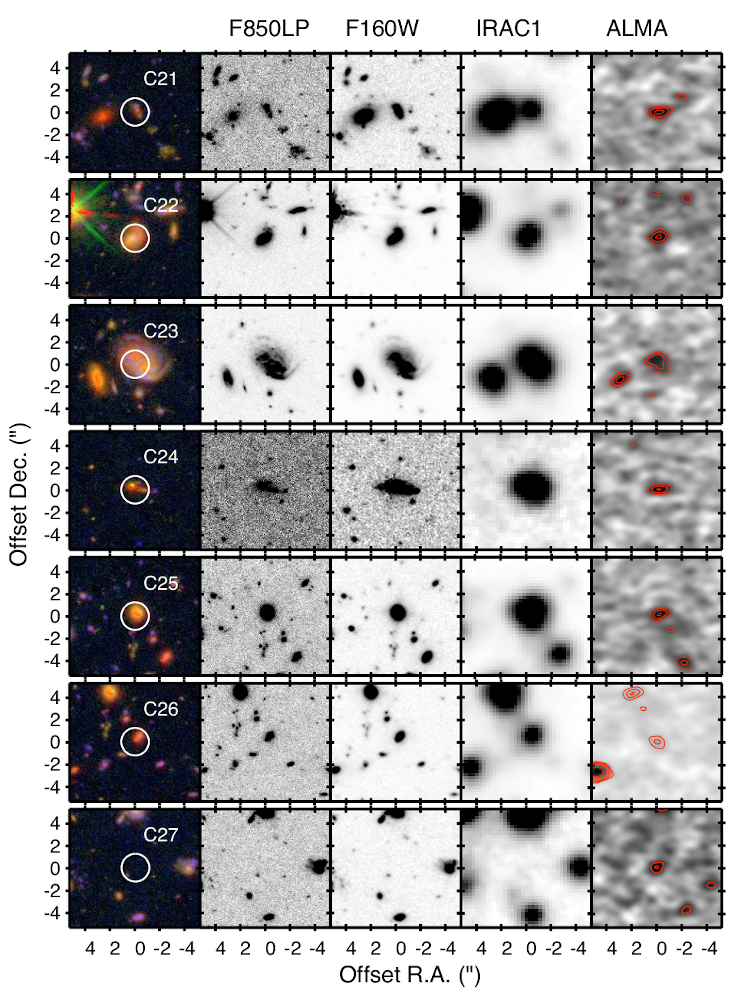}
     \figurenum{\ref{fig:postage1}} 
     \caption{Continued.} 
     \end{figure}

\begin{figure}[H] 
\centering 
\includegraphics[scale=2.3]{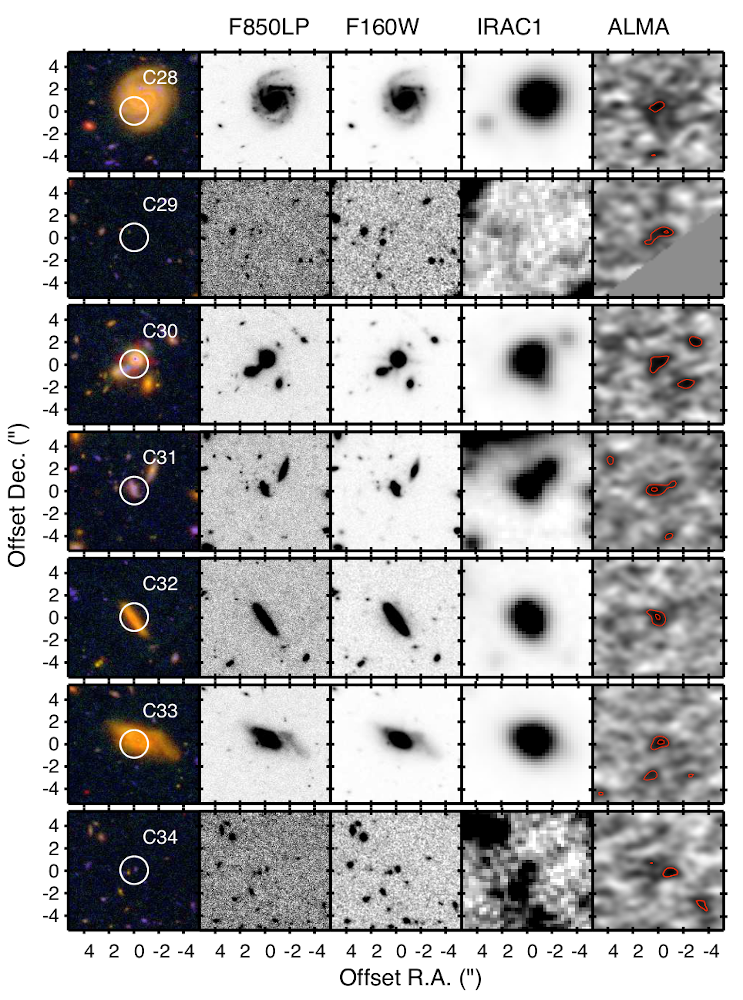} 
\figurenum{\ref{fig:postage1}} 
\caption{Continued.} 
\end{figure}

\newpage
\section{Spectral energy distribution fits}
\label{app_c}

Figures \ref{fig:seds} and \ref{fig:seds2} show the multi-wavelength photometry (from UV to  mm wavelengths) and the best-fit SEDs obtained with MAGPHYS for 1.2 mm galaxies in the main and secondary samples, respectively.

\begin{figure}[H]
    \centering
    \includegraphics[scale=1.4]{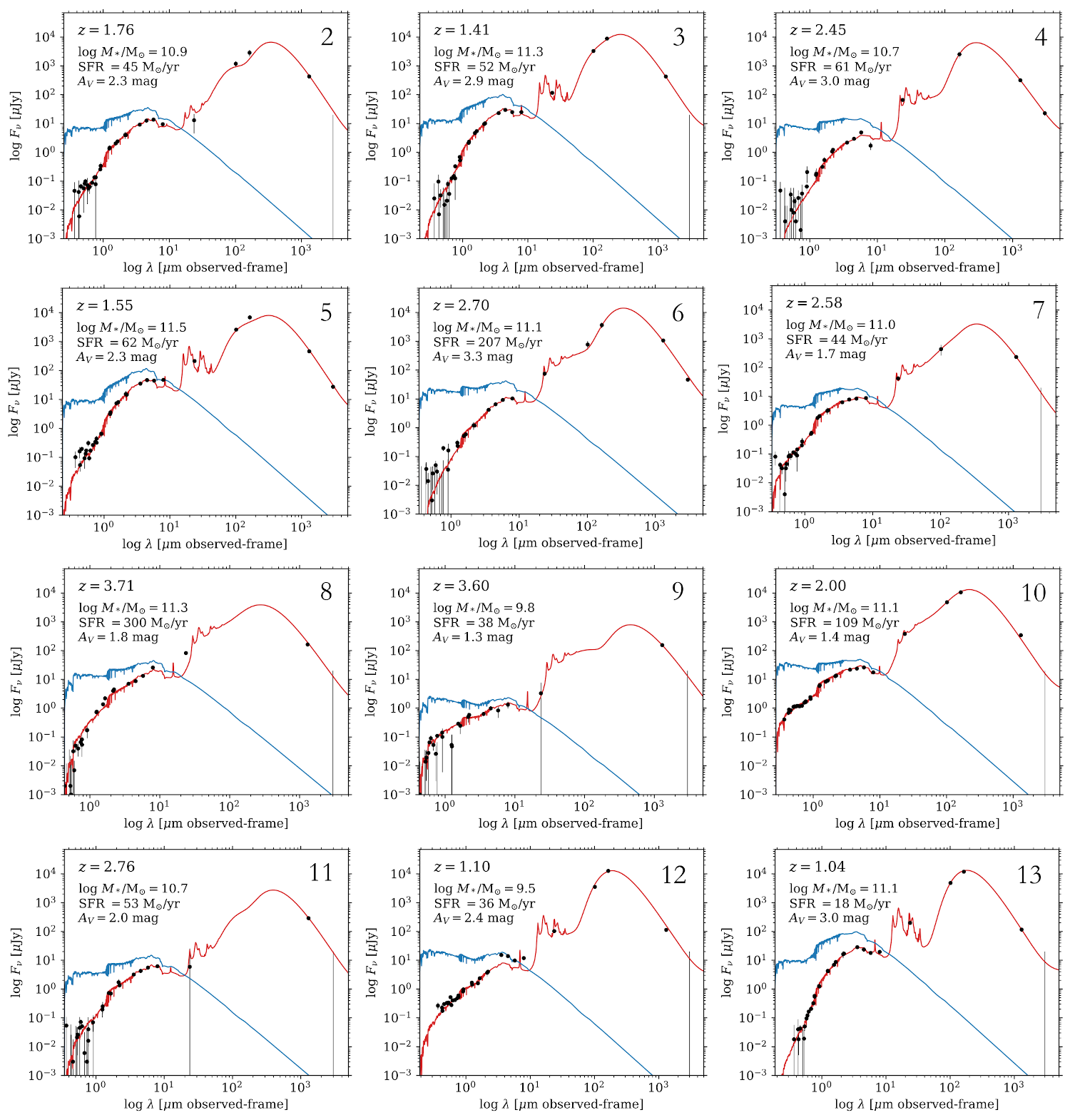}
    \caption{SEDs of the ASPECS 1.2 mm galaxies for sources in the main sample, except for source ID C01, which is shown in Fig. \ref{fig:sed1}. The black circles show the observed photometry. The solid red curves represent the best fit SED obtained with \textsc{magphys}. The blue curves show the unattenuated stellar emission. The redshift, stellar mass, SFR, and optical attenuation are shown for each case ($A_{\rm V}$).}
    \label{fig:seds}
\end{figure}

\begin{figure}[H]
    \centering
    \includegraphics[scale=1.4]{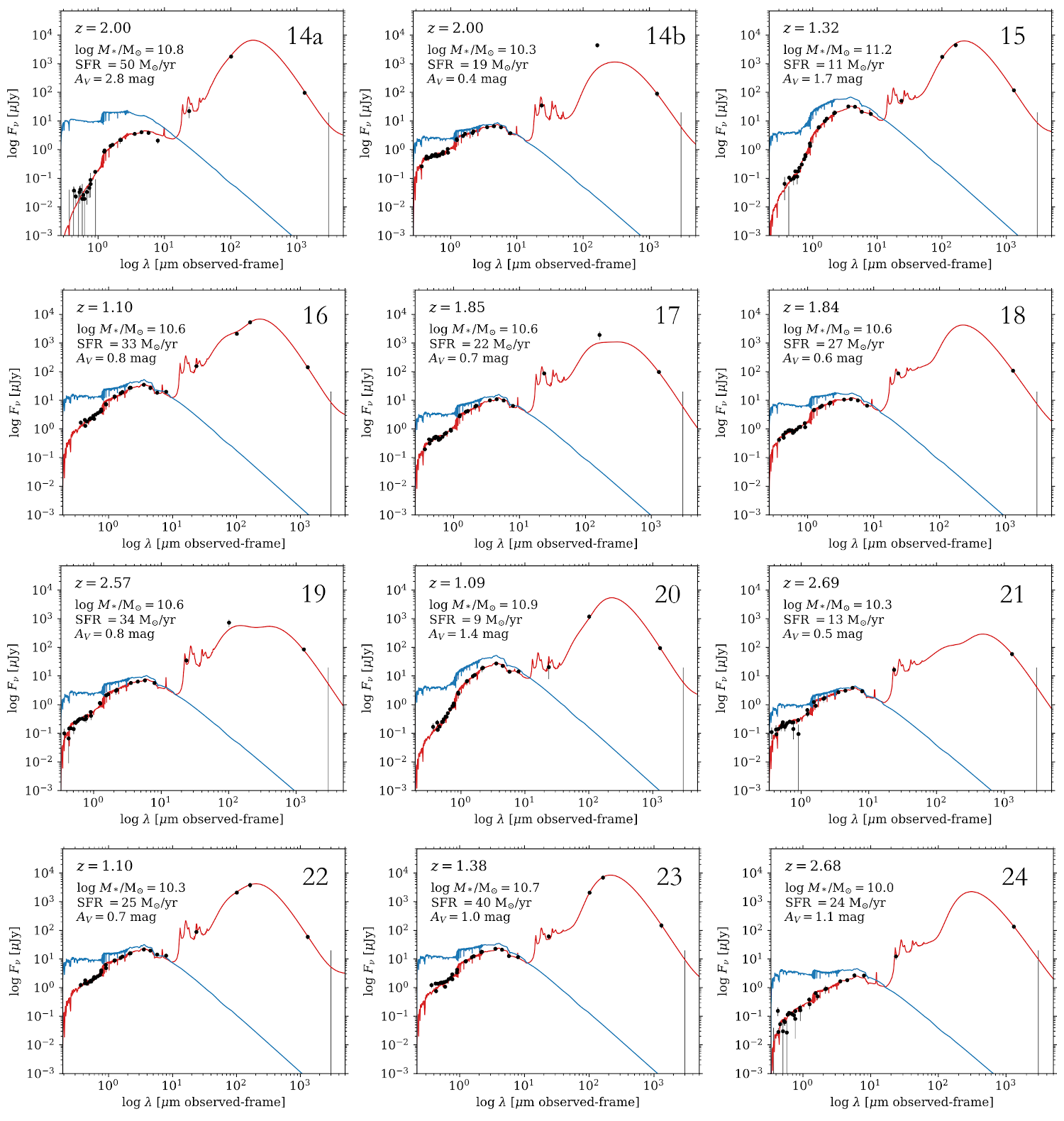}\figurenum{\ref{fig:seds}}
    \caption{Continued}
\end{figure}

\begin{figure}[H]
    \centering
    \includegraphics[scale=1.4]{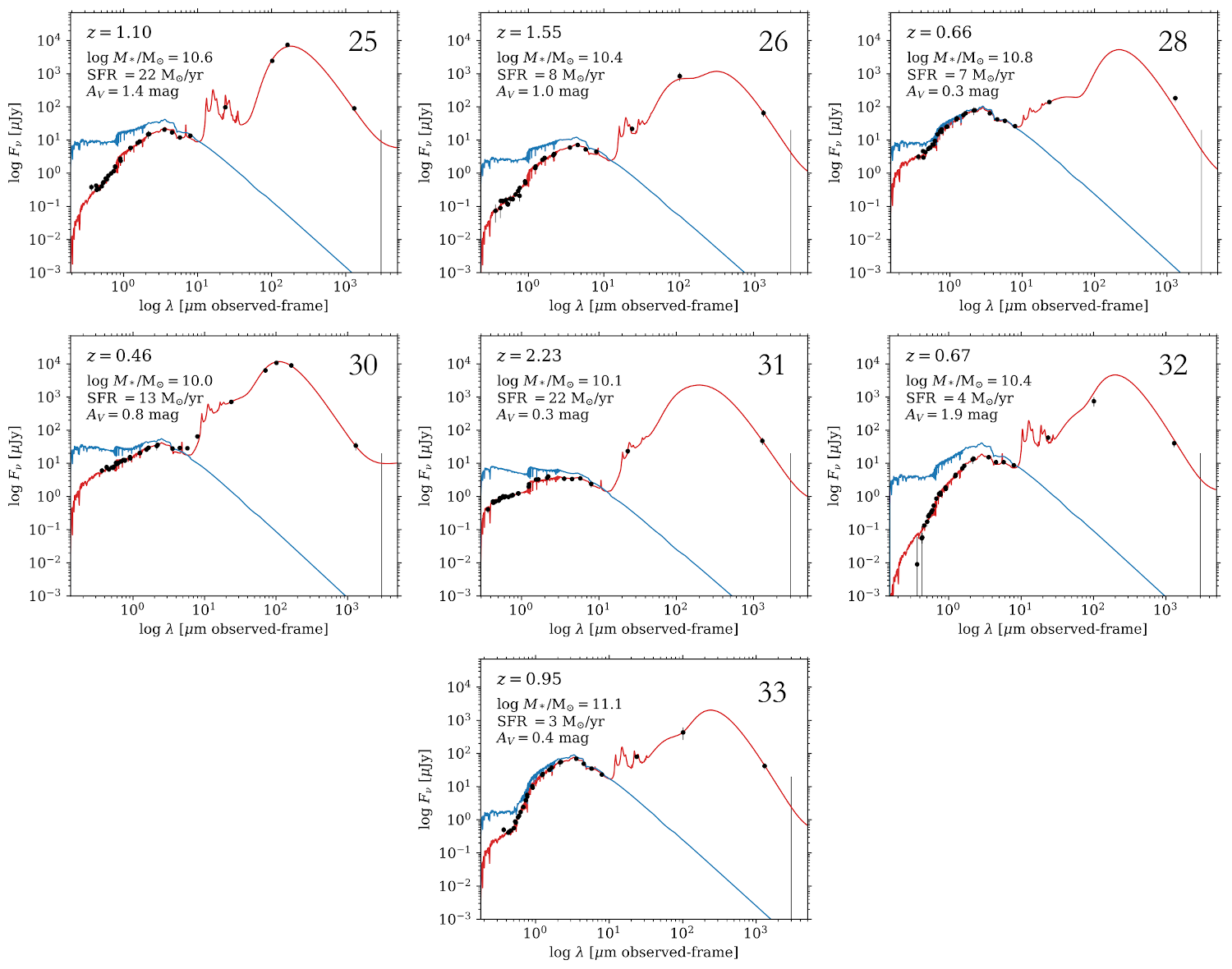}\figurenum{\ref{fig:seds}}
    \caption{Continued}
\end{figure}

\begin{figure}[H]
    \centering
    \includegraphics[scale=1.4]{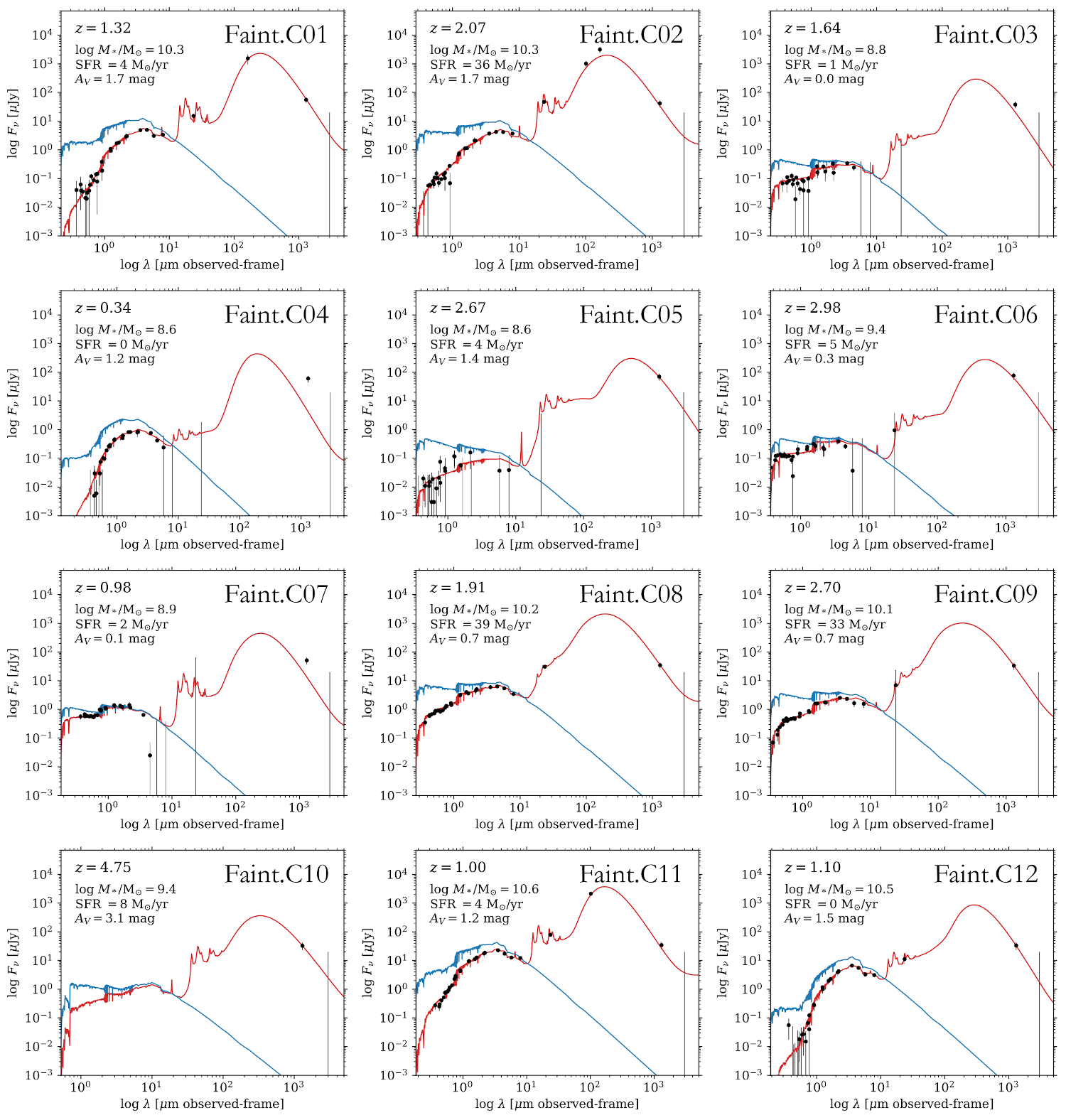}
    \caption{SEDs of the ASPECS 1.2 mm galaxies for sources in the faint sample. The black circles show the observed photometry. The solid red curves represent the best fit SED obtained with \textsc{magphys}. The blue curves show the unattenuated stellar emission. The redshift, stellar mass, SFR, and optical attenuation are shown for each case ($A_{\rm V}$).}
    \label{fig:seds2}
\end{figure}

\begin{figure}[H]
    \centering
    \includegraphics[scale=1.4]{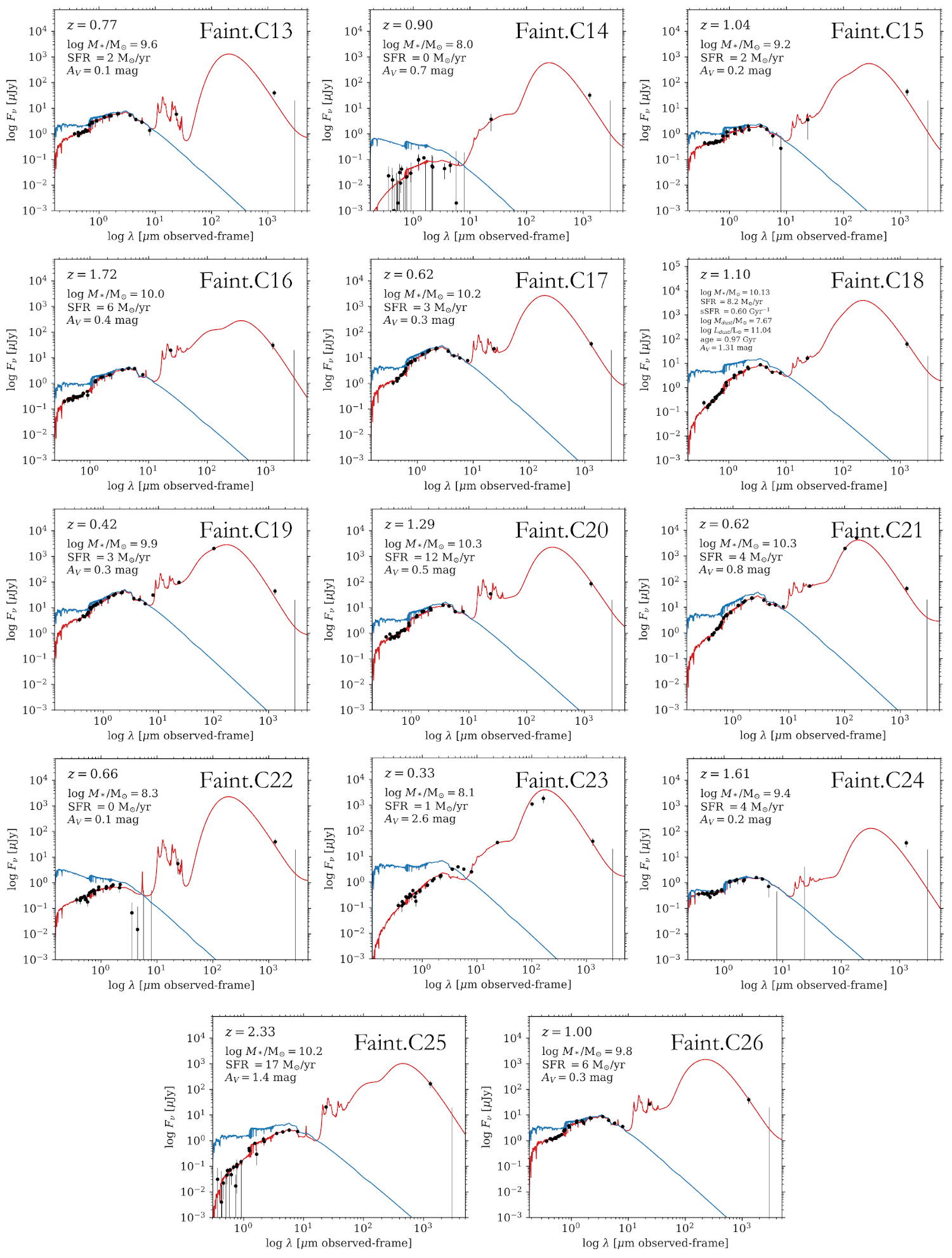}\figurenum{\ref{fig:seds2}}
    \caption{Continued}
\end{figure}

\section{Properties of Faint Prior-based Sample}
\label{app_d}
Tables \ref{tab:3} and \ref{tab:4} list the main derived properties for the ASPECS 1.2 mm galaxies in the secondary sample. 

\begin{table*}[h!]
\centering 
\caption{Secondary sample of ASPECS 1.2 mm galaxies. \label{tab:3}} 
\begin{tabular}{lccccccccc}
\hline 
ID & R.A. & Dec. & S/N & $F$ & $S_{\rm 1.2mm}$ & $S_{\rm 3mm}$ & CO ID & OIR? & Reference\\ 
Faint. & (J2000)  & (J2000)   &   &      & ($\mu$Jy)  & ($\mu$Jy) &       & (Y/N) \\ 
 (1)    &    (2)   &     (3) &   (4) &  (5)  &   (6)    &   (7)     &    (8)        &  (9)  &  (10) \\ 
\hline
FC01  &  03:32:34.66  &  $-$27:47:21.2  &    3.9  &  0.91   &  $ 55.6\pm 13.7$  &  $<20.0$  &  $\ldots$  & Y & M \\ 
FC02  &  03:32:35.74  &  $-$27:46:39.6  &    3.8  &  0.90   &  $ 41.7\pm 10.7$  &  $<20.0$  &  $\ldots$  & Y & M  \\ 
FC03  &  03:32:41.32  &  $-$27:47:06.6  &    3.7  &  0.91   &  $ 38.0\pm  9.8$  &  $<20.0$  &  $\ldots$  & Y & 3D  \\  
FC04  &  03:32:41.47  &  $-$27:47:29.2  &    3.7  &  0.92   &  $60.4\pm 15.8$  &  $<20.0$  &  $\ldots$  &  Y & M \\   
FC05  &  03:32:37.51  &  $-$27:47:56.6  &    3.6  &  0.90   &  $ 70.4\pm 18.9$  &  $<20.0$  &  $\ldots$ &  Y &  M \\   
FC06  &  03:32:41.63  &  $-$27:46:25.8  &    3.6  &  0.90   &  $ 76.2\pm 20.5$  &  $<20.0$  &  $\ldots$  &  Y & M  \\  
FC07  &  03:32:40.01  &  $-$27:47:51.2  &    3.5  &  0.83   &  $ 51.3\pm 14.0$  &  $<20.0$  &  $\ldots$  &  Y & M \\   
FC08  &  03:32:35.85  &  $-$27:47:18.6  &    3.5  &  0.90   &  $ 34.6\pm  9.5$  &  $<20.0$  &  $\ldots$  &  Y & M \\   
FC09  &  03:32:38.56  &  $-$27:47:30.6  &    3.4  &  0.90   &  $33.3\pm  9.3$  &  $<20.0$  &  $\ldots$  &  Y & 3D \\   
FC10  &  03:32:38.62  &  $-$27:47:34.4  &    3.4  &  0.85   &  $ 32.7\pm  9.3$  &  $<20.0$  &  $\ldots$ &  Y & M \\   
FC11  &  03:32:36.66  &  $-$27:46:31.2  &    3.3  &  0.87   &  $ 34.7\pm 10.0$  &  $<20.0$  &  $\ldots$  &  Y & M \\  
FC12  &  03:32:37.17  &  $-$27:46:26.2  &    3.3  &  0.85   &  $ 33.4\pm  9.8$  &  $<20.0$  &  $\ldots$  &  Y & M \\ 
FC13  &  03:32:37.85  &  $-$27:47:51.8  &    3.2  &  0.85   &  $ 39.5\pm 11.7$  &  $<20.0$  &  $\ldots$  &  Y & M \\   
FC14  &  03:32:35.37  &  $-$27:47:17.0  &    3.2  &  0.81   &  $32.1\pm  9.8$  &  $<20.0$  &  $\ldots$  &  Y & 3D \\   
FC15  &  03:32:38.36  &  $-$27:46:00.2  &    3.1  &  0.81   &  $ 44.1\pm 13.5$  &  $<20.0$  &  $\ldots$ &  Y & M \\   
FC16  &  03:32:35.79  &  $-$27:46:55.4  &    3.1  &  0.82   &  $ 30.6\pm  9.4$  &  $<20.0$  &  $\ldots$  &  Y & 3D \\ 
FC17  &  03:32:38.56  &  $-$27:46:31.0  &    3.0  &  0.80   &  $ 34.8\pm  9.6$  &  $<20.0$  &  $\ldots$  &   Y & M \\ 
FC18  &  03:32:37.33  &  $-$27:45:57.8  &    3.0  &  0.80   &  $ 62.5\pm 19.9$  &  $<20.0$  &  MP01  & Y & CM  \\  
FC19  &  03:32:38.98  &  $-$27:46:31.0  &    3.8  &  0.80   &  $ 43.9\pm11.7$  &  $<20.0$  &  $\ldots$  &  Y & M \\   
FC20  &  03:32:39.89  &  $-$27:46:07.4  &    3.6  &  0.82   &  $ 85.8\pm 23.8$  &  $<20.0$  &  16  &  Y &  CM  \\
FC21  &  03:32:41.35  &  $-$27:46:52.0  &    3.5  &  0.84   &  $ 54.0\pm 15.2$  &  $<20.0$  &  $\ldots$  &  Y & M \\ 
FC22  &  03:32:37.60  &  $-$27:47:40.6  &    3.4  &  0.85   &  $ 39.7\pm 11.9$  &  $<20.0$  &  $\ldots$  &  Y & M \\ 
FC23  &  03:32:42.37  &  $-$27:46:57.8  &    3.0  &  0.81   &  $ 38.8\pm 12.7$  &  $<20.0$  &  $\ldots$  &  Y & M \\ 
FC24  &  03:32:36.86  &  $-$27:46:35.0  &    3.0  &  0.82   &  $ 35.9\pm 11.8$  & $<20.0$  &  $\ldots$  &  Y & M \\   
FC25  &  03:32:41.80  &  $-$27:47:39.0  &    3.0  &  0.83   &  $165.2\pm 54.2$  &  $<20.0$  &  $\ldots$  &  Y & 3D \\
FC26$^\dagger$  &  03:32:38.09  &  $-27:46:14.1$  &    3.0  &  0.50   &  $ 39.5\pm 12.0$  &  $<20.0$  &  $\ldots$  & N & M \\
\hline\hline \end{tabular}\\
\flushleft \noindent {\bf Note.} Column (1): source name, ASPECS-LP.1mm.Faint.xx; an ``F'' has been added to differentiate these sources from the main sample. Columns (2) and (3): position of the continuum detection in the ALMA 1.2 mm map. Column (4): S/N of the 1.2 mm detection. Column (5): fidelity ($F$) of the 1 mm detection, as defined in the text \citep[for details see][]{gonzalezlopez20}. Column (6): flux density at 1.2 mm, corrected for PB. Column (7): flux density at 3 mm. Column (8): CO source ID. Column (9): is there an optical counterpart identification for this source? (yes/no). Column (10): redshift code. M: MUSE \citep[spec-z, from][]{inami17}, CO: CO line confirmed \citep[spec-z;][]{aravena19,boogaard19}, CM: CO and MUSE joint redshift determination \citep[spec-z;][]{boogaard19}, 3D: 3D-HST \citep[photo-z][]{momcheva15, skelton14}, GS: other HST redshifts \citep[photo-z;][]{rafelski15, morris15}.  
\flushleft \noindent $^\dagger$ Identification made based on IR Herschel PACS prior.
\end{table*}

\begin{table*} [h!]
\centering 
\caption{Properties of the secondary sample of ASPECS 1.2 mm galaxies.} \label{tab:4}
\begin{tabular}{lcccccccccc} 
\hline 
ID & $z$ & $m_{\rm 160}$ & SFR & $M_{\rm stars}$ & $\Delta_{\rm MS}$ &  $T_{\rm d}$ & $L_{\rm IR}$ & $M_{\rm Mol, SED}$ &  $M_{\rm Mol, RJ}$ & $M_{\rm Mol, CO}$\\ 
   Faint.      &      & (AB mag)    &  ($M_\sun$ yr$^{-1}$) &  ($10^{9} M_\sun$) &   & (K)        & ($10^{10} L_\sun$) &  ($10^{9} M_\sun$) & ($10^{9} M_\sun$) & ($10^{10} M_\sun$) \\
(1) & (2) & (3) & (4) & (5) & (6) & (7) & (8) & (9) & (10) & (11) \\ 
\hline\hline 
FC01  &  1.315  &  23.3  &  $  4_{-  1}^{+  2}$  &  $20.0_{-  2.9}^{+ 3.6}$  &  $-0.72_{- 0.18}^{+ 0.18}$  &  $33_{- 4}^{+ 5}$  &  $ 7.8_{-  1.8}^{+ 2.8}$  &  $  8.2_{-   3.4}^{+  4.1}$  &  $  8.0\pm  2.0$  &  $\ldots$  \\
FC02  &  2.067  &  23.8  &  $ 36_{- 11}^{+ 14}$  &  $22.4_{-  2.7}^{+ 3.1}$  &  $-0.01_{- 0.27}^{+ 0.14}$  &  $50_{- 5}^{+ 6}$  &  $37.1_{-  7.9}^{+ 6.6}$  &  $  3.5_{-   1.0}^{+  1.5}$  &  $  5.8\pm  1.5$  &  $\ldots$  \\
FC03  &  1.640  &  25.4  &  $  0.5_{-  0.1}^{+  0.1}$  &  $ 0.7_{-  0.1}^{+ 0.1}$  &  $-0.26_{- 0.11}^{+ 0.18}$  &  $48_{-12}^{+15}$  &  $ 0.14_{-  0.05}^{+ 0.07}$  &  $  0.1_{-   0.1}^{+  0.2}$  &  $  5.5\pm  1.4$  & $\ldots$   \\
FC04  &  0.338  &  24.1  &  $  0.1_{-  0.1}^{+  0.1}$  &  $ 0.4_{-  0.1}^{+ 0.2}$  &  $-1.46_{- 0.32}^{+ 0.22}$  &  $34_{- 7}^{+ 6}$  &  $ 0.04_{-  0.02}^{+ 0.03}$  &  $  0.04_{-   0.02}^{+  0.03}$  &  $  3.4\pm  0.9$  &  $\ldots$  \\
FC05  &  2.674  &  27.0  &  $  4_{-  2}^{+  3}$  &  $ 0.4_{-  0.2}^{+ 0.3}$  &  $ 0.63_{- 0.41}^{+ 0.14}$  &  $37_{- 7}^{+13}$  &  $ 5.1_{-  3.8}^{+ 4.5}$  &  $  4.8_{-   4.3}^{+  5.6}$  &  $  9.2\pm  2.5$  & $\ldots$   \\
FC06  &  2.981  &  25.1  &  $  5_{-  1}^{+  1}$  &  $ 2.3_{-  0.4}^{+ 0.2}$  &  $-0.07_{- 0.10}^{+ 0.10}$  &  $35_{- 6}^{+13}$  &  $ 4.8_{-  0.9}^{+ 0.5}$  &  $  4.7_{-   3.7}^{+  3.8}$  &  $  9.7\pm  2.6$  &  $\ldots$  \\
FC07  &  0.980  &  23.6  &  $  1.6_{-0.2}^{+  0.2}$  &  $ 0.9_{-  0.3}^{+ 0.1}$  &  $ 0.33_{- 0.10}^{+ 0.14}$  &  $45_{-11}^{+16}$  &  $ 1.2_{-  0.6}^{+ 0.1}$  &  $  0.2_{-   0.2}^{+  0.6}$  &  $  6.9\pm  1.9$  &  $\ldots$  \\
FC08  &  1.906  &  22.4  &  $ 39_{-  7}^{+  4}$  &  $14.1_{-  1.4}^{+ 4.3}$  &  $ 0.23_{- 0.22}^{+ 0.10}$  &  $55_{- 7}^{+ 5}$  &  $39_{-  6}^{+ 7}$  &  $  3.4_{-   0.9}^{+  1.2}$  &  $  4.9\pm  1.3$  &  $\ldots$  \\
FC09  &  2.700  &  23.4  &  $ 33_{- 10}^{+  3}$  &  $11.7_{-  1.2}^{+ 3.3}$  &  $ 0.07_{- 0.14}^{+ 0.10}$  &  $53_{- 7}^{+ 7}$  &  $37_{- 15}^{+ 4}$  &  $  2.6_{-   0.7}^{+  1.0}$  &  $  4.3\pm  1.2$  & $\ldots$   \\
FC10  &  4.752  &  $\ldots$  &  $  8_{-  4}^{+  9}$  &  $ 2.3_{-  1.4}^{+ 3.7}$  &  $-0.11_{- 0.27}^{+ 0.36}$  &  $43_{-9}^{+12}$  &  $ 9.8_{-  5.1}^{+ 9.8}$  &  $  2.4_{-   0.9}^{+  1.8}$  &  $  3.7\pm  1.1$  & $\ldots$   \\
FC11  &  0.997  &  21.3  &  $  4_{- 1}^{+1}$  &  $41.7_{-  4.2}^{+ 4.3}$  &  $-0.89_{- 0.10}^{+ 0.10}$  &  $38_{- 2}^{+ 3}$  &  $12.9_{-  1.3}^{+ 1.3}$  &  $  2.2_{-   0.8}^{+  1.1}$  &  $  4.7\pm  1.3$  &  $\ldots$  \\
FC12  &  1.096  &  23.2  &  $  0.5_{-  0.1}^{+  0.1}$  &  $31.6_{-  4.7}^{+ 5.7}$  &  $-1.77_{- 0.11}^{+ 0.14}$  &  $28_{- 3}^{+ 4}$  &  $ 2.4_{-  0.3}^{+ 0.4}$  &  $  5.5_{-   2.8}^{+  2.7}$  &  $  4.6\pm  1.4$  &  $\ldots$  \\
FC13  &  0.768  &  22.1  &  $  1.5_{-  0.3}^{+  0.3}$  &  $ 3.6_{-  0.4}^{+ 0.7}$  &  $-0.25_{- 0.10}^{+ 0.14}$  &  $40_{- 9}^{+16}$  &  $ 0.7_{-  0.2}^{+ 0.6}$  &  $  0.3_{-   0.2}^{+  1.1}$  &  $  4.7\pm  1.4$  &  $\ldots$  \\
FC14  &  0.900  &  26.2  &  $  0.1_{-  0.1}^{+  0.2}$  &  $ 0.1_{-  0.0}^{+ 0.1}$  &  $-0.09_{- 0.81}^{+ 0.51}$  &  $41_{- 8}^{+11}$  &  $ 0.07_{-  0.06}^{+ 0.16}$  &  $  0.1_{-   0.1}^{+  0.2}$  &  $  4.2\pm  1.3$  &  $\ldots$  \\
FC15  &  1.036  &  23.3  &  $  2_{-  1}^{+  1}$  &  $ 1.5_{-  0.2}^{+ 0.2}$  &  $ 0.11_{- 0.11}^{+ 0.14}$  &  $40_{- 8}^{+16}$  &  $ 1.1_{-  0.1}^{+ 0.7}$  &  $  0.7_{-   0.5}^{+  1.7}$  &  $  6.0\pm  1.8$  &  $\ldots$  \\
FC16  &  1.720  &  23.3  &  $  6_{-  1}^{+  1}$  &  $ 9.2_{-  1.4}^{+ 1.1}$  &  $-0.33_{- 0.11}^{+ 0.14}$  &  $37_{- 6}^{+13}$  &  $ 5.3_{-  1.2}^{+ 0.6}$  &  $  3.5_{-   2.0}^{+  2.1}$  &  $  4.4\pm  1.3$  & $\ldots$   \\
FC17  &  0.622  &  20.8  &  $  3_{-  1}^{+  1}$  &  $14.5_{-  1.9}^{+ 1.4}$  &  $-0.47_{- 0.18}^{+ 0.11}$  &  $33_{- 3}^{+ 7}$  &  $ 2.8_{-  1.2}^{+ 0.3}$  &  $  3.0_{-   1.6}^{+  2.0}$  &  $  3.6\pm  1.0$  &  $\ldots$  \\
FC18  &  1.096  &  22.3  &  $  8_{-  2}^{+  3}$  &  $13.5_{-  1.5}^{+ 2.4}$  &  $-0.21_{- 0.14}^{+ 0.14}$  &  $36_{- 5}^{+ 6}$  &  $11.0_{-  2.9}^{+ 3.3}$  &  $  9.3_{-   5.9}^{+  4.4}$  &  $  8.7\pm  2.8$  &  $0.9\pm0.2$ \\
FC19  &  0.419  &  20.2  &  $  3_{-  1}^{+  1}$  &  $ 8.6_{-  1.3}^{+ 0.9}$  &  $-0.19_{- 0.10}^{+ 0.11}$  &  $34_{- 5}^{+13}$  &  $ 2.0_{-  0.2}^{+ 0.5}$  &  $  2.5_{-   1.2}^{+  2.0}$  &  $  3.2\pm  0.8$  &  $\ldots$  \\
FC20  &  1.294  &  21.8  &  $ 12_{-  2}^{+  2}$  &  $21.4_{-  3.8}^{+ 3.8}$  &  $-0.31_{- 0.11}^{+ 0.14}$  &  $35_{- 5}^{+12}$  &  $10.7_{-  1.8}^{+ 3.3}$  &  $ 11.6_{-   5.9}^{+  6.7}$  &  $ 12.3\pm  3.4$  &  $0.8\pm0.2$   \\
FC21  &  0.620  &  20.9  &  $  4_{-  1}^{+  1}$  &  $18.2_{-  1.8}^{+ 1.8}$  &  $-0.45_{- 0.10}^{+ 0.10}$  &  $30_{- 2}^{+ 2}$  &  $ 4.4_{-  0.4}^{+ 0.5}$  &  $  4.6_{-   1.6}^{+  3.0}$  &  $  5.6\pm  1.6$  &  $\ldots$  \\
FC22  &  0.664  &  24.2  &  $  0.3_{-  0.1}^{+  0.6}$  &  $ 0.2_{-  0.0}^{+ 0.0}$  &  $ 0.34_{- 0.14}^{+ 0.36}$  &  $44_{-10}^{+14}$  &  $ 0.12_{-  0.03}^{+ 0.66}$  &  $  0.1_{-   0.1}^{+  0.6}$  &  $  4.3\pm  1.3$  & $\ldots$   \\
FC23  &  0.332  &  23.8  &  $  1_{-  1}^{+  1}$  &  $ 0.1_{-  0.0}^{+ 0.0}$  &  $ 1.29_{- 0.10}^{+ 0.10}$  &  $25_{- 1}^{+13}$  &  $ 0.9_{-  0.1}^{+ 0.1}$  &  $  2.6_{-   2.0}^{+  0.3}$  &  $  2.2\pm  0.7$  &  $\ldots$  \\
FC24  &  1.610  &  23.7  &  $  4_{-  2}^{+  2}$  &  $ 2.6_{-  0.4}^{+ 2.1}$  &  $ 0.00_{- 0.51}^{+ 0.22}$  &  $39_{- 8}^{+14}$  &  $ 3.0_{-  2.2}^{+ 2.7}$  &  $  1.3_{-   1.1}^{+  3.8}$  &  $  5.2\pm  1.7$  &   $\ldots$ \\
FC25  &  2.330  &  24.2  &  $ 17_{-  5}^{+  9}$  &  $14.8_{-  2.9}^{+ 2.6}$  &  $-0.21_{- 0.22}^{+ 0.22}$  &  $36_{- 5}^{+ 9}$  &  $21.4_{-  6.3}^{+10.5}$  &  $ 18.8_{-  10.8}^{+ 14.3}$  &  $ 22.4\pm  7.4$  &  $\ldots$  \\
FC26  &  0.997  &  22.0  &  $  6_{-  1}^{+  2}$  &  $ 6.5_{-  0.8}^{+ 1.0}$  &  $ 0.02_{- 0.14}^{+ 0.18}$  &  $35_{- 4}^{+12}$  &  $ 4.7_{-  0.9}^{+ 2.3}$  &  $  4.0_{-   2.2}^{+  2.7}$  &  $  5.3\pm  1.6$  &  $\ldots$  \\
\hline 
\end{tabular}\\
\flushleft \noindent {\bf Notes.} For all \textsc{magphys}-derived parameters, an additional 0.1 dex error has been added in quadrature to the original MAGPHYS uncertainties. This is particularly important in cases with excellent SED fits, where low uncertainty values are produced owing to the discrete sampling spacing of the underlying SED templates. Column (1): source ID, ASPECS-LP.1mm.Faint.xx. Column (2): best redshift available (see Table \ref{tab:3} for redshift references). Column (3): AB magnitude in the F160W HST band. Columns (4) to (8): SFR, stellar mass, normalized specific SFR ($\Delta_{\rm MS}$), dust temperature ($T_{\rm d}$) and IR luminosity ($L_{\rm IR}$), derived from \textsc{magphys} SED fitting. Column (9): molecular gas mass derived from the dust mass delivered by MAGPHYS and a gas-to-dust ratio $\delta_{\rm GDR}=200$. Column (10): molecular gas mass obtained from the 1.2 mm flux and the calibrations from \citet{scoville14}. Column (11): molecular gas mass obtained from the CO line emission detected by ASPECS 3 mm spectroscopy. To convert the CO $J>1$ to the ground transition, we use the average line ratios derived for the ASPECS sample itself \citep[][]{boogaard20}, with $r_{21}=0.83\pm0.12$, $r_{31}=0.58\pm0.10$ and $r_{41}=0.30\pm0.08$ for galaxies at $z<1.7$ and $r_{21}=1.02\pm0.18$, $r_{31}=0.92\pm0.17$ and $r_{41}=0.76\pm0.16$ for galaxies at $z>1.7$ (see Appendix \ref{app_a}). For consistency with previous ASPECS work, a fixed $\alpha_{\rm CO}=3.6$ $M_\odot$ (K km s$^{-1}$)$^{-1}$ is used. This represents a systematic underestimation of $<$0.1 dex with respect to values obtained when using a metallicity-dependent $\alpha_{\rm CO}$ scheme. \end{table*}

\label{lastpage}

\end{document}